\documentclass[aps,prb,twocolumn,superscriptaddress]{revtex4-1}

\usepackage{graphicx}  
\usepackage{subfigure}
\usepackage{dcolumn}   
\usepackage{bm}        
\usepackage{amssymb}   
\usepackage{comment}
\usepackage{hyperref}
\usepackage{color}
\usepackage{amsmath}
\usepackage{mathrsfs}
\usepackage{mathcomp}
\usepackage{textcomp}
\usepackage{dsfont}
\usepackage{esint}
\usepackage{braket}
\usepackage{textgreek}
\usepackage{lipsum}

\begin{document}

\title{Electric transport in three-dimensional Skyrmion/monopole crystal}
\author{Xiao-Xiao Zhang}
\email{zhang@appi.t.u-tokyo.ac.jp}
\affiliation{Department of Applied Physics, The University of Tokyo, 7-3-1 Hongo, Bunkyo-ku, Tokyo 113-8656, Japan}
\author{Andrey S. Mishchenko}
\affiliation{RIKEN Center for Emergent Matter Science (CEMS), 2-1 Hirosawa, Wako, Saitama 351-0198, Japan}
\author{Giulio De Filippis}
\affiliation{SPIN-CNR and Dipartimento di Fisica, Universit\`a di Napoli Federico II, I-80126 Napoli, Italy}
\author{Naoto Nagaosa}
\email{nagaosa@ap.t.u-tokyo.ac.jp}
\affiliation{Department of Applied Physics, The University of Tokyo, 7-3-1 Hongo, Bunkyo-ku, Tokyo 113-8656, Japan}
\affiliation{RIKEN Center for Emergent Matter Science (CEMS), 2-1 Hirosawa, Wako, Saitama 351-0198, Japan}


\newcommand\dd{\mathrm{d}}
\newcommand\ii{\mathrm{i}}
\newcommand\ee{\mathrm{e}}
\makeatletter
\def\ExtendSymbol#1#2#3#4#5{\ext@arrow 0099{\arrowfill@#1#2#3}{#4}{#5}}
\newcommand\LongEqual[2][]{\ExtendSymbol{=}{=}{=}{#1}{#2}}
\newcommand\LongArrow[2][]{\ExtendSymbol{-}{-}{\rightarrow}{#1}{#2}}
\newcommand{\cev}[1]{\reflectbox{\ensuremath{\vec{\reflectbox{\ensuremath{#1}}}}}}
\newcommand{\myred}[1]{\textcolor{red}{#1}} 
\newcommand{\red}[1]{} 
\makeatother

\begin{abstract}
We study theoretically the transport properties of a three-dimensional spin texture made from three orthogonal helices, which is essentially a lattice of monopole-antimonopole pairs connected by Skyrmion strings. This spin structure is proposed for MnGe based on the neutron scattering experiment as well as the Lorentz transmission electron microscopy observation. Equipped with a sophisticated spectral analysis method, we adopt finite temperature Green's function technique to calculate the longitudinal dc electric transport in such system. We consider conduction electrons interacting with spin waves of the topologically nontrivial spin texture, wherein fluctuations of monopolar emergent magnetic field enter. We study in detail the behavior of electric resistivity under the influence of temperature, external magnetic field and a characteristic monopole motion, especially a novel magnetoresistivity effect describing the latest experimental observations in MnGe, wherein a topological phase transition signifying strong correlation is identified.

\end{abstract}

\maketitle

\section{Introduction}
Although Skyrmion, mathematically being a topologically nontrivial soliton solution of an $\mathrm{O}(3)$ nonlinear sigma model\cite{Rajaraman,Ng}, is originally proposed as a hadron model decades ago\cite{Skyrme}, its revival came with condensed matter systems in the end, including liquid crystal\cite{LQ}, Bose-Einstein condensate\cite{BEC1,BEC2}, 2D electron gas of integer quantum Hall effect\cite{QHE}, etc. For example, the low-energy theory of the integer quantum Hall system possesses a similar structure of a quantum ferromagnet whose elementary excitations are Skyrmion-like. This implicitly raised the question whether it is achievable in real magnetic systems. Pioneering predictions\cite{Skprediction1,Skprediction2} studied the mean-field theory of easy-axis ferromagnets with chiral spin-orbit interaction. Afterwards, magnetic Skyrmions were finally realized not only in $P2_13$ space group chiral magnets of metals\cite{MnSi1,MnSi2,FeGe}, semiconductors\cite{FeCoSi1,FeCoSi2}, and multiferroic insulator\cite{CuOSeO}, but also in a one-atomic-layer Fe thin film on a Ir substrate as tiny nano-Skyrmions\cite{Fe_film}. Affluent new phenomena have been experimentally discovered and theoretically investigated, including the topological Hall effect (THE)\cite{THE1,THE2}, the Skyrmion Hall effect\cite{Zang,SkHall}, the non-Fermi liquid behavior in a wide temperature regime\cite{NFL}, the ultralow-current-driven motion\cite{ultralow1,ultralow2}, the quantized topological Hall effect\cite{Hamamoto}, and so on, paving the way for 'Skyrmionics' and even applications in magnetic information storage and processing\cite{memofunc0,memofunc1,memofunc2,memofunc3}.

Not only can isolated Skyrmions be excited by means such as local heating\cite{heating} and applying electric currents\cite{Iwasaki2,memofunc0,creation0}, but more common Skyrmion crystal (SkX) has also been observed in $k$-space by neutron scattering\cite{MnSi1,FeCoSi2} and in real space by Lorentz transmission electron microscopy (LTEM)\cite{MnSi2,FeCoSi1,FeGe} and magnetic force microscopy \cite{SkMerge}. Contrary to the thin film realization, SkX only exists within a narrow region of temperature and external magnetic field in the bulk material. However, a metastable SkX state can extend over a wide temperature region\cite{FeGe}, which is procured by cooling without changing the magnetic field. Typically in the bulk, Skyrmion tubes with translational symmetry along the cylindrical axis can form. One is then naturally urged to contemplate the intriguing possibility of the coalescence of Skyrmion tubes at certain singular points in three dimensions (3D). These singularities must be hedgehog spin textures that can stepwise alter the topological number, reminding us of a more ordinary realization of this type of mapping, the Dirac monopole. In fact, those singular points can be identified as a variant of Dirac magnetic monopole in terms of the so-called emergent electromagnetic field (EEMF)\cite{EEMF0,EEMF1,EEMF2,Zang}, which has been confirmed experimentally\cite{SkMerge}.

Since then, there have been several theoretical works in regards to emergent magnetic monopoles driven by the foregoing energetic instability in the bulk. The evolution of Skyrmion number under external magnetic field was studied\cite{monopolegrowth} in a system similar to the experiment\cite{SkMerge}. The effect of the coalescence on electric current was calculated in a postulated two-Skyrmion-merging model based on a soliton solution of the nonlinear sigma model\cite{monopole3}. Making use of micromagnetic simulations based on the stochastic Landau-Lifshitz-Gilbert equation, people studied the dynamics and energetics of monopoles created by thermal fluctuations \cite{monopole1} and the dynamics of monopoles and Dirac string-like objects under an electric current drive\cite{monopole2}. Monopoles acquiring electric charges via the $\theta\vec{E}\cdot\vec{B}$ term in the Witten effect can also be driven by an electric field to induce a SkX phase in an insulating helimagnet\cite{Watanabemonopole}. These studies are in a way concerned with accidental monopole defects in the Skyrmion tube background. Here comes a further question -- Can we realize an arrangement of emergent monopoles in a deterministic way? This was partly answered by a theoretical prediction in a 3D SkX phase, i.e., there resides a simultaneous monopole crystal\cite{SkX1,SkX2}.

In a bulk polycrystal of $B20$-type MnGe, a much larger and magnetic-field-dependent distinctive THE signal, in contrast to the ones for other $B20$-type Skyrmion-hosting chiral magnets like MnSi, was detected\cite{Kanazawa1} and then tentatively explained\cite{Kanazawa2} by the foregoing 3D SkX model composed of tilted Skyrmion strings and a periodic array of points where spin moment $\vec{S}=\vec{0}$. Small-angle neutron scattering\cite{Kanazawa2} further confirmed the cubic symmetry of the magnetic texture therein. On the other hand, difficulty in the single-crystal synthesis and sub-nanometer resolution LTEM obstructed real-space analysis of this material until a very recent study on thin film MnGe (thickness $\sim 30nm$) clearly revealed the magnetic moment configuration and the underlying atomic crystal lattice through high-resolution LTEM\cite{Kanazawa3}. Despite a minor difference between intensities of different spirals possibly due to the thin film setting, both an anomalous temperature dependence of the SkX period and a magnetic texture comprising three orthogonal spin spirals (see the model in section \ref{SkX}) were undoubtedly confirmed. 

At those vanishing points of spin moment, the directional vector \[\vec{n}=\vec{S}/|\vec{S}|\] becomes singular. There is a crucial difference between these two viewpoints. The former, $\vec{S}(\vec{r})$, mathematically being a mapping to a 3-ball $B^3$, is trivial in the sense that any configuration can be smoothly deformed to $\vec{S}(\vec{r})=\vec{0}$. And the latter, which is the orientational field $\vec{n}(\vec{r})$, is topologically characterized by the homotopy group of a 2-sphere $S^2$. This is the more appropriate way to explain localized spins' influence on conduction electrons in a strongly correlated system because of the prohibition of vast variation in the length of spin moments. We associate the MnGe in the experiments with this strong correlation picture and indeed, besides a reduced bandwidth, its saturated magnetization is several times larger than that of MnSi. Accordingly, we identify the singular points as pairs of magnetic monopole and antimonopole in terms of EEMF (See Sec.~\ref{SkX}, Sec.~\ref{resistivity2} and our paper\cite{Nii2}). And since the electron correlation and spin-orbit interaction are enhanced, the 3D spin texture containing the Skyrmion strings is formed even without the external magnetic field. 

One of the significant physical aspects in such a system turns out to be that thermally excited spin waves should couple with itinerant electrons and hence affect the resistivity massively. Especially, we expect novel phenomena originated from the resultant fluctuation of the nontrivial monopolar magnetic field. To this end, we adopted finite temperature Green's function technique to calculate the correlation functions for attaining longitudinal dc resistivity, since the transverse anomalous behavior has been well described by the THE. The dependence on both temperature $T$ and magnetization $m_z$ along the external magnetic field were considered. To compare with and support our resistivity calculation, a study of magnetic susceptibility was conducted as well. 
Fortunately, our magnetoresistivity predictions have been confirmed by the latest experimental advances\cite{Nii2}. Readers are referred to that publication for a detailed comparison between experiment and theory.

This paper is organized as follows. In Sec.~\ref{models}, we introduce the physical models, the effective Hamiltonian for conduction electrons, the 3D SkX, and spin waves in SkX. Then we present a derivation of our calculation formalism for resistivity in Sec.~\ref{methods} and describe and discuss the results of asymptotic analysis and magnetoresistivity in Sec.~\ref{resistivity}. In Sec.~\ref{conclusion}, we conclude and comment on this work. Some development and calculation of the model and formalism are organized into several appendices.

\section{Theoretical Models}\label{models}
\subsection{Effective model of itinerant electrons in SkX}\label{EEMF}
A powerful and elegant theoretical framework, EEMF, was invented based on an adiabatic approximation for the real space description of Berry phases produced by the non-collinear spin textures\cite{EEMF0,EEMF1,EEMF2,Zang}. This is valid when the size of a Skyrmion is much larger than the Fermi wavelength and in between the no-spin-flip mean free path and spin-flip mean free path and the time to traverse a Skyrmion is much larger than the inverse of band-splitting
. When strong coupling with itinerant electrons is present, the constraint drawn by the localized spins produces the EEMF, which elegantly explains the topological Hall effect (THE)\cite{THE1,THE2}. We also mention the exceptional largeness of the emergent magnetic fields (about $1000$T, $30$T, $1$T in MnGe, MnSi, and FeGe, respectively) that makes the external magnetic field typically of $0.1$T negligible. This is easily estimated from the magnetic length data of the SkXs\cite{Kanazawa1,THE1,FeGe}.
\begin{figure*}
\begin{center}
  \scalebox{0.5}{\includegraphics{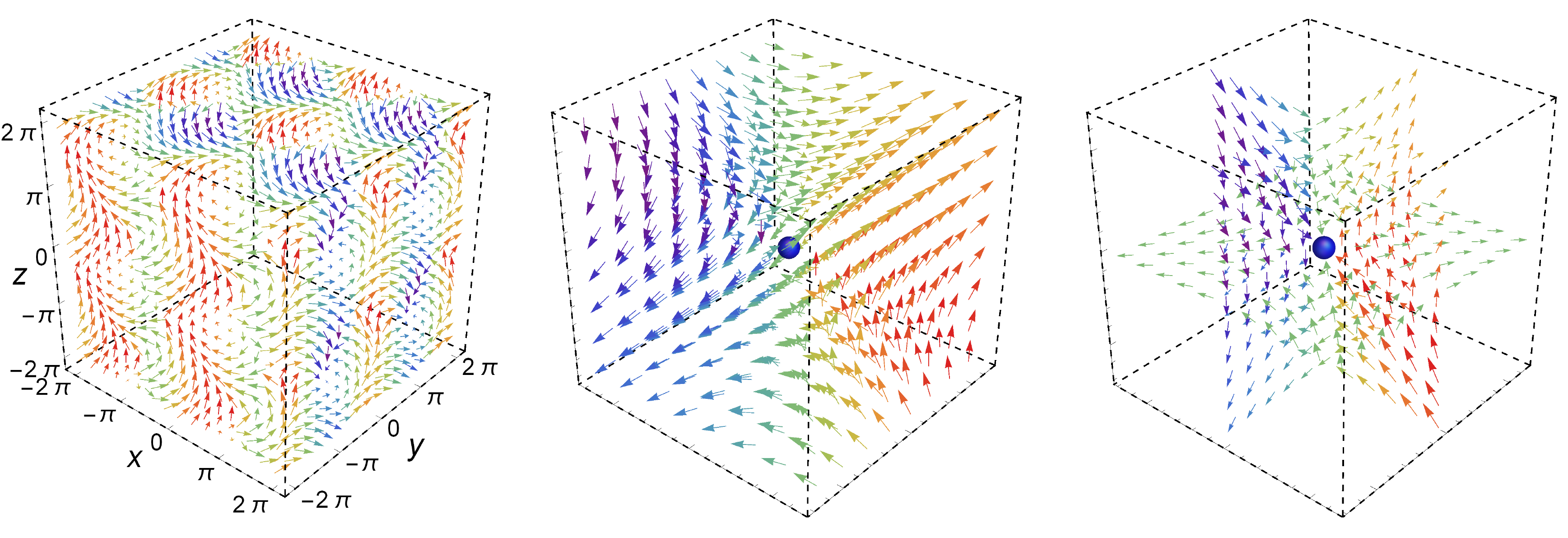}}
  \caption{(Color online) (LEFT) Spin texture $\vec{n}(\vec{r})$ at the boundary of a $2\times2\times2$ unit cell of the SkX/monopole crystal. 
  (MIDDLE, RIGHT) Spin texture $\vec{n}(\vec{r})$ around an antimonopole (blue point) in the SkX/monopole crystal explores all the possible directions wrapping up a sphere. (RIGHT) We show for clearness only the in-plane component of the spin texture on three mutually orthogonal planes cutting the antimonopole. Uniform magnetization $m_z=0$. Rainbow colors encode $n_z$ as red (blue) means more polarized up (down).
  }\label{fig_SkX}
\end{center}
\end{figure*}

We consider a double-exchange model comprising both itinerant electrons and magnetic textures in 3D, in which conduction electrons are coupled with localized spins ferromagnetically via an $sd$-type Hund's rule coupling\cite{Review}
\begin{equation}\label{L_ele-spin1}
\begin{split}
&\tilde{\mathcal{L}}_{\textrm{ele--spin}} \\
&= \Psi^\dag(\ii\hbar\partial_0+\varepsilon_F)\Psi + \frac{1}{2m} (\hat{\vec{p}}\Psi)^\dag \cdot (\hat{\vec{p}}\Psi) + \frac{J_H}{2}S \vec{n}\cdot\Psi^\dag\vec{\sigma}\Psi,
\end{split}
\end{equation}
wherein $\Psi = (\Psi_1,\Psi_2)^\mathrm{T}$ and $\vec{\sigma} = (\sigma_x,\sigma_y,\sigma_z)^\mathrm{T}$ are the spinor field and the 3-vector of spin-$\frac{1}{2}$ Pauli matrices of conduction electrons respectively. When the Hund's rule coupling strength $J_H$ is strong enough, the antiparallel spinor component of minority population has very large energy and spin-flip transition to this state driven by off-diagonal matrix elements in Hamiltonian scarcely occurs. Thus one is able to make an adiabatic approximation to drop that component and corresponding off-diagonal terms
, which defines the $\mathrm{U}(1)$ gauge fields. Therefore, as shown in Appendix~\ref{App:H_eff}, the constraint drawn by the background spin texture yields emergent electromagnetic fields (EEMF, signified by lowercase) seen by itinerant electrons
\begin{eqnarray}
b_i &= (\nabla\times\vec{a})_i=\frac{1}{2}\frac{\hbar}{q_\mathrm{e}}\epsilon_{ijk}\vec{n}\cdot(\partial^j \vec{n}\times\partial^k \vec{n})\label{eq_b}\\
e_i &= (-\partial_0\vec{a} - \nabla a_0)_i = \frac{\hbar}{q_\mathrm{e}} \vec{n}\cdot(\partial_i \vec{n}\times\partial_0 \vec{n})\label{eq_e}
\end{eqnarray}
and the low-energy effective Hamiltonian
\begin{equation}\label{eq_Heff}
\mathcal{H}_{\textrm{eff}} = \frac{1}{2m}(\hat{\vec{p}}-q_\mathrm{e}\vec{a})^2 + V(\vec{r},t),
\end{equation}
where the potential field $V$ and the gauge potential $\vec{a}$ are given in Appendix~\ref{App:H_eff}.
Note that the emergent gauge charge $q_\mathrm{e}$, which should not be confused with the elementary electric charge $e$, does not really enter Eq.~\eqref{eq_Heff} simply because $\vec{a}$ has a $\frac{1}{q_\mathrm{e}}$ factor by definition.

\subsection{Three-dimensional Skyrmion crystal/monopole crystal}\label{SkX}
A magnetic Skyrmion
is defined as a unit-norm mapping $\vec{n}(\vec{r})\equiv \frac{\vec{S}}{|\vec{S}|}$ from a 2D compact base manifold (real space) to the target manifold (directional space), which wraps around the latter certain times, rigorously characterized by the homotopy group $\pi_2(S^2)=\mathds{Z}$.
Explicitly, this winding number, or the topological Skyrmion number for a 2D compact manifold parametrized by $(u,v)$ reads\cite{Rajaraman,CMFT,Ng}
\begin{equation}\label{eq_SkN1}
N_\mathrm{Sk} = \frac{1}{4\pi} \iint{ \dd u \dd v \vec{n}\cdot(\frac{\partial  \vec{n}}{\partial u} \times \frac{\partial  \vec{n}}{\partial v}) }.
\end{equation}
In a 3D chiral magnet, the Skyrmion number Eq.~\eqref{eq_SkN1} for a (compactified) region in $x_\beta,x_\gamma$-plane consequently becomes a function of $x_\alpha$ coordinate: 
\begin{equation}\label{eq_SkN2}
N_\mathrm{Sk}^\alpha(x_\alpha) = \frac{1}{4\pi} \epsilon^{\alpha\beta\gamma} \iint{ \dd x_\beta \dd x_\gamma \vec{n}\cdot(\partial_\beta \vec{n}\times\partial_\gamma \vec{n}) }.
\end{equation}
This corresponds to the observed 2D SkX and aforementioned columnar Skyrmion tubes in 3D. The latter can be viewed as piling up 2D SkXs.

In general, a periodic non-collinear or non-coplanar spin configuration can be viewed as a hybridized state of multiple, say, $N$ independent spiral spin textures\cite{SkX1} of wave vectors $\vec{k}_\alpha$ 
\begin{equation}\label{eq_SkX}
\vec{S}(\vec{r}) = \vec{m}+ \sum_{\alpha=1}^N{(\vec{M}_\alpha\ee^{\ii \vec{k}_\alpha\cdot\vec{r}} + \vec{M}_\alpha^*\ee^{-\ii \vec{k}_\alpha\cdot\vec{r}})}
\end{equation}
where $\vec{m}$ is the uniform magnetization in proportion to applied external magnetic field. Trivially, when $N=1$, i.e., there is no hybridization at all, one obtains the ordinary helical or conical state. On the other hand, topologically protected magnetic Skyrmions in chiral magnets can be well characterized by the $N>1$  scenario. To this end, one can retain solely the lowest order Fourier components and assume that all $\vec{k}_\alpha$'s ($\vec{M}_\alpha$'s) are equal in norm and without loss of generality, complex phases in $\vec{M}_\alpha = |\vec{M}_\alpha|\ee^{\ii\phi_\alpha}$'s are locked to be the same. This description, for instance, can give us a hexagonal SkX in 2D or a simple cubic one in 3D when $N=3$. The former for MnSi reads $\vec{k}_1=k(1,0,0)\,,\vec{k}_2=k(-\frac{1}{2},\frac{\sqrt{3}}{2},0)\,,\vec{k}_3=k(-\frac{1}{2},-\frac{\sqrt{3}}{2},0)$ and $\vec{M}_1=(\hat{z}+\ii\hat{y})/2\,,\vec{M}_2=(\hat{z}-\ii\frac{\sqrt{3}}{2}\hat{x}-\ii\frac{1}{2}\hat{y})/2\,,\vec{M}_3=(\hat{z}+\ii\frac{\sqrt{3}}{2}\hat{x}-\ii\frac{1}{2}\hat{y})/2.$ The latter for MnGe reads $\vec{k}_1=(k,0,0)\,,\vec{k}_2=(0,k,0)\,,\vec{k}_3=(0,0,k)$ and $\vec{M}_1=(\hat{y}-\ii\hat{z})/2\,,\vec{M}_2=(\hat{z}-\ii\hat{x})/2\,,\vec{M}_3=(\hat{x}-\ii\hat{y})/2$. Henceforward, we study the latter and  set $\left|\vec{k}_\alpha\right|=1\,,\alpha=1,2,3$ and $\vec{m}=m_z$ for simplicity, which amounts to
\begin{equation}\label{texture}
\vec{S} (\vec{r})= (\sin y+\cos z,\cos x+\sin z,m_z+\sin x+\cos y).
\end{equation}
We show the corresponding spin texture $\vec{n}(\vec{r})$ in Fig.~\ref{fig_SkX}.

The conventional exchange interaction (EXI) originated from the Coulomb interaction and the Fermion statistics, usually yields ferromagnetic or antiferromagnetic order. Those helical, conical or multi-spiral states can be generated by various mechanisms\cite{Review}, e.g., frustrated exchange interactions, spin-orbit interactions, long-range magnetic dipolar interactions, magnetic anisotropy, and so on. An important example of the relativistic spin-orbit case is the Dzyaloshinskii-Moriya interaction (DMI)\cite{Dzyaloshinskii,Moriya,Fert&LevyDMI}. This work deals with B20-type material without inversion symmetry that can host DMI (including both MnSi and MnGe). The minimal Hamiltonian in $d$ spatial dimensions
\begin{equation}\label{H_SkX}
\begin{split}
&\mathcal{H}_\mathrm{SkX}= \\
&\int{ \dd^d\vec{r}  \left[ \frac{J\hbar^2}{a_0^{d-2}} \left(\nabla\vec{S}\right)^2  + \frac{D\hbar^2}{a_0^{d-1}} \vec{S}\cdot\left(\nabla\times\vec{S}\right)  - \frac{\hbar}{a_0^{d}} \mu \vec{S}\cdot\vec{B} \right]},
\end{split}
\end{equation}
includes the EXI, the Bloch-type DMI, and the Zeeman energy, wherein and henceforth dimensionless $\vec{S}$ of the spatial configuration of spin moments is defined without the $\hbar$ factor. The ratio of the magnitude of the DMI to the EXI, $\frac{D}{J}$, is supposed to be small enough to justify the continuum approximation to be used, since $a_0=\frac{D}{J}a_\mathrm{SkX}$, where $a_0$ ($a_\mathrm{SkX}$) is the microscopic lattice constant of the material (the size of the magnetic unit cell or the period of the incommensurate SkX). From the scalar triple product form of DMI, one can see it energetically favors circularly polarized spiral modes, i.e., the spin plane remains perpendicular to spin density wave vector. Such a configuration, remaining spiral texture inside (DM energy gain) and ferromagnetism outside (EX and Zeeman energy gain), is a compromise between different magnetic energies.

From Hamiltonian Eq.~\eqref{H_SkX}, one can estimate the characteristic length and energy scales in the system by plugging in $\vec{S}\propto \ee^{\ii\vec{k}\cdot\vec{r}}$ and minimize the energy in $k$-space, which results in $k\sim\frac{D}{Ja_0}$ hence magnetic energy density $\sim\frac{D^2}{J{a_0}^2}$.  This is why the (critical) magnetic fields of different phases (and their differences) are of the order $\frac{D^2}{J}$. Notwithstanding, the area of a Skyrmion of the order $\left(\frac{J}{D}\right)^2{a_0}^2$ compensates and makes the melting temperature of a SkX modestly as high as $J$, which is the energy scale to destroy a Skyrmion (SkX) by various means. As aforementioned, Skyrmion coalescence or bisection is ascribed to singular points in the spin texture, around which hedgehog/anti-hedgehog spin configuration (Fig.~\ref{fig_SkX}) with an energy of the order $J$ is indeed formed\cite{monopole1} and can naturally create or annihilate a Skyrmion.

\begin{figure*}
\begin{center}
  \scalebox{0.5}{\includegraphics{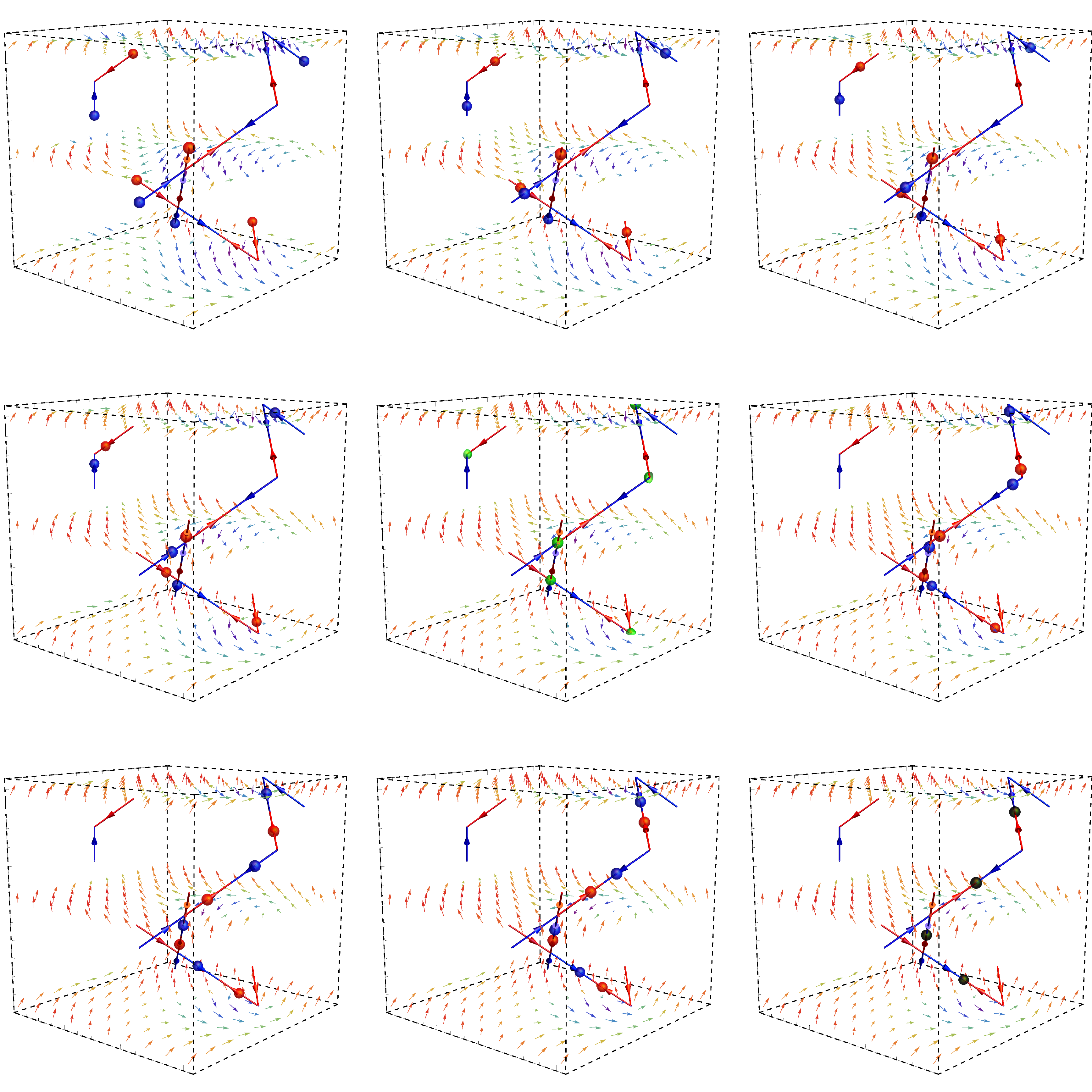}}
  \caption{(Color online) Spin textures $\vec{n}(\vec{r})$ on three successive planes of $z=0,\pi,2\pi$ and evolution of four pairs of monopole (red point/trajectory) and antimonopole (blue point/trajectory) in a unit cell of the SkX/monopole crystal. Monopoles and antimonopoles collide at green points while annihilate at black points. From left to right and up to down: uniform magnetization $m_z=0,0.3,0.5,0.75,1.0,1.2,1.37,\sqrt{2}$. Coordinates and vector colors are the same as Fig.~\ref{fig_SkX}.
  }\label{fig_mpEvo}
\end{center}
\end{figure*}

Further, the most prominent feature of the SkX in MnGe is that it contains not only Skyrmion strings but also a periodic array of singularities, identified as pairs of magnetic monopole and antimonopole in terms of EEMF, whose magnetic flux quantization can be shown by applying the generic $N_\mathrm{Sk}$ formula Eq.  \eqref{eq_SkN1} and Eq.~\eqref{eq_b} to an $S^2$ base manifold: 
\begin{equation}
\begin{split}
\oiint{\dd \vec{S}\cdot\vec{b}} 
\red{= \oiint{\dd S_\alpha b^\alpha} }
= \frac{\hbar}{2q_\mathrm{e}} \epsilon^{\alpha\beta\gamma} \iint{ \dd S_\alpha \vec{n}\cdot(\partial_\beta \vec{n}\times\partial_\gamma \vec{n}) } = \mathcal{Z}\phi_0
\end{split}
\end{equation} wherein $\mathcal{Z}\in\mathds{Z}$ and $\phi_0$ 
 is the magnetic flux quantum. We also analytically confirmed this flux quantization in this simple cubic monopole crystal, where each emergent monopole has magnetic flux $\pm\frac{h}{q_\mathrm{e}}$. A detailed inspection of the (anti)monopoles' motion under magnetization process is presented in Fig.~\ref{fig_mpEvo} and in Sec.~\ref{fluctuation}. The readers are also referred to our paper\cite{Nii2} for some alternative discussion.

\subsection{Low-energy spin-wave theory of SkX}\label{spinwave}
We study a low-energy spin-wave theory for the localized spins. This will affect electrons' motion via the vector potential $\vec{a}$ and potential field $V$ in Eq.~\eqref{eq_Heff} since spin waves in SkX render the Berry phase hence the EEMF produced by the spin texture fluctuating all along.

For quantum spins $\vec{S}$ in a spin helix along, say, the $z$-axis, one can use the spherical angle to construct the action (setting $\hbar=1$ henceforth)
\begin{equation}\label{spinfield_action}
\mathcal{S} = \int_0^\beta{ \dd\tau \int{\dd^d\vec{r} (-\ii) S_z\partial_\tau\phi_z} }  + \int_0^\beta{ \dd\tau H(\vec{S}(\tau))}
\end{equation}
where $(\phi_z,S_z)$ are a canonical conjugate pair of fields. Note that $\phi_z$ is the azimuth with respect to the generic $z$-direction, which simply denotes the direction of the rotation axis of a certain spin spiral, being arbitrary actually. These are constructed from scratch in Appendix~\ref{App:spin_action}.
Phenomenologically, we include two quadratic terms $\int{ \dd^3\vec{r} \left[\chi{S_z}^2 + \rho(\nabla\phi_z)^2\right]}$ into $H(\vec{S})$, penalizing fluctuations due to the rigidity gained after spontaneous symmetry breaking (formation of helical texture).
There are still other possible terms like $(\nabla S_z)^2$. Nonetheless, the two we include are the energetically most relevant ones allowed by symmetry and are sufficient to characterize the physics in the interested low-energy regime.
We can then march on to the scenario of plural independent spirals ($N>1$). When $N=3$ that accounts for either MnSi or MnGe, one has three conjugate pairs of fields $\left(\phi_i(\vec{r},\tau)\,,S_i(\vec{r},\tau)\right)\,,i=1,2,3$ and consequently three similar parts in the action.

Interestingly, because of the nontrivial real-space spin Berry phase, as shown in Appendix~\ref{App:spin_action}, the modes of these three spirals will mingle with each other as a result of finite Skyrmion density in space. 
Together with the EEMF Eq.~\eqref{eq_b}, this implies new canonical conjugate pairs and their commutation relations $[\hat{\phi}_i,\hat{\phi}_j] \propto \varepsilon_{ijk} b_k$, whereupon a very similar new crucial term must be added to the Lagrangian. Here we define $b_k$ as the component of the emergent magnetic field $\vec{b}$ that is parallel to $\vec{k}_i\times \vec{k}_j$. One thing to point out is that our action describes the Gaussian fluctuation of the fields $\left(\phi_i\,,S_i\right)$, that is actually $\left(\phi_i\,,m_i\right)$ in terms of Eq.~\eqref{eq_SkX}, away from their mean field values. And $\phi_i$ corresponds to the displacement field of the SkX along $i$-th direction. Without loss of generality, we set the static mean field value of any $\phi_i$ to $0$ and denote fluctuation in $m_i$ by $\delta m_i$. Combining Eq.~\eqref{spinfield_action}\red{\eqref{spin_H}} and our discussion above, we are ready to write down the low-energy spin-wave Lagrangian density for SkX
\begin{equation}\label{eq_Lspinwave0}
\begin{split}
&\mathcal{L}_\mathrm{SW} = \\
&\sum_{i} \left[ \ii \epsilon^{ijk}  A b_i \phi_j\dot{\phi}_k + B(-\ii)\delta m_i\dot{\phi}_i + \chi{\delta m_i}^2 + \rho (\nabla\phi_i)^2 \right],
\end{split}
\end{equation}
wherein $A = -2q_\mathrm{e} S\frac{1}{k_j k_k} \frac{1}{a_0^{d}}\,, B = \frac{1\red{\hbar}}{a_0^{d}}\,, \chi = \frac{D^2}{J a_0^{d}} \red{\frac{\hbar^2}{a_0^{d}}}\,, \rho = \frac{J}{a_0^{d-2}}\red{J\frac{\hbar^2}{a_0^{d-2}}}$ and $S$ and $b_i$ are substituted for by their spatial averages since the spin-wave fields are presumably slowly varying
. According to the helical configurations introduced in section \ref{SkX}, we have only two pairs of effective degrees of freedom $\phi_\alpha,\delta m_\alpha\,,\alpha=x,y$ in the Lagrangian density for MnSi due to the phase locking among the three helices. And for MnGe, it takes the form
\begin{equation}\label{eq_Lspinwave}
\begin{split}
&\mathcal{L}_\mathrm{SW} = \sum_{\alpha=x,y,z} \left[ \ii \epsilon^{\alpha\beta\gamma}  A b_\alpha \phi_\beta\dot{\phi}_\gamma + B(-\ii)\delta m_\alpha\dot{\phi}_\alpha \right. \\
&\left.
+ \chi{\delta m_\alpha}^2 + \rho (\nabla\phi_\alpha)^2 
\vphantom{A b_\alpha \phi_\beta\dot{\phi}_\gamma}
\right],
\end{split}
\end{equation}
In terms of the properties of Skyrmion, especially Eq.~\eqref{eq_SkN2}\eqref{eq_SkN_ave}, discussed in section \ref{SkX} and to be discussed in section \ref{resistivity2}, we notice the spatial average $\braket{b_i} \propto \bar{N}_\mathrm{Sk}^i$, which, within SkX phases, is always nonzero for MnSi and is nonzero for MnGe when uniform magnetization appears. This first term is characteristic of Skyrmion's nontrivial topology. Note also that we inject concrete values to the phenomenological rigidity constants, anisotropy energy $\chi$ and stiffness $\rho$, according to the underlying Hamiltonian Eq.~\eqref{H_SkX}.

\section{Calculation Methods}\label{methods}
\subsection{Memory function method}\label{memofun}
From Sec.~\ref{SkX}, we understand that itinerant electrons described by the Hamiltonian Eq.~\eqref{eq_Heff} are actually moving in a background of magnetic monopoles. For a spin spiral, the $\phi$ field introduced in Sec.~\ref{spinwave} is the phase of the constituent spin density wave, signifying the shift of the SkX or more specifically, the deviation of monopoles away from their equilibrium points. The aftermath is that one has to introduce Dirac strings or patches of gauge choices for the vector potential, i.e., failure in constructing a global description of the gauge field in $\mathds{R}^3$ space because of the nontrivial U(1) bundle with monopole present\cite{CNYang1}. In order to overcome this and to retain gauge invariance in a succinct manner, rather than involving a cumbersome recovery of Ward-Takahashi identity\cite{PatrickLee}, we adopt the memory function approach\cite{MemoFunc,Mori1,Mori2}, calculating $\dot{j}\textrm{-}\dot{j}$ correlator, which is in a sense similar to a force-force correlator\cite{Mahan}. \red{From now on, we in general set $\hbar=1$ and consider a system of unit volume.}

According to the Kubo formula, the optical conductivity tensor can be expressed as
$\sigma(z,T) = \frac{\ii e^2}{z} (\frac{n_e}{m}+\frac{\Pi(z,T)}{V})$,
wherein $n_e$ is electron concentration, $z$ is the (complex) frequency, lying in the complex upper half-plane, and $\Pi$ is the 
$j\textrm{-}j$ correlation function
\begin{equation}
\Pi_{\alpha\mu}(z,T) = -\ii\int_{-\infty}^\infty { \dd t \; \ee^{\ii zt} \Theta(t-0) \braket{[j_\alpha(t), j_\mu(0)]} }.
\end{equation} 
 Note that $\sigma,\Pi$ and $M,\phi,A$ below are 3D rank-2 tensors and matrix inverse is understood accordingly. Henceforth, $\Braket{\quad}$ abbreviates the thermodynamic average at certain temperature $T$ and we omit the argument $T$ for simplicity, i.e., $\braket{*} \equiv \mathrm{Tr}[\ee^{-\beta(\mathcal{K}-\Omega)}*]$ with macroscopic thermodynamic potential $\Omega$ given by $\ee^{-\beta\Omega} = \mathrm{Tr}\;\ee^{-\beta \mathcal{K}}$ and $\mathcal{K} = \mathcal{H} - \mu \mathcal{N}$ in grand canonical ensemble. Here $\mathcal{H}$ and $\mathcal{N}$ are generic Hamiltonian operator and particle number operator respectively and inverse temperature $\beta=\frac{1}{k_BT}$. To facilitate resistivity calculation, one can express conductivity as 
\begin{equation}\label{eq_cond}
\sigma(z,T) = \frac{\ii e^2n_e/m}{z+M(z,T)},
\end{equation} 
using the memory function $M$. Within the lowest order of coupling this memory function with built-in resonance structure is approximated as\cite{MemoFunc} $M(z) = \frac{m}{n_eV\red{\hbar^3}} \frac{\phi(z) - \phi(0)}{z}$, using the finite temperature $\dot{j}\textrm{-}\dot{j}$ correlator defined in imaginary time by
\begin{equation}\label{eq_jdotjdot}
\phi_{\alpha\mu}(\tau) = -\braket{\mathrm{T}_\tau [j_\alpha,\mathcal{H}](\tau)[j_\mu,\mathcal{H}](0)},
\end{equation}
where $\tau\in[0,\beta\red{\hbar}]$. This corresponds to a partial sum of infinite diagrams including self-energy and vertex corrections. Then we relate them to the retarded Green's function $\phi^\mathrm{R}(\omega)$, given by  $\lim_{\eta\rightarrow 0^+} \phi(z\rightarrow\omega+\ii\eta)$, in which physical responses are embedded in. We henceforth consider solely longitudinal conductivity ($\alpha=\mu$, but for completeness and notational consistency we will keep using $\alpha$ and $\mu$), then $[j_\alpha,\mathcal{H}]^\dag = -[j_\mu,\mathcal{H}]$. Lehmann representation $\phi_{\alpha\mu}(\ii\omega_n) = \int_{-\infty}^\infty { \frac{\dd\omega}{2\pi} \frac{-2\Im\phi_{\alpha\mu}^\mathrm{R}(\omega)}{\ii\omega_n-\omega} }$ can be attained, whereupon $\phi_{\alpha\mu}(\tau)$ can be further expressed by an integration of a spectral function 
$A_{\alpha\mu}(\omega)=\Im \frac{(\phi^\mathrm{R}(\omega) - \phi^\mathrm{R}(0))_{\alpha\mu}}{\omega} = -\red{\hbar^2} \Im \omega\Pi_{\alpha\mu}^\mathrm{R}(\omega)$
  weighted by a positive kernal $\mathit{K}(\tau,\omega) = \frac{1}{\pi} \frac{\omega\ee^{-\omega\tau}}{1-\ee^{-\beta\red{\hbar}\omega}}$
\begin{equation}\label{eq_Fredholm}
\phi_{\alpha\mu}(\tau) = \int_{-\infty}^\infty { \dd \omega \mathit{K}(\tau,\omega) A_{\alpha\mu}(\omega) },
\end{equation}
from which $\phi_{\alpha\mu}(\tau)\in\mathds{R}$ becomes obvious. This kernal is conventional for optical conductivity calculation\cite{BayesianInference}.
We restrict ourselves to dc resistivity. Then 
$\rho_{\alpha\mu}(\omega=0) = {\sigma^{-1}}_{\alpha\mu}(\omega=0) = \frac{m}{e^2n_e} \Im(\omega+M_{\alpha\mu}(\omega))\vert_{\omega\rightarrow 0} \propto A_{\alpha\mu}(0)$.
Further, 
based on the cyclic property of the trace, we can also obtain a useful symmetry (see Appendix~\ref{App:symmetry})
\begin{equation}\label{eq_phiSymmetry}
\phi_{\alpha\mu}(\tau) = \phi_{\mu\alpha}(\beta\red{\hbar}-\tau).
\end{equation}

In this study, focusing on the lowest order contribution, we evaluate this $\phi$ function over a non-interacting system of electrons and bosonic fluctuations of the EEMF (spin waves), i.e., $\mathcal{H}_\textrm{non-int} = \mathcal{H}_{\textrm{{ele}}} + \mathcal{H}_{\textrm{{SW}}}$
. Notwithstanding, the coupling between electrons and spin waves is \textit{de facto} accounted for by plugging the $\mathcal{H}_\textrm{eff}$ Eq.~\eqref{eq_Heff} to Eq.~\eqref{eq_jdotjdot}. After a long derivation presented in Appendix~\ref{App:jdot}, we are able to obtain a simple form of the $\dot{j}\textrm{-}\dot{j}$ correlator Eq.~\eqref{eq_jdotjdot}
\begin{equation}\label{eq_phi1}
\begin{split}
&\phi_{\alpha\mu}(\tau) = \sum_{\vec{k}\vec{q}}  \mathcal{D}_{\textrm{e}}(\vec{k},\vec{q},\tau) \times 
\left\{ 
\frac{\red{\hbar^2}1}{m^2} q_\alpha q_\mu  \mathcal{D}_{VV}(\vec{q},\tau) \right. \\
&\left. - \left( \frac{\red{\hbar^2} q_\mathrm{e}}{2m^2} \right)^2  \varepsilon^{\alpha\beta\gamma}\varepsilon^{\mu\nu\sigma} (2k+q)_\beta (2k+q)_\sigma \mathcal{D}_{b_\gamma b_\nu}(\vec{q},\tau) \right. \\
 &\left. + \frac{\ii\red{\hbar^3} q_\mathrm{e}}{2m^2}  \varepsilon^{\alpha\beta\gamma} q_\alpha (2k+q)_\beta \left[ \mathcal{D}_{b_\gamma V}(\vec{q},\tau) - \mathcal{D}_{V b_\gamma}(\vec{q},\tau) \right]  
\vphantom{\frac{\hbar^2}{m^2}}
\right\}
\end{split}
\end{equation}
wherein we introduce several Matsubara Green's functions. For instance, $\mathcal{D}_{b_\gamma V}(\vec{q},\tau) = -\braket{\mathrm{T}_\tau b_\gamma(\vec{q},\tau)  V(-\vec{q},0)}$ is for the fluctuations of EEMF $b_\gamma$ and potential $V$, and $\mathcal{D}_{\textrm{e}}(\vec{k},\vec{q},\tau) = -\braket{ \mathrm{T}_\tau D_1(\tau)  D_2(0)  }$ is for the electrons, in which $D_1(\tau) = c_{\vec{k}_1+\vec{q}_1}^\dag(\tau) c_{\vec{k}_1}(\tau) , D_2(0) = c_{\vec{k}_2+\vec{q}_2}^\dag(0) c_{\vec{k}_2}(0) $. And similarly, we also have $\mathcal{D}_{b_\alpha b_\beta},\mathcal{D}_{V b_\alpha},\mathcal{D}_{VV}$. The reason why we prefer Matsubara Green's functions to directly calculating retarded Green's functions in real time lies in the fact that, in the latter, a Green's function not among the six conventional Green's functions\cite{Mahan} appears and requires clumsy Fourier transformations back and forth. 

Now the task turns out to be extracting $A_{\alpha\mu}$, i.e., solving Eq.~\eqref{eq_Fredholm}, a Fredholm integral equation of the first kind, once $\phi(\tau)$ is known (calculated) at imaginary times. This numerical \textit{analytic continuation} problem belongs to the category of ill-posed problems and is ubiquitous when dealing with quantum Monte Carlo data\cite{BayesianInference}. Among various techniques aiming at this, we adopted a hybrid of Stochastic Optimization\cite{Mishchenko1,Mishchenko2,Mishchenko3} and consistent constraints\cite{ConsistentConstraint} methods
, that does not depend on any \textit{a priori} expectation of the result, avoids artificial smoothening, and solves the discretized version of Eq.~\eqref{eq_Fredholm}.

\subsection{Electron Green's function}\label{electron_correlators}
The original material should have produced an electronic band structure of characteristic wavenumber $\frac{\pi}{a_0}$ if it was not for the formation of the SkX. Now it is reconstructed such that the first Brillouin zone is folded to have length $\frac{2\pi}{a_\mathrm{SkX}}$ (see Sec.~\ref{spinwave}). Considering the smoothness of the skyrmion structure, we did not take into account other possible modification due to the new band structure. Therefore, to describe the itinerant electrons, we used an oversimplified free electron model for $\mathcal{H}_\textrm{ele}$, that is a parabolic dispersion relation $\xi(\vec{k}) = \frac{|\vec{k}|^2}{2m} - \mu$. This should be regarded as a low-energy approximation around the new Fermi surface.

For free electrons, field operators are given by $c_{\vec{k}}(\tau) = \ee^{\tau(\mathcal{H}_{\textrm{ele}}-\mu\mathcal{N})} c_{\vec{k}} \ee^{-\tau(\mathcal{H}_{\textrm{ele}}-\mu\mathcal{N})} = \ee^{-\xi_{\vec{k}}\tau}c_{\vec{k}} \,, c_{\vec{k}}^\dag(\tau) = \ee^{\xi_{\vec{k}}\tau}c_{\vec{k}}^\dag$ and $c_{\vec{k}}(t) = \ee^{-\ii\xi_{\vec{k}}t}c_{\vec{k}} \,, \; c_{\vec{k}}^\dag(t) = \ee^{\ii\xi_{\vec{k}}t}c_{\vec{k}}^\dag$ in imaginary and real time, respectively. Applying Wick's theorem, we can calculate the previously defined electron's 4-point Green's function 
\begin{equation}
\begin{split}
\mathcal{D}_{\textrm{e}} (\tau) &= -\braket{ \mathrm{T}_\tau D_1(\tau)  D_2(0)  } \\
&= -\ee^{(\xi_{\vec{k}_1+\vec{q}_1} - \xi_{\vec{k}_1})\tau} \times 
\left( 
\delta_{\vec{q}_1,0}\delta_{\vec{q}_2,0} n_F(\xi_{\vec{k}_1}) n_F(\xi_{\vec{k}_2} )  \right. \\
&\left.
+ \delta_{\vec{k}_1+\vec{q}_1,\vec{k}_2}\delta_{\vec{k}_2+\vec{q}_2,\vec{k}_1} n_F(\xi_{\vec{k}_1+\vec{q}_1}) (1- n_F(\xi_{\vec{k}_1}))  
\vphantom{\delta_{\vec{k}_2+\vec{q}_2,\vec{k}_1}}
\right),
\end{split}
\end{equation}
in which the second term is physically relevant and can be directly obtained by analytically continuating $\mathcal{D}_{\textrm{e}} (\tau)$'s retarded counterpart $D_{\textrm{e}}^{\mathrm{R}}(t) = \ee^{\ii (\xi_{\vec{k}_1+\vec{q}_1} - \xi_{\vec{k}_1})t} \braket{[c_{\vec{k}_1+\vec{q}_1}^\dag c_{\vec{k}_1} , c_{\vec{k}_2+\vec{q}_2}^\dag c_{\vec{k}_2} ]} 
= \ee^{\ii (\xi_{\vec{k}_1+\vec{q}_1} - \xi_{\vec{k}_1})t} (n_F(\xi_{\vec{k}_1+\vec{q}_1}) - n_F(\xi_{\vec{k}_1}))$ and thereafter summing up Matsubara frequencies using bosonic weight $n_B(z)+1$. $n_{F}$ ($n_B$) is ordinary fermionic (bosonic) function. Thus, we will use 
\begin{equation}
\begin{split}
&\mathcal{D}_{\textrm{e}}(\vec{k},\vec{q},\tau) \\
&= - \ee^{ (\beta-\tau) ( \xi_{\vec{k}} - \xi_{\vec{k}+\vec{q}})} n_B(\xi_{\vec{k}} - \xi_{\vec{k}+\vec{q}}) (n_F(\xi_{\vec{k}+\vec{q}}) - n_F(\xi_{\vec{k}})),
\end{split}
\end{equation}
who has the symmetry 
\begin{equation}\label{eq_DeSymmetry}
\mathcal{D}_{\textrm{e}}(\vec{k},\vec{q},\beta-\tau) = \mathcal{D}_{\textrm{e}}(\vec{k}+\vec{q},-\vec{q},\tau).
\end{equation}


\subsection{Spin-wave Green's function}\label{spinwave_correlators}
We introduced in Eq.~\eqref{eq_phi1} the Green's functions of bosonic fluctuations of EEMF $b_\alpha$ or $V$. In conjunction with the Gaussian fluctuation spin-wave model in Sec.~\ref{spinwave}, they are treated up to the first order deviation away from the ground state. For instance, the $b_\alpha$ field is expanded as $b_\alpha (k_i r_i + \phi_i(\vec{r},\tau), \vec{m}(\vec{r},\tau)) = b_\alpha^{(0)}(k_i r_i,\vec{m}_0) + (\partial_{\varphi_\mu} b_\alpha)^{(0)}  \varphi_\mu(\vec{r},\tau)$, wherein superscript $(0)$ signifies the ground state value, $\varphi$ field is defined as $\varphi_\mu = (\vec{\phi},\delta\vec{m})^{\mathrm{T}}$ and only in this sense $\mu=1,\dots,6$. In momentum space, we have $b_\alpha(\vec{q},\tau) = b_\alpha^{(0)}(\vec{q}) + \sum_{\vec{l}} { (\partial_{\varphi_\mu} b_\alpha)^{(0)}(\vec{l}) \varphi_\mu(\vec{q}-\vec{l},\tau) } $, where $\vec{l}$ is an integer-valued 3-vector. This is a variant of the conventional convolution theorem since $(\partial_{\varphi_\mu} b_\alpha)^{(0)}$ is $2\pi$-periodic in real space in our study (see Sec.~\ref{SkX}). Therefore, representatively, we have 
\begin{eqnarray}\label{eq_SWcorr1}
\begin{split}
&\mathcal{D}_{b_\alpha V}(\vec{q},\ii\omega_n) = \int_0^\beta { \dd\tau \ee^{\ii\omega_n\tau} \mathcal{D}_{b_\alpha V}(\vec{q},\tau-0) } \\
&= \sum_{\vec{l}\vec{l}'}  \int_0^\beta  \dd\tau \ee^{\ii\omega_n\tau} (-1) \langle \mathrm{T}_\tau \; (\partial_{\varphi_\mu} b_\alpha)(\vec{l}) \varphi_\mu(\vec{q}-\vec{l},\tau)  \\
& \times(\partial_{\varphi_\nu} V)(\vec{l}') \varphi_\nu(-\vec{q}-\vec{l}',0) \rangle   \\
&= \sum_{\vec{l}} { (\partial_{\varphi_\mu} b_\alpha)(-\vec{l}) (\partial_{\varphi_\nu} V)(\vec{l}) \; \mathcal{G}_{\mu\nu}(\vec{q}+\vec{l},\ii\omega_n) }.
\end{split}
\end{eqnarray}
wherein we neglect the superscript $(0)$ and the newly defined spin-wave correlator 
\begin{equation}\label{eq_SWg}
\mathcal{G}_{\mu\nu} (\vec{q},\ii\omega_n) = \int_0^\beta { \dd\tau \ee^{\ii\omega_n\tau} (-1) \braket{\mathrm{T}_\tau \; \varphi_\mu(\vec{q},\tau)\varphi_\nu(-\vec{q},0)} }
\end{equation}
 will be discussed below. Note that the ground state static configuration does not contribute. We henceforth neglect all the $\vec{l}\neq \vec{0}$ terms, i.e., Umklapp scattering involving large momentum transfer, in the summation except the $\vec{l}=\vec{0}$ one since we mainly concern in the long wavelength limit, which results in 
\begin{equation}\label{eq_SWcorr2}
\mathcal{D}_{b_\alpha V}(\vec{q},\ii\omega_n) = (\partial_{\varphi_\mu} b_\alpha)(-\vec{l}=\vec{0}) (\partial_{\varphi_\nu} V)(\vec{l}=\vec{0}) \; \mathcal{G}_{\mu\nu}(\vec{q},\ii\omega_n),
\end{equation}
in which the zeroth harmonics $(\partial_{\varphi_\mu} b_\alpha)(-\vec{l}=\vec{0}) $ and $(\partial_{\varphi_\nu} V)(\vec{l}=\vec{0})$ are real.

The spin-wave model Eq.~\eqref{eq_Lspinwave} can be exactly solved in momentum space. We introduce Fourier transformation $\varphi_\mu(\vec{r},\tau) = (\beta V)^{-\frac{1}{2}} \sum_{\vec{q},\ii\omega} {\ee^{-\ii\omega\tau + \ii\vec{q}\cdot\vec{r}} \varphi_\mu(\vec{q},\ii\omega)  }$, wherein $\omega$ is bosonic Matsubara frequency $\omega_n= 2\pi n/\beta$ when periodic boundary condition $\varphi_\mu(\vec{r},0) = \varphi_\mu(\vec{r},\beta)$ is imposed. Then the action of Eq.~\eqref{eq_Lspinwave} is transformed to
\begin{equation}
\begin{split}
\mathcal{S}_\mathrm{SW} &= \int_0^\beta {\dd\tau  \int {\dd^d \vec{r}  \mathcal{L}_\mathrm{SW}  }  } \\
&= \sum_{\vec{q},\ii\omega} {\varphi^\mathrm{T}(\vec{q},\ii\omega) M(q,\ii\omega) \varphi(-\vec{q},-\ii\omega)},
\end{split}
\end{equation}
in which $6\times6$ matrix $M$ takes the block form 
$M = \left[
\begin{array}{cc}
M_1 & M_2 \\ 
-M_2 & M_3
\end{array} \right] $,
wherein $(M_1)_{jk} = \rho q^2 \delta_j^i\delta_{ik} - \omega A b_i \varepsilon_{jk}^i \,,(M_2)_{jk} = -\frac{1}{2}B\omega\delta_{jk} \,, (M_3)_{jk} = \chi\delta_{jk}$ and we use $q = |\vec{q}|$. This diagonalization in $\vec{q},\ii\omega$-space explains the $\delta$-function implicitly used in Eq.~\eqref{eq_SWcorr1}. Following the functional derivative approach\cite{Nagaosa,CMFT}, correlator $\mathcal{G}_{\mu\nu} (q,\ii\omega_n)$ is given by $-(M^{-1})_{\mu\nu}(q,\ii\omega_n)$. Certainly, all the 36 spin-wave correlators contain a common denominator 
\begin{equation*}\label{eq_denominator1}
\begin{split}
&\mathrm{Det}(M) = \frac{1}{64} \left(-B^2 z^2+4 q^2 \rho  \chi \right) \left(16 \chi ^2 \left(-A^2 z^2 b^2 + q^4 \rho ^2\right) \right. \\
&\left.
+ B^4 z^4 - 8 B^2 q^2 \rho  \chi  z^2
\vphantom{A^2 z^2 b^2}
\right) \\
&= -\frac{B^6}{64} (z-z_1) (z+z_1) (z-z_2) (z+z_2) (z-z_3) (z+z_3),
\end{split}
\end{equation*}
wherein 
\begin{equation}\label{eq_dispersion}
\begin{split}
z_1 = \sqrt{\rho  \chi}\frac{2 q }{B},
z_2 = \frac{2 A b \chi  + 2 \sqrt{A^2 b^2 \chi ^2+B^2 q^2 \rho  \chi }}{B^2},\\
z_3 = \frac{-2 A b \chi + 2 \sqrt{A^2 b^2 \chi ^2+B^2 q^2 \rho  \chi }}{B^2}
\end{split}
\end{equation}
 and we denote $b = \sqrt{b_x^2+b_y^2+b_z^2}$ henceforth. Note that we did substitution $\omega \rightarrow -\ii z$ for the sake of analytic continuation $\ii\omega_n \rightarrow \omega+\ii\delta$ to retarded Green's functions. Correlator matrix $\mathcal{G}(\vec{q},z)$ defined in Eq.~\eqref{eq_SWg}, which actually depends solely on $(|\vec{q}|,z)$, has a property that $\Re\mathcal{G}$ ($\Im\mathcal{G}$) is (anti-)symmetric when $z\in\mathds{R}$. Combining this with Eq.~\eqref{eq_SWcorr2}, one can readily prove that Eq.~\eqref{eq_phi1} is reduced to 
\begin{eqnarray}\label{eq_phi2}
\begin{split}
&\phi_{\alpha\alpha}(\tau)
 = \sum_{\vec{k}\vec{q}}  \mathcal{D}_{\textrm{e}}(\vec{k},\vec{q},\tau) \times 
\left\{ 
\frac{\red{\hbar^2}1}{m^2} q_\alpha q_\alpha  \Re \mathcal{D}_{VV}(q,\tau) \right. \\
&\left. - \left( \frac{\red{\hbar^2} q_\mathrm{e}}{2m^2} \right)^2  \varepsilon^{\alpha\beta\gamma}\varepsilon^{\alpha\nu\sigma} (2k+q)_\beta (2k+q)_\sigma \Re\mathcal{D}_{b_\gamma b_\nu}(q,\tau) \right. \\
&\left.
 - \frac{\red{\hbar^3} q_\mathrm{e}}{2m^2}  \varepsilon^{\alpha\beta\gamma} q_\alpha (2k+q)_\beta \Im \left[ \mathcal{D}_{b_\gamma V}(q,\tau) - \mathcal{D}_{V b_\gamma}(q,\tau) \right]  
\vphantom{\frac{\red{\hbar^2}1}{m^2}}
\right\},
\end{split}
\end{eqnarray}
which is consistent with $\phi_{\alpha\alpha}(\tau)\in\mathds{R}$ stated alongside Eq.~\eqref{eq_Fredholm} in Sec.~\ref{memofun}.
Finally, we still need to carry out Matsubara frequency summation to get $\mathcal{G}_{\mu\nu} (q,\tau)$. This and a reconfirmation of the symmetry Eq.~\eqref{eq_phiSymmetry} $\phi_{\alpha\alpha}(\tau) = \phi_{\alpha\alpha}(\beta-\tau)$ are sketched in Appendix~\ref{App:symmetry}.

\subsection{Numerical aspects}\label{numerics}
In our calculation, we set physical constants, electron mass $m$, elementary electric charge $e$, reduced Planck constant $\hbar$, and Boltzmann constant $k_B$ to unity, $D$, strength of DMI, to unity, $J$, strength of EXI, to $10D$, SkX lattice constant $a_\mathrm{SkX}$ to $2\pi$ since we set the magnetic wave vector $k$ to unity, and electron chemical potential $\mu$ to one third of the energy at the boundary of the first Brillouin zone of the parabolic electronic band.
The noninteracting spin-wave theory Eq.~\eqref{eq_Lspinwave} is in principle more suitable for the long wavelength limit, i.e., when the magnon momentum is small. A natural momentum cutoff for this continuum theory comes from the SkX lattice structure, which is taken to be $q_0 = \frac{\pi}{a_\mathrm{SkX}}$ in our calculation. Therefore, we multiply an exponential decay factor $\ee^{-\frac{q}{q_0}}$ to any spin-wave correlators. In addition, due to this lattice nature we also introduce an auxiliary small enough constant to the spin magnitude, i.e., $S=\sqrt{{S_x}^2+{S_y}^2+{S_z}^2+0.05^2}$, in all calculations, unless otherwise stated, so as to cut off the monopolar singularities.

The calculation of the Matsubara Green's function $\phi(\tau)$ is reduced to a 6D integral of two 3-vectors $\vec{k},\vec{q}$ (see Eq.~\eqref{eq_phi2}), for which one has to set the integration region. Due to the complex magnetic structure and the subtly high dimensionality, this integral appears to be tractable by neither conventional numerical integration methods suitable for lower dimensions nor well-established Monte Carlo integration methods like MISER or VEGAS\cite{NumericalRecipes,GSL}. We found and employed a deterministic recursive algorithm\cite{Genz1,Genz2,Cubature}, which can also handle the integrable singularity at $\vec{q} = \vec{0}$ in spin-wave correlators, to carry out the numerical integration in a 6D hypercube $[-p_{\textrm{max}}, p_{\textrm{max}}]^6$, wherein $p_{\textrm{max}}$, the numerically determined momentum boundary up to which the integral converges, monotonously increases with temperature as it should do. As for the $\vec{b}$-$\vec{b}$ correlation calculation in Sec.~\ref{resistivity2} that reduces to a 1D integration of the magnitude of spin-wave momentum $\vec{q}$, we used the CQUAD routine
\cite{GSL} to handle the integrable singularity.

Practically, for each temperature, we calculated $\phi$ at $N=320$ nonuniformly distributed $\tau$'s in $[0,\beta/2]$ (c.f. symmetry \eqref{eq_phiSymmetry}), wherein more is located among small $\tau$ region since $\phi(\tau)$ decreases rapidly therein, however, becomes flatter and flatter near $\tau=\beta/2$. This is realized by the $\tau$-generating formula $\tau_i = \frac{c_1 i + c_2 i^2}{c_1 (N-1) + c_2 (N-1)^2} \frac{\beta}{2}\,,i=0,1,2,\ldots,N-1$, wherein we set $c_1=40\,,c_2=1.0$. Resistivity errorbars were determined in the numerical analytic continuation algorithm. All integrations were performed with relative error no larger than $10^{-4}$ (inconstant due to the implementation of the algorithm). 

\section{Main Results}\label{resistivity}
\subsection{Asymptotic behavior at low energy}\label{resistivity1}
The three positive roots in Eq.~\eqref{eq_dispersion} of $\mathrm{Det}(M)=0$  actually give us the magnon spectrum. When $m_z \neq 0$, $z_1\,,z_2$ give rise to two gapless modes $\omega\propto Dq\,,\omega\propto Jq^2$ when $q$ is small while $z_3$ corresponds to an excitation with an energy gap proportional to $\frac{D^2}{J}$
. We should owe the noteworthy $Jq^2$ mode to the nonzero Skyrmion number that brings about the anomalous coupling, i.e., the $\phi$-quadratic term in Eq.~\eqref{eq_Lspinwave}. This coupling of different $\phi$ fields as canonical conjugate pairs mixes the transverse and longitudinal phonon-like lattice waves of a SkX, partially corresponding to the rotational motion of Skyrmions.
These three modes degenerate into the first gapless mode $\omega\propto Dq$ when $m_z=0$. Certainly, the gapless ones correspond to Nambu-Goldstone bosons that in a way restore the spontaneously broken symmetries.

\subsubsection{Temperature dependence of resistivity $\rho(T)$}
In the following, we estimate the relaxation time $\tau$ of conduction electron to attain the low-energy asymptotic behavior of resistivity $\rho(T)$ by a Fermi-golden-rule-type analysis.
When the temperature is low, an energy shell of the scale $k_BT$ around the Fermi surface is active for quasiparticle scattering and only magnons of $\hbar \omega (\vec{q}) \lesssim k_BT$ can be absorbed or emitted. One readily gains an order of magnitude estimation $\hbar\omega(\vec{q}) \sim k_BT$. The predominant magnon dispersion relation at small $\vec{q}$ takes the form $\omega(\vec{q}) = c q^n$: $n=1$ ($m_z = 0$), $n=2$ ($m_z \neq 0 $). For the coupling with Nambu-Goldstone boson fields in a SkX, 
the derivative form of the emergent Berry connection renders the vertex $|g_{\vec{k},\vec{k}+\vec{q}}|^2 \sim q^2$ for small momentum transfer\cite{WatanabeSkX,WatanabeNFL}. The relevant $q$-subsurface that massively contributes to magnon exchange is of a linear dimension proportional to $T^{1/n}$. Because of energy-momentum conservation $\hbar \omega (\vec{q}) = \pm (\xi_{\vec{k}+\vec{q}} - \xi_{\vec{q}})$, the permissible $q$-space is restrained from 3D to 2D, giving rise to an relevant area proportional to $T^{2/n}$ in a 2D $q$-subsurface. In addition, the scattering rate responsible for transport property should be $\tau_{\textrm{tr}}^{-1} \sim (1-\cos\theta)\tau^{-1}$ in the Boltzmann equation, wherein $\theta$ is the angle between $\vec{k}$ and $\vec{k}+\vec{q}$, and $1-\cos\theta = (q/k_F)^2/2 \propto T^{2/n}$ for small-$q$ scattering near the Fermi surface. Therefore, $1/\tau \sim T^{2/n} |g_{\vec{k},\vec{k}+\vec{q}}|^2 \sim T^{4/n}$ and hence, $\rho \sim 1/\tau_{\textrm{tr}} \sim T^{2/n}/\tau \sim T^{6/n}$. Then we attain $T^6$ and $T^3$ dependences of $\rho(T)$ for zero and nonzero $m_z$, respectively, controlled by applying external magnetic field. 
Both of the two cases satisfy the Landau criterion $\omega\tau \rightarrow \infty$ when $\omega \rightarrow 0$, which means electronic quasiparticle remains valid although we have such anomalous exponents. At very low temperature in this metallic material, along with possible residue resistivity due to quenched disorder, normal Fermi liquid contribution in proportion to $T^2$ arising from particle-hole excitation presumably dominates, to which our result had better be taken as a correction. 

\subsubsection{Frequency-dependent spin relaxation $\Im\chi(\omega)$}
\begin{figure}
  \scalebox{0.4}{\includegraphics{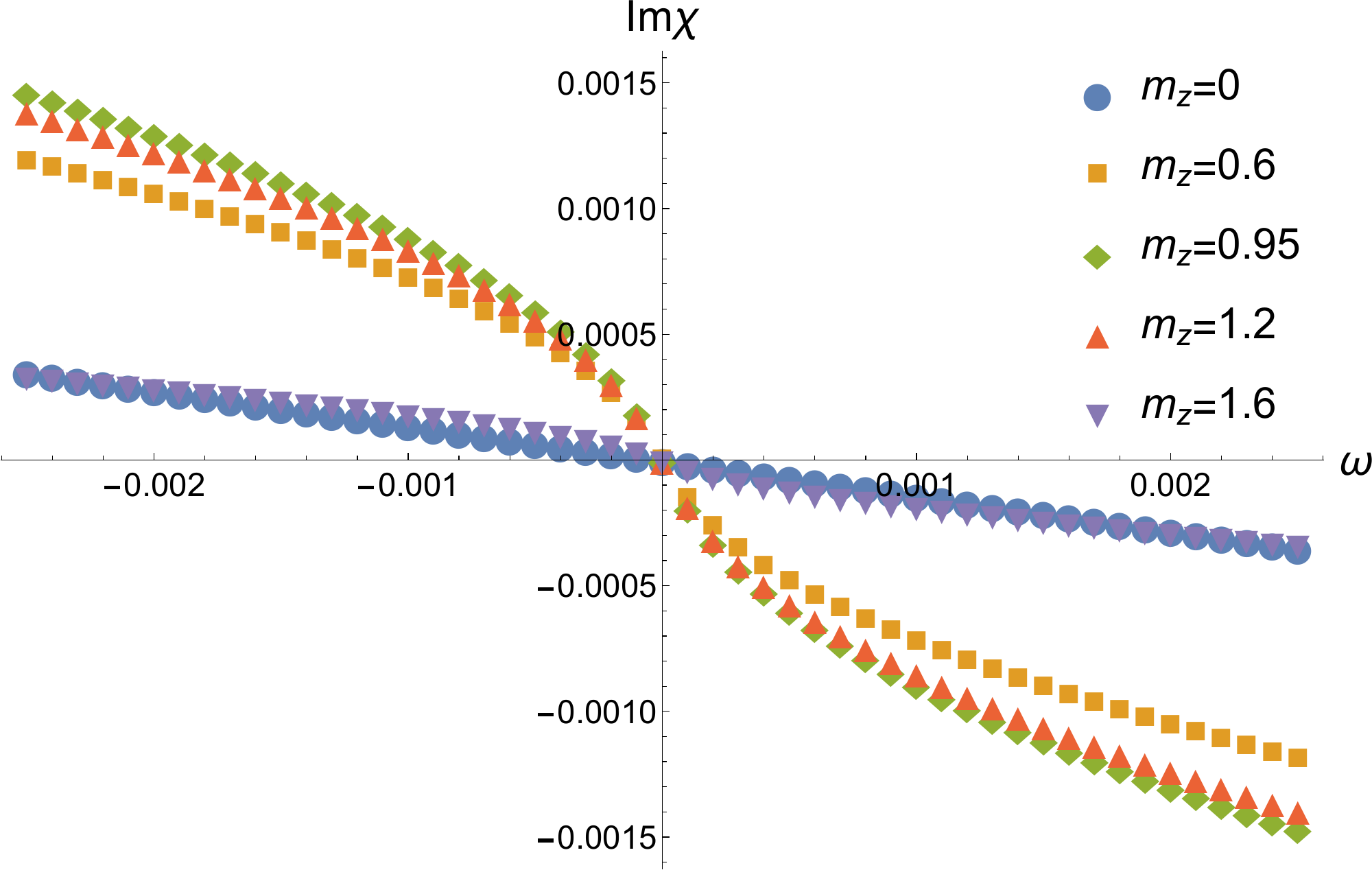}}
  \caption{(Color online) Imaginary part of the magnetic susceptibility $\Im\chi(\omega)$ for various uniform magnetization $m_z$ values.}\label{fig_chi}
\end{figure}

Because of the different low-energy magnon excitation spectra, the imaginary part of magnetic susceptibility $\Im\chi(\omega)$ at low-energy scale, corresponding to the $1/T_1T$ signal\cite{Slichter} in nuclear magnetic resonance (NMR) or muon spin resonance ($\mu$SR) experiments, as well ought to behave distinctly for $m_z \neq 0$ and $m_z=0$ cases. We can check by calculating the temporal Matsubara correlators of spin moment $\vec{S}(\vec{r},z)$: 
\begin{equation}
\chi_{ii}(z) = \int_0^\beta {\dd \tau \ee^{\ii z \tau} (-1)\braket{\mathrm{T}_\tau S_i(\vec{r},\tau)S_i(\vec{r},0)}  } \,,i=x,y,z
\end{equation}
 and analytically continuate it to the retarded one. Since we already have the analytic expressions of the Green's functions for spin waves (Sec.~\ref{spinwave_correlators}) responsible for the quantum and thermal fluctuations in spin moments, we substitute $\omega+\ii\delta$ in the first place and in the same manner as Eq.~\eqref{eq_SWcorr1}, we have
\begin{equation}
\begin{split}
&\Im \chi_{ii}(\omega+\ii\delta) \\
&= \int{\dd \vec{q} 
\sum_{\vec{l}} { (\partial_{\varphi_\mu} S_i)(-\vec{l}) (\partial_{\varphi_\nu} S_i)(\vec{l})  \Im\mathcal{G}_{\mu\nu}(\vec{q}+\vec{n},\omega+\ii\delta) }
}.
\end{split}
\end{equation}
Here we do not involve any approximation since the $\vec{l}$-summation contains finite terms for the static spin configuration. We used the same multidimensional integration method in Sec.~\ref{numerics} to evaluate such 3D integral with $\delta=1\times 10^{-4}$ and $|\omega|<0.0025 \ll \frac{D^2}{J}$ for various magnetization $m_z$'s. We show several typical cases in Fig.~\ref{fig_chi}, in which $\Im \chi(\omega)$ is always an odd function as expected. By extracting the power law dependence on $\omega$, one obtains a drastic change from linear power ($m_z=0$) to some power quite near $0.5$ ($m_z>0$) and a recovery to linearity when $m_z>\sqrt{2}$. This is just what one should expect from the distinct magnon spectra and the destruction of SkX where total Skyrmion number vanishes and equals the zero magnetization case.

We alternatively give a power law estimation valid for low-energy scale. In the spin-wave correlators, 
  we take $\braket{\phi_\alpha\phi_\alpha}$ for it is in general larger than other. Neglecting high order terms of momentum $\vec{q}$ and Matsubara frequency $z$, we obtain $\left\langle \phi_\alpha \phi_\alpha\right\rangle (q,\omega +\text{i$\delta $})\sim 
\left\{
\begin{array}{cc}
\begin{array}{cc}
 \frac{q^2}{q^4-(\omega +\ii\delta )^2}\approx \frac{q^2}{q^4-\omega ^2-2 \ii\delta \omega } & m_z \neq 0 \\
 \frac{1}{q^2-(\omega +\ii\delta )^2}\approx \frac{1}{q^2-\omega ^2-2 \ii\delta \omega } & m_z = 0 \\
\end{array} \\
\end{array}
\right.
$ after analytic continuation.
We picturesquely approximate the EEMF as being purely produced by the periodic array of vibrating magnetic monopoles (mp), whereupon the total spin configuration might crudely be regarded as comprising many fluctuating spin textures $\vec{S}_{\textrm{mp}}$ around singular points $|\vec{S}| = 0$ responsible for monopoles (c.f. Sec.~\ref{SkX}), $\vec{S}(\vec{r}) = \sum_n \vec{S}_{\text{mp}} (\vec{r} - \vec{R}_n(t))$, in which temporal dependence of the position of $n$th singularity $\vec{R}_n(t) = \vec{R}_n^{(0)}+ \vec{u}_n(t)$ is reflected in its deviation $\vec{u}_n(t)$ away from the static position $\vec{R}_n^{(0)}$. Neglecting directional dependence, we use the ansatz $\vec{S}_\textrm{mp}(\vec{r}) \sim \vec{r}$ which is analytically confirmed and whose Fourier transformation is $\vec{S}_{\text{mp}}(\vec{q}) \propto \ii \nabla_{\vec{q}} \delta(\vec{q})$. On the other hand,
$\vec{S}(\vec{q}) = \sum_n \int \dd \vec{r}\vec{S}_{\text{mp}} (\vec{r}- \vec{R}_n(t)) \ee^{-\ii \vec{q}\cdot \vec{r}} = \sum _n \vec{S}_{\text{mp}}(\vec{q}) \ee^{-\ii \vec{q}\cdot (\vec{R}_n^{(0)}+ \vec{u}_n(t))} \approx \sum _n \ee^{-\ii \vec{q}\cdot \vec{R}_n^{(0)}} \left(1-\ii \vec{q}\cdot \vec{u}_n(t)\right) \vec{S}_{\text{mp}}(\vec{q})$. The part relevant to quantum fluctuation reads  $-\ii \vec{S}_{\text{mp}}(\vec{q}) \sum _n \ee^{-\ii \vec{q}\cdot \vec{R}_n^{(0)}} \vec{q}\cdot \vec{u}_n(t) 
= - \ii\vec{q}\cdot \vec{u}_{\vec{q}}(t) \vec{S}_{\text{mp}}(\vec{q})$. 
And we can conclude that the asymtotic behavior in terms of $q$ of the fluctuating part in $\vec{S}(\vec{q})$ takes the form $\sim q^0 \vec{\phi }_{\vec{q}}(t)$, wherein we replaced deviation $\vec{u}$ by $\vec{\phi}$. Therefore, assuming isotropy for simplicity, the quantity is roughly given by $\Im \chi_{{ii}}(\omega) \sim \int \dd q q^2 \lim\limits_{\delta\rightarrow 0} \Im\left\langle \phi_i \phi_i \right\rangle (q,\omega +\ii\delta)$. We readily obtain the asymtotic power law dependences at an energy scale $\ll\frac{D^2}{J}$ as summarized in Table \ref{table}, which confirms our Green's function calculation nicely.
In summary, via asymptotic analysis and Green's function calculation, mutual corroboration of our resistivity and magnetic susceptibility studies is obtained.
\begin{table}
\begin{tabular}{c|c|c}
& $\rho(T)$ & $\Im\chi(\omega)$\\
\hline
$m_z=0$ & $T^6$ & $\omega$\\
\hline
$m_z\ne 0$ & $T^3$ & $\sqrt{\omega}$\\
\end{tabular}
\caption{Magnon spectra's influence on resistivity and susceptibility}
\label{table}
\end{table}


\subsection{Magnetoresistivity $\rho(m_z)$ profile at low temperature}\label{resistivity2}
\subsubsection{Comparison between theory and experiment}
\begin{figure*}
  \scalebox{0.35}{\includegraphics{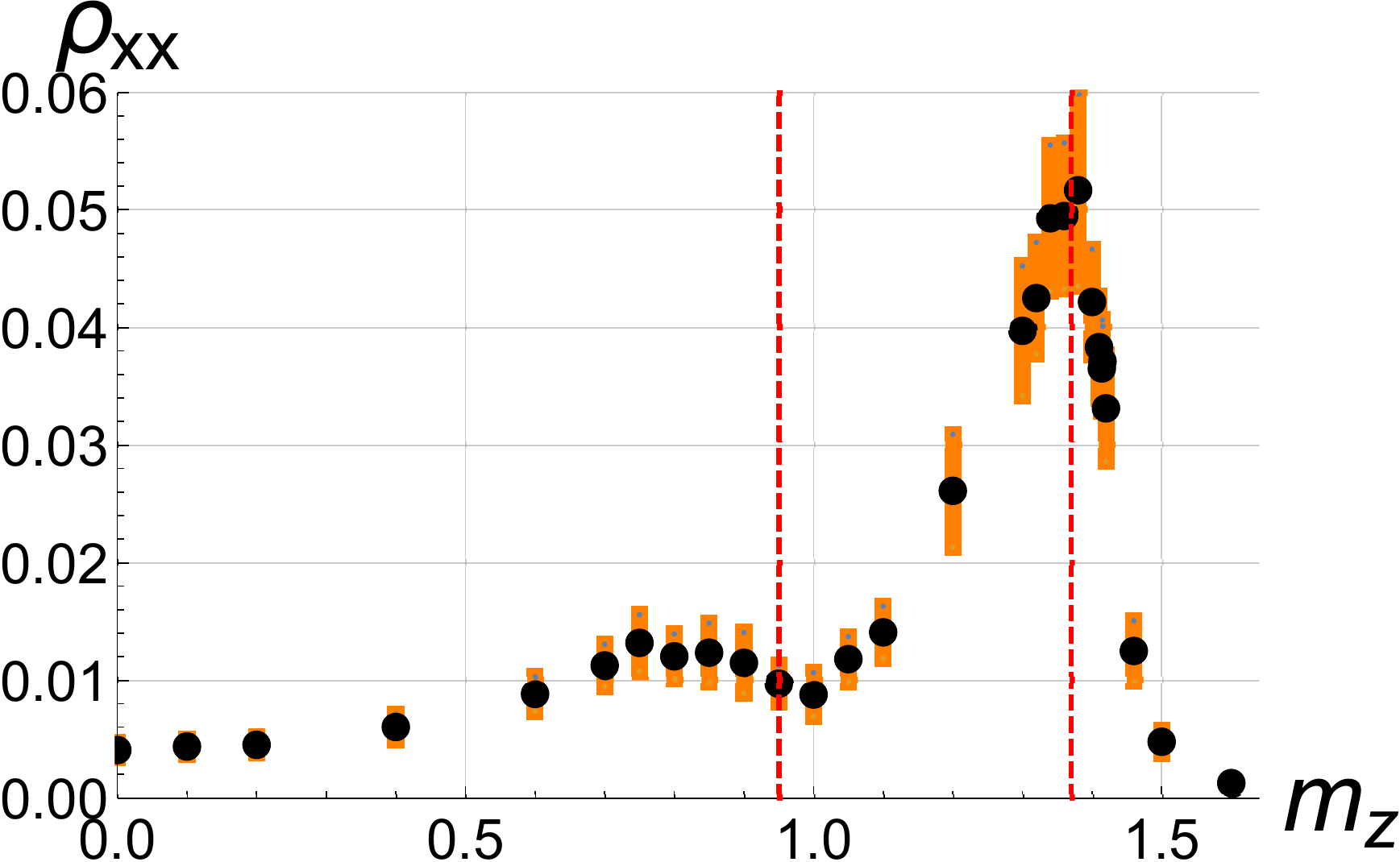}}
  \scalebox{0.35}{\includegraphics{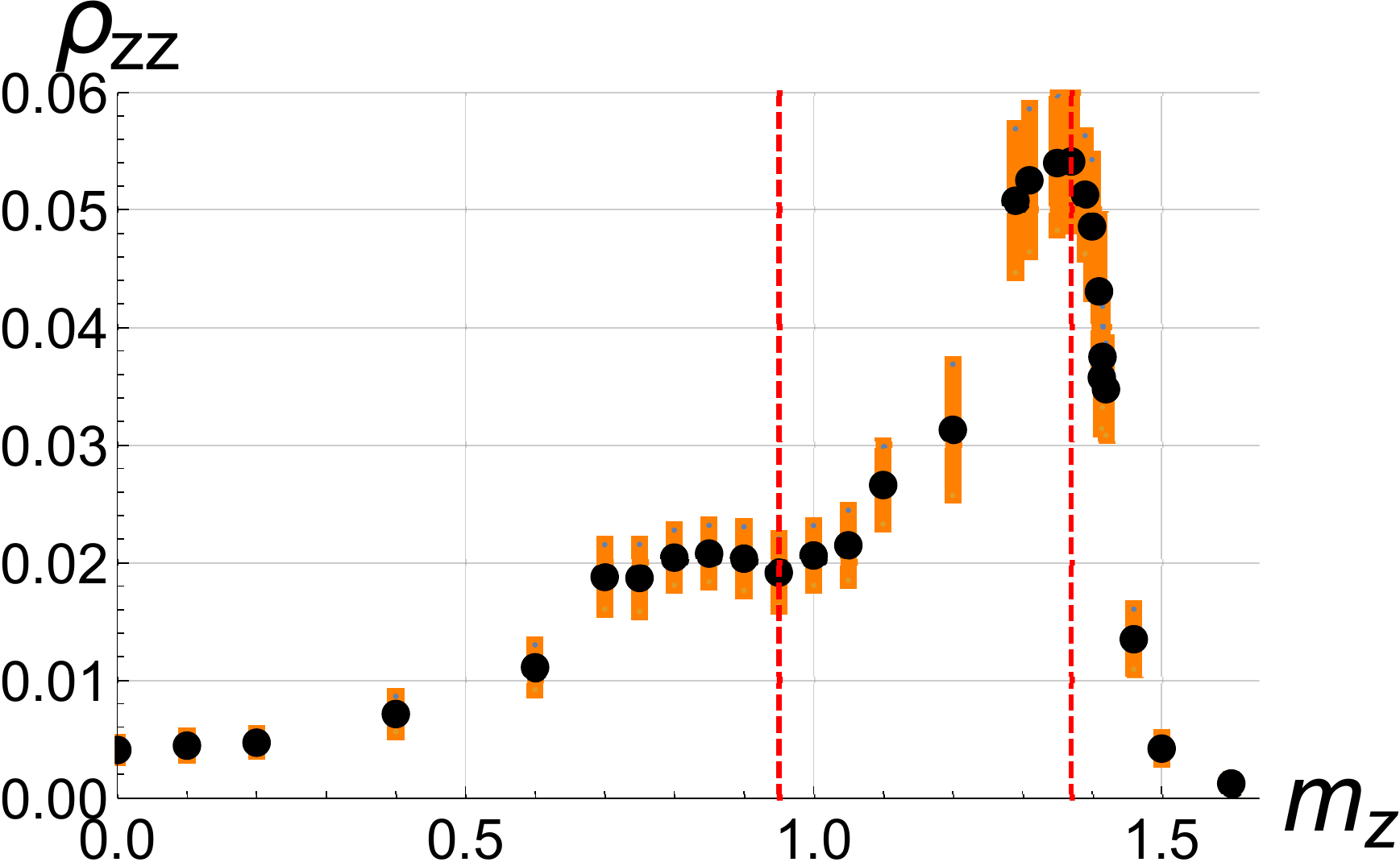}}
  \scalebox{0.35}{\includegraphics{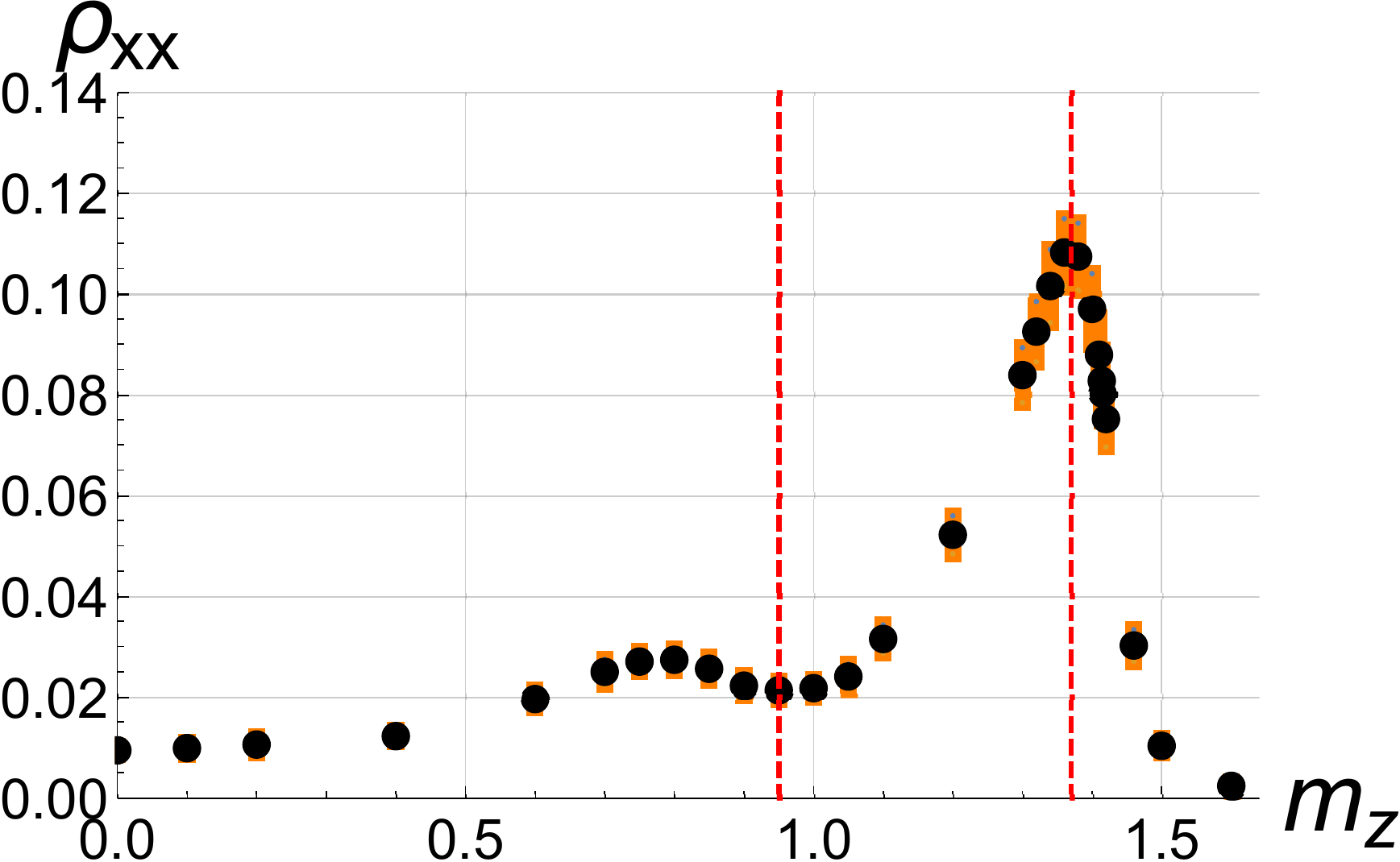}}
  \scalebox{0.35}{\includegraphics{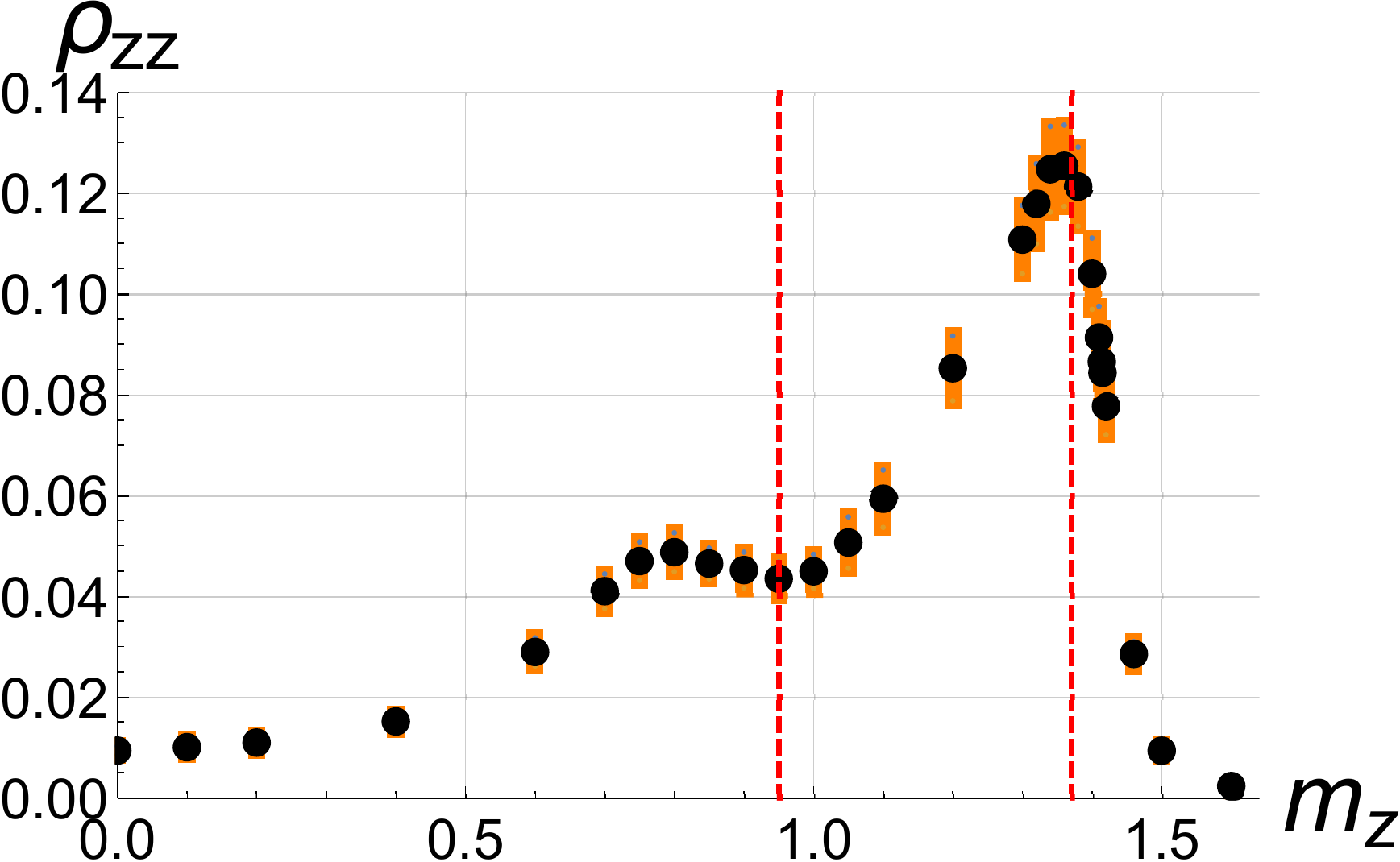}}
  \caption{(Color online) $\rho_\mathrm{xx}(m_z)$ and $\rho_\mathrm{zz}(m_z)$ plots with orange vertical errorbars show similar but anisotropic hump-dip-peak profile. Red lines indicate the dip around $m_z=0.95$ and the peak around $m_z=1.37$. Note the different scales of the vertical axis between the top panel ($\beta=\frac{10}{D\red{\hbar^2}}$) and the bottom panel ($\beta=\frac{7}{D\red{\hbar^2}}$).}\label{fig_rho}
\end{figure*}

Following the method stated in Sec.~\ref{memofun}, we carefully studied resistivity's variation with magnetization $m_z$ under different temperatures of typical energy scales from $\frac{D^2}{J}$ to $J$. They exhibited the same characteristic profile without exception and resistivity increases with temperature since at higher temperatures thermal fluctuations hence the excitation of spin waves becomes larger, rendering the inelastic scattering of electrons severer. However, for the high temperatures, the calculation suffers from numerical instability for too small $\beta$s while rather low temperatures call for much more time consumption because the relevant momentum region has to be swept more intricately. Therefore, we henceforth focus on the more interesting magnetoresistivity at some fixed and reasonably small temperatures of the order of $\frac{D^2}{J}$. In Fig.~\ref{fig_rho} we show for instance the resistivities at low temperatures of $\beta=\frac{10.0}{D\red{\hbar^2}}$ and $\beta=\frac{7.0}{D\red{\hbar^2}}$ as a function of uniform magnetization $m_z$, wherein $\rho_{ii}$ signifies longitudinal dc resistivity along $i$-axis. Firstly, numerically we confirmed our expectation of anisotropy that $\rho_\mathrm{xx} = \rho_\mathrm{yy} \neq \rho_\mathrm{zz}$ always holds because the cubic symmetry is broken solely by the application of magnetic field along $z$-axis as reflected by $m_z$. Therefore we only show $\rho_\mathrm{xx}$ and $\rho_\mathrm{zz}$. The characteristic features comprise a conspicuous hump-dip-peak structure in both of them and that $\rho_\mathrm{zz}$ is in gross larger than $\rho_\mathrm{xx}$. A small hump occurs near $m_z=0.8$, followed by a shallow dip slightly deviated leftwards from $m_z=1.0$ and a drastic peak in the vicinity of $m_z=1.37$.
      We compared a part of our theoretical results with experimental data and discussed the consistency in a separate paper\cite{Nii2}. The hump-dip-peak structure can be clearly seen in the $\rho_{zz}$ plots of low enough temperatures (around $20K$) while the hump and dip are relatively obscured in the $\rho_{xx}$ plots. At these low temperatures corresponding to evident hump-dip-peak structure, also one can notice that $\rho_{xx}$ is obviously lower than or about half the height of $\rho_{zz}$ around the hump-and-dip region while the global largeness of $\rho_{zz}$ than $\rho_{xx}$ holds as well.

\subsubsection{Fluctuation of the emergent magnetic field and topological phase transition of the monopole crystal}\label{fluctuation}
\begin{figure*}
\subfigure[]{
\label{fig:bbcorr} 
\begin{minipage}[c]{0.5\textwidth}
\centering
  \scalebox{0.6}{\includegraphics{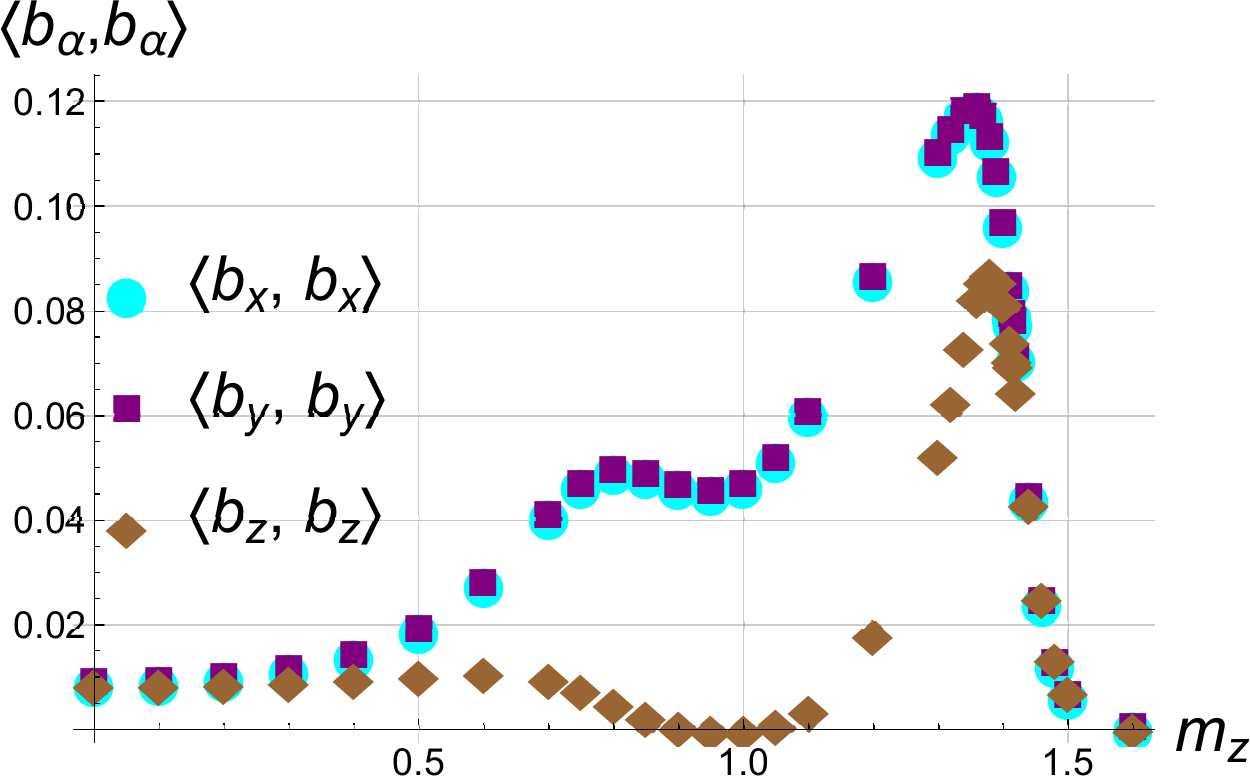}}
\end{minipage}}
\subfigure[]{
\label{fig:SkN} 
\begin{minipage}[c]{0.5\textwidth}
\centering
  \scalebox{0.46}{\includegraphics{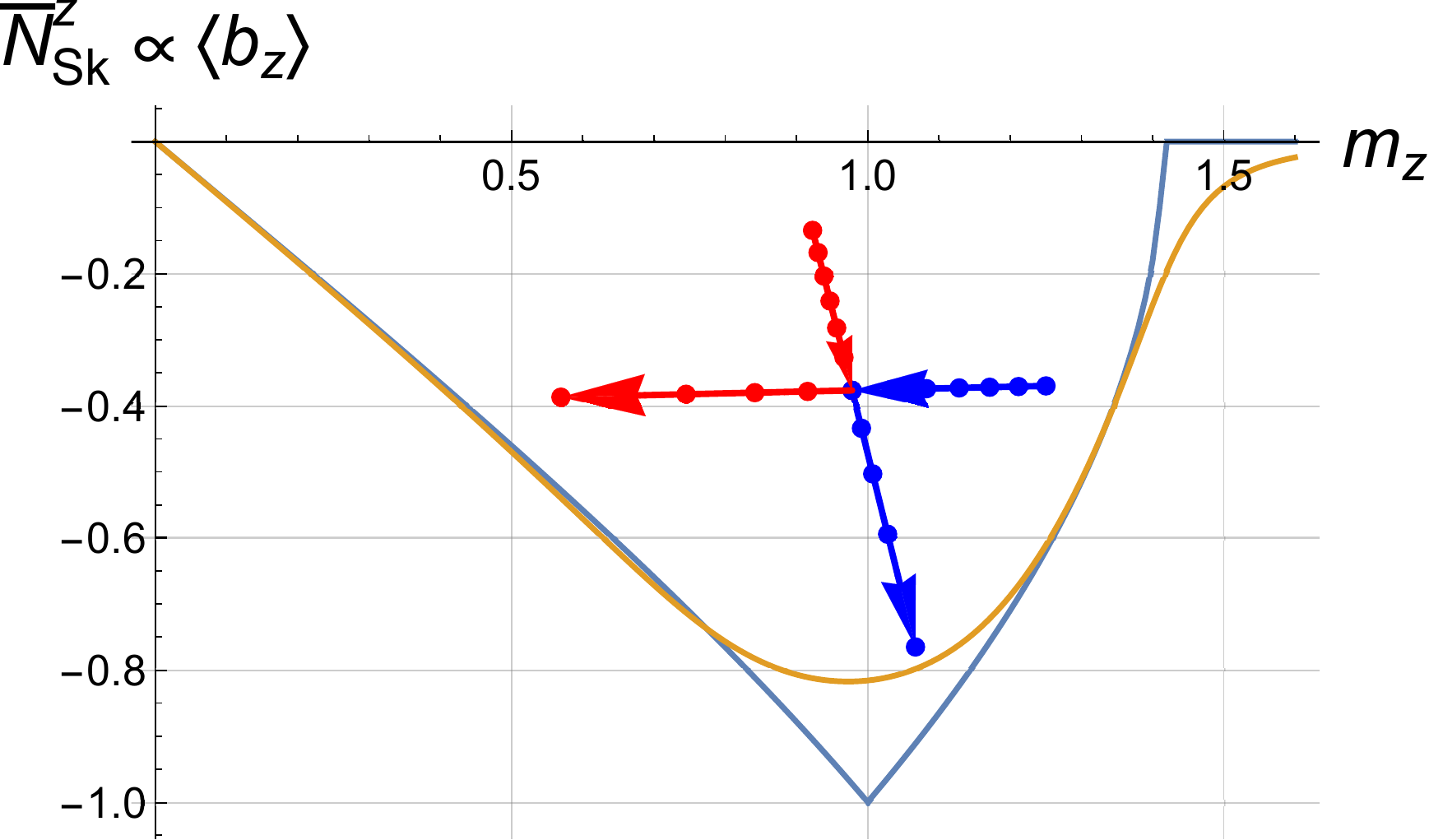}}
\end{minipage}}\\[-10pt]
  \caption{(Color online) (a) Equal time correlation functions of emergent $\vec{b}$-field vary with respect to uniform magnetization $m_z$. (b) $\bar{N}_\mathrm{Sk}^z$ ($\propto\braket{b_z}$) v.s. $m_z$. Yellow: Natural cutoff of monopolar singularity incorporated. Blue: Analytic value without cutoff. Inset: monopole-antimonopole pair collision process.}\label{fig:SkNALL}
\end{figure*}

The key to the interpretation of the anisotropy lies in the fact that the spin waves obstructing electrons' free motion entail fluctuations of the emergent $\vec{b}$ field felt by the conduction electrons. Indeed, the other contribution in an equal-time calculation of correlation function Eq.~\eqref{eq_phi2} is at most 5\% of the $\vec{b}$-only part. Intuitively, this lies in the fact that the $\vec{b}$ field reflects the most singular monopolar field in contrast to the $V$ part that turns out to be a nonsingular potential energy. Thus, the correlation functions of fluctuating $\vec{b}$ field should considerably reflect the intensity of scattering. To this end, we calculated the relevant real part of equal-time $\vec{b}$-field correlators $\braket{b_\alpha b_\alpha} = \int \dd q \Re\mathcal{D}_{b_\alpha b_\alpha}(q,\tau=0) $ as shown in Fig.~\ref{fig:bbcorr}. $\braket{b_x b_x}$ and $\braket{b_y b_y}$ coincide with each other and exhibit a profile very similar to the magnetoresistivity while $\braket{b_z b_z}$ shows a more pronounced dip near $m_z=1.0$ and is much smaller than $\braket{b_x b_x}$ in a wide region. Also the hump, dip and peak positions coincide with $\rho(m_z)$ plots up to $5\%$ precision in $m_z$. Thinking of Lorentz force, electrons traversing in (emergent) magnetic fields are mainly deflected by the fields perpendicular to their motion. Consequently, it is the fluctuations of $b_y,b_z$ and $b_x,b_y$ that massively contribute to $\rho_\mathrm{xx}$ and $\rho_\mathrm{zz}$ respectively. Thus, by taking into account of different contributions in Fig.~\ref{fig:bbcorr}, one can understand why in general $\rho_\mathrm{xx}$ is smaller than $\rho_\mathrm{zz}$ and especially around the hump-and-dip region we observe $\rho_\mathrm{xx}\approx\frac{1}{2}\rho_\mathrm{zz}$, which is grabbed pretty well by the contrasting behaviors in $\braket{b_\alpha b_\alpha}$'s.

In order to understand the nature of the hump-dip-peak structure occurring in both $\rho$ and $\braket{b_\alpha b_\alpha}$, it is necessary to inspect the ground state spin configuration carefully, on which the fluctuations in $\vec{b}$ are largely dependent.
We then scrutinize the monopole crystal structure. In the light of Skyrmion number formula (\ref{eq_SkN2}), one can calculate its spatial average along $\hat{z}$ direction within a cubic magnetic unit cell\cite{SkX2}
\begin{equation}\label{eq_SkN_ave}
  \bar{N}_\mathrm{Sk}^z  \equiv \frac{1}{2\pi}\int_0^{2\pi} { \dd z  N_\mathrm{Sk}^z(z) } = \begin{cases}
      -\frac{4}{\pi}\eta & (0 \leq m_z \leq 1)\\
          \frac{4}{\pi}(\eta-\frac{\pi}{2}) & (1 < m_z \leq \sqrt{2}),
  \end{cases}
\end{equation}
where $\eta\equiv \arcsin\frac{m_z}{\sqrt{2}}$. This, along with Eq.~\eqref{eq_b}, implies its relation to the spatial average of EEMF $\bar{N}_\mathrm{Sk}^z = 2\pi \braket{b_z}$. In Fig.~\ref{fig:SkN}, we show the $\bar{N}_\mathrm{Sk}^z(m_z)$ plot of Eq.~\eqref{eq_SkN_ave}, which is an analytic result for the ideal SkX with genuine monopoles (Eq.~\eqref{texture}), and another one with a cutoff of the singular monopolar field (see Sec.~\ref{numerics}), which is natural and necessary for a lattice system. Also note that $\braket{b_x},\braket{b_y}$ are always equal to zero. The blue line's profile recurs in the yellow one with two cusps at $m_z=1.0$ and $m_z=\sqrt{2}$, presumably corresponding to the extremum and the inflection point on the right in the yellow line, smoothed and slightly shifted leftwards. Notably, the average Skyrmion number undergoes two inverse monotonous variations with respect to increasing $m_z$, reaching its extremum around $m_z=1$ and tending to zero at zero or large enough $m_z$. We owe the decline near $m_z=\sqrt{2}$ to the destruction of the SkX, above which Eq.~\eqref{eq_SkN_ave} fails and residual monopoles gradually bocome connected to form some helicoid state and end in induced ferromagnetism at large enough $m_z$.

The following analysis is for the ideal simple cubic SkX/monopole crystal depicted in Fig.~\ref{fig_SkX} and Fig.~\ref{fig_mpEvo}, i.e., the blue line in Fig.~\ref{fig:SkN}, which ought to reflect the essential features of a realistic one. 
There are four pairs of monopole and antimonopole in a magnetic unit cell when $0\le m_z<\sqrt{2}$, i.e., four monopoles when $m_z<1$ (antimonopoles when $m_z>1$) at $(\frac{\pi}{4}-\eta,\frac{5\pi}{4}-\eta,\frac{7\pi}{4}-\eta)\,,(\frac{3\pi}{4}+\eta,\frac{3\pi}{4}+\eta,\frac{3\pi}{4}-\eta)\,,(\frac{7\pi}{4}-\eta,\frac{\pi}{4}+\eta,\frac{5\pi}{4}-\eta),(\frac{5\pi}{4}+\eta,\frac{7\pi}{4}-\eta,\frac{\pi}{4}-\eta)$ together with four antimonopoles when $m_z<1$ (monopoles when $m_z>1$) at $(\frac{\pi}{4}-\eta,\frac{3\pi}{4}+\eta,\frac{5\pi}{4}+\eta)\,,(\frac{3\pi}{4}+\eta,\frac{5\pi}{4}-\eta,\frac{\pi}{4}+\eta)\,,(\frac{5\pi}{4}+\eta,\frac{\pi}{4}+\eta,\frac{3\pi}{4}+\eta),(\frac{7\pi}{4}-\eta,\frac{7\pi}{4}-\eta,\frac{7\pi}{4}+\eta)$. Dissimilar to ordinary Dirac monopole, in spite of the aforementioned charge quantization, calculation shows that these (anti)monopoles are neither isotropic nor homogeneous, i.e., exact $r^{-2}$ divergence of EEMF $\vec{b}(\vec{r})$ only holds in the vicinity of one such (anti)monopole and the strength varies with direction. As one can observe in Fig.~\ref{fig_mpEvo}, in a cubic unit cell, there exist four monopoles and four antimonopoles. The salient point is that as uniform magnetization traverses the $m_z=1$ point, any (anti)monopole can be viewed as belonging to a monopole-antimonopole pair which undergoes a collision whose trajectory (Fig.~\ref{fig_mpEvo} and inset of Fig.~\ref{fig:SkN}) resembles an elastic collision of two point masses. The monopole and antimonopole coincide exactly when $m_z=1$. Moreover, the $r^{-2}$ divergence transforms to $r^{-1}$ at this crucial point. However, as $m_z$ increases to $\sqrt{2}$, each monopole (antimonopole) approaches another antimonopole (monopole) that is different from the one once 'collided' with and finally annihilates altogether.

It is plausible to owe the minute shift of the dip (peak) from $m_z=1$ ($m_z=\sqrt{2}$) in either $\rho(m_z)$ or $\braket{b_\alpha,b_\alpha}$ plots to the two smoothed and leftwards shifted cusps in Fig.~\ref{fig:bbcorr}, which actually originates from the monopole cutoff. And now we can relate the dip to the monopole-antimonopole collision motion at $m_z=1$. The fluctuation effect around this point is expected to be relatively suppressed to a low level since the spin texture just before and after the collision is quite similar to each other, which can be notably altered by neither a slight increase nor decrease in the magnetic field. Thus it is a mild, albeit intriguing change rather than any phase transition. This can also be roughly traced to the maximum in Skyrmion number and its flatness in the proximity shown in Fig.~\ref{fig:SkN}. On the other hand, the drastic peaks around $m_z=1.37$ are naturally attributed to a phase transition of massive change in topology, i.e., the destruction of the SkX or the monopole-antimonopole pair annihilation in the monopole crystal occurring a bit below $m_z=\sqrt{2}$, during which, the dramatic structural change in spin configuration, especially the disappearance of singularities, entails large fluctuation in $\vec{b}$.

We further comment on the topological aspect of this phase transition. It is exactly the length-fixed (unit-norm) spin texture $\vec{n}$, rather than the bare spin moment $\vec{S}$ itself, that yields the topological feature of the emergent $\vec{b}$ field characterized by the second homotopy group. Besides the strong Hund's rule coupling that makes the original length insignificant in some sense, this length constraint should also be understood as coming from the strong electron correlation herein, which renders the variation in length hard since the double occupation of relevant orbits is suppressed. Thus, topology here manifests strong correlation. 
Furthermore, the ordinary 2D triangular SkX\cite{MnSi1}, composed of three spirals whose wave vectors $\vec{k}_1=k(1,0,0),\vec{k}_2=k(-\frac{1}{2},\frac{\sqrt{3}}{2},0),\vec{k}_3=k(-\frac{1}{2},-\frac{\sqrt{3}}{2},0)$ subtend $\pi/3$ angles with each other and no singularity in $\vec{n}$ at all, gives nonzero Skyrmion number even when one uses $\vec{S}$ to calculate \eqref{eq_SkN2}. This is because the spatial integral extracts the zeroth Fourier component, which does not vanish since $\vec{k}_1+\vec{k}_2+\vec{k}_3 = \vec{0}$ is fulfilled. As for our 3D SkX/monopole crystal, although $\vec{0}+\vec{k}_\alpha+(-\vec{k}_\alpha) = \vec{0}$, the two spatial derivatives make it zero in the end. Thus in stark contrast, the spatially averaged Skyrmion number \eqref{eq_SkN_ave} vanishes unless $\vec{n}$ is used, singling out the singularity contribution from the monopoles. Being not special at all from the point of view of superposition of spin density waves, these singular points indeed gain significance from the strong correlation generated nontrivial topology. In this spirit, the peak is finally attributed to such a nontrivial topological phase transition.

\section{Concluding remark}\label{conclusion}
We have studied theoretically the novel magnetoresistance in the three-dimensional topological spin texture composed of magnetic monopoles and antimonopoles connected by the Skyrmion strings. This topological nature is the manifestation of the strong correlation which leads to the saturated magnetic moment with fixed length whose direction is defined as ${\vec n}$ while the superposition of the three helices ${\vec S}$ with variable magnitude exhibits trivial topology only. This nontrivial topology results in the topological phase transition characterized by the onset of finite Skyrmion number associated with the creation of monopole-antimonopole pairs as the uniform magnetization is reduced. This phase transition is accompanied by the critical fluctuation of the emergent magnetic field, which scatters the conduction electrons and enhances the resistivity.

While it is always the most fundamental issue whether the system of interest belongs to the weak correlation regime or the strong one, it often remains an ambiguous and quantitative problem and no sharp criterion can be found. The clear difference in the topological nature between the weak and the strong correlation limits discussed in this paper will offer a qualitative criterion for this issue and the comparison with the experiment on MnGe indicates that this material corresponds to the strong correlation regime. This is consistent with the appearance of the three-dimensional spin texture in the ground state, which requires enhanced magnetic moments and associated spin-orbit interaction and spin anisotropy. It is an intriguing issue to explore other consequences of this topological phase transition. The ultrasonic absorption is one possibility already reported\cite{Nii2} and will be discussed elsewhere. For example, the spin wave dynamics near the transition is an interesting issue but left for future studies.

\section*{Acknowledgments}
We thank Naoya Kanazawa for useful discussions and the indispensable experimental results. X.-X.Z is grateful 
 to Fei Xue for helpful discussions and to Shaoyu Wang for prompt help on coding. X.-X.Z was partially supported by the Panasonic Scholarship and by Japan Society for the Promotion of Science through Program for Leading Graduate Schools (ALPS) and Grant-in-Aid for JSPS Fellows (No.~16J07545). This work was supported by JSPS Grant-in-Aid for Scientific Research (No.~24224009) and JSPS Grant-in-Aid for Scientific Research on Innovative Areas (No.~26103006) from the Ministry of Education, Culture, Sports, Science and Technology (MEXT) of Japan and ImPACT Program of Council for Science, Technology and Innovation (Cabinet office, Government of Japan).
\\
\\
\appendix
\section{Derivation of the effective Hamiltonian for itinerant electrons}\label{App:H_eff}
To derive the effective model where electrons are coupled to the $\mathrm{U}(1)$ gauge field, we firstly choose the spin quantization axis oriented along the direction $\vec{n} = \vec{S}/|\vec{S}| = (\sin{\theta}\cos{\phi},\sin{\theta}\sin{\phi},\cos{\theta})$ of a local spin $\vec{S}(\vec{r},t)$. This is done by a gauge transformation $\Psi = G \Psi'$ satisfying \[G^\dag\vec{n}\cdot\vec{\sigma}G = \sigma_z,\] wherein $\Psi' = (\psi_1,\psi_2)^\mathrm{T}$ is the transformed spinor field, $G(\vec{r},t) = \vec{d}\cdot\vec{\sigma}$ and \[\vec{d} = (\sin{\frac{\theta}{2}}\cos{\phi},\sin{\frac{\theta}{2}}\sin{\phi},\cos{\frac{\theta}{2}}).\] A redundant $\mathrm{U}(1)$ gauge factor $\ee^{\ii\chi(\vec{r},t)\sigma_z}$ ($\ee^{-\ii\chi(\vec{r},t)\sigma_z}$) can be attached to $G$ ($\Psi'$). The pure $\mathrm{SU}(2)$ gauge fields can be readily read off from the covariant derivative \[\partial_\mu\Psi = G(\partial_\mu+G^\dag\partial_\mu G)\Psi',\] which results in
\[A_\mu^a\sigma_a \equiv -\ii\frac{\hbar}{q_\mathrm{e}} G^\dag\partial_\mu G = \frac{\hbar}{q_\mathrm{e}} (\vec{d}\times\partial_\mu\vec{d})\cdot\vec{\sigma}\,,a=x,y,z.\]
Here we shortly use a 4D index $\mu=0,1,2,3$ for this $\mathrm{SU}(2)$ gauge field.
At this stage, we feed $\Psi'$ to Eq.~\eqref{L_ele-spin1}
\begin{widetext}
\begin{equation}\label{L_ele-spin2}
\begin{split}
&\mathcal{L}_{\textrm{ele--spin}}\\
&= \Psi'^\dag (\ii\hbar\partial_0 + \varepsilon_F + \ii\hbar G^\dag\partial_0G + \frac{J_H}{2}S\sigma_z) \Psi'
-\frac{\hbar^2}{2m} \left[ \nabla\Psi'^\dag\cdot\nabla\Psi' + \Psi'^\dag\nabla G^\dag\cdot\nabla G\Psi' + (\nabla\Psi'^\dag G^\dag\cdot\nabla G\Psi' + \textrm{h.c.})\right]\\
 &= \psi^\dag \left[ \ii\hbar\partial_0 + \varepsilon_F + \ii\hbar(G^\dag\partial_0G)_{11} + \frac{J_H}{2}S(\sigma_z)_{11} \right] \psi
-\frac{\hbar^2}{2m} \left[ \nabla\psi^\dag\cdot\nabla\psi + \psi^\dag(\nabla G^\dag\cdot\nabla G)_{11}\psi + \left(\nabla\psi^\dag (G^\dag\cdot\nabla G)_{11}\psi + \textrm{h.c.}\right)\right]\\
&= \psi^\dag \left[ \ii\hbar\partial_0 + \varepsilon_F - q_\mathrm{e}A_0^z + \frac{J_H}{2}S \right] \psi
-\frac{\hbar^2}{2m} \left[ \nabla\psi^\dag\cdot\nabla\psi + \frac{q_\mathrm{e}^2}{\hbar^2}\psi^\dag\left(|\vec{A}^z|^2 + |\vec{A}^x + \ii\vec{A}^y|^2\right)\psi + \left(\nabla\psi^\dag \cdot (-\ii\frac{q_\mathrm{e}}{\hbar}\vec{A}^z)\psi + \textrm{h.c.}\right)\right]\\
&= \psi^\dag \left[ \ii\hbar\partial_0 - V(\vec{r},t) + \varepsilon_F - q_\mathrm{e}A_0^z + \frac{J_H}{2}S \right] \psi
+ \frac{1}{2m} (\hat{\vec{p}}+q_\mathrm{e}\vec{A}^z)\psi^\dag \cdot (\hat{\vec{p}}-q_\mathrm{e}\vec{A}^z)\psi,
\end{split}
\end{equation}
\end{widetext}
wherein we drop the $\psi_2$ component in $\Psi'$ and rename $\psi_1$ by $\psi$ to obtain the second equality
and we also define \[V \equiv \frac{\hbar^2}{8m} \left( (\nabla\theta)^2 +\sin^2\theta(\nabla\phi)^2 \right) = \frac{\hbar^2}{8m}(\nabla\vec{n})^2.\] Now the emergent U(1) gauge field and concomitant electromagnetic minimal coupling manifest while the two off-diagonal SU(2) fields $\vec{A}^x\,,\vec{A}^y$ enter the potential term $V$ only. Henceforth in the main text, we rename $\vec{A}^z$ by $\vec{a}$. We then retain the significant $\vec{a}$ and $V$ terms in Eq.~\eqref{L_ele-spin2} who have nonzero static mean field values, and after Legendre transformation, we finally attain the low-energy effective Hamiltonian Eq.~\eqref{eq_Heff}.

\section{Action for spin helices}\label{App:spin_action}
For a quantum spin $\hat{\vec{S}} = (\hat{S}_x,\hat{S}_y,\hat{S}_z)$ defined without the $\hbar$ factor, we have the commutation relation \[[\hat{S}_z,\hat{S}_x\pm\ii\hat{S}_y] = \pm(\hat{S}_x\pm\ii\hat{S}_y).\]
Noticing the natural spherical coordinate representation of a 3-vector, we have $\hat{\vec{S}} = S\hat{\vec{n}} = S(\sin{\hat{\theta}}\cos{\hat{\phi}},\sin{\hat{\theta}}\sin{\hat{\phi}},\cos{\hat{\theta}})$ where we promoted $\theta\,,\phi$ to quantum operators. Then the commutation relation can be cast in the form \[[\hat{S}_z,\ee^{\pm\ii\hat{\phi}}] = \pm\ee^{\pm\ii\hat{\phi}}.\] Adopting the ansatz $[\hat{\phi},\hat{S}_z]=\textrm{c-number}$, we readily get \[[\hat{\phi},\hat{S}_z]=\ii.\] This means, there exits a canonical conjugate pair $(\hat{\phi}\,,\hat{S}_z)$ that fully characterizes the algebra of a quantum spin, in the same manner as $(\hat{x},\hat{p})$ does for a particle's orbital degree of freedom. In the imaginary-time path integral formalism, the action of a quantum spin is given by 
\begin{equation}\label{spin_action}
\begin{split}
\mathcal{S} = &- \int_0^\beta{ \dd\tau \braket{\dot{\tau}|\tau} } + \int_0^\beta{ \dd\tau \braket{\tau|\hat{H}|\tau} }\\
= &\;\ii S \Omega + \int_0^\beta{ \dd\tau H(\vec{S}(\tau)) },
\end{split}
\end{equation}
wherein the first term is the spin Berry phase and \[\Omega = \int_0^\beta{ \dd\tau (1-\cos\theta)\dot{\phi}}\] is the solid angle subtended by the closed locus of $\vec{n}$.
In order to facilitate the description of helical spin textures, we promote spin $\hat{\vec{S}}$ to a field for the sake of continuum limit and hence two independent fields $\phi_z(\vec{r},\tau)$ and $S_z(\vec{r},\tau)$. Then the partition function and action in $(d+1)$ dimensions are given by
\begin{equation}\label{spinfield_action_append}
\begin{split}
&\mathcal{Z} = \int { \mathscr{D}S_z(\vec{r},\tau)  \mathscr{D}\phi_z(\vec{r},\tau)  \ee^{-\mathcal{S}\red{/\hbar}} }\\
&\mathcal{S} = \int_0^\beta{ \dd\tau \int{\dd^d\vec{r} (-\ii) S_z\partial_\tau\phi_z} }  + \int_0^\beta{ \dd\tau H(\tau)},
\end{split}
\end{equation}
wherein we licitly dropped the total $\tau$ differential in $\Omega$.

For the multi-spiral case, we first variate the solid angle in action Eq.~\eqref{spin_action}
\begin{equation}\label{spin_BerryPhase}
\delta\Omega = \int_0^\beta{ \dd\tau \delta\vec{n} \cdot (\partial_\tau\vec{n}\times\vec{n})}.
\end{equation}
If we write the spin orientation texture of Eq.~\eqref{eq_SkX} in an abstract form $\vec{n} = \vec{n}(\vec{k}_i\cdot \vec{r}+\phi_i(\vec{r},\tau))$, Eq.~\eqref{spin_BerryPhase} becomes
\begin{equation}\label{spinfield_BerryPhase}
\begin{split}
\delta\Omega = &\int_0^\beta{ \dd\tau \int{\dd^d\vec{r} \; \delta\vec{n} \cdot (\partial_\tau\vec{n}\times\vec{n}) } }\\
= &\int_0^\beta{ \dd\tau \int{\dd^d\vec{r} \;
\frac{\partial\vec{n}}{\partial\phi_i} \delta\phi_i \cdot \left( \frac{\partial\vec{n}}{\partial\phi_j} \partial_\tau \phi_j \times \vec{n} \right)
}  }\\
= &\int_0^\beta{ \dd\tau \int{\dd^d\vec{r} \;\frac{1}{k_i k_j}
\vec{n} \cdot \left( \frac{\partial\vec{n}}{\partial r_i} \times \frac{\partial\vec{n}}{\partial r_j} \right) \delta\phi_i \dot{\phi}_j
}  },
\end{split}
\end{equation}
wherein $k_i=\left|\vec{k}_i\right|$ and $r_i=\vec{r}\cdot\vec{k}_i/k_i$ (Latin indices) should not be confused with their spatial components like $k_\alpha\,,r_\alpha$ (Greek indices). Comparing this with the Skyrmion number Eq.~\eqref{eq_SkN1}, we realize the mixing between the $\phi$ fields as a result of the nontrivial real-space spin Berry phase.

\section{Some proofs for the symmetry of the $\dot{j}\textrm{-}\dot{j}$ correlation function}\label{App:symmetry}
The symmetry property Eq.~\eqref{eq_phiSymmetry} in Sec.~\ref{memofun} can be proved as follows 
\begin{eqnarray*}
\begin{split}
&\ee^{-\beta\Omega} \braket{A(\tau)B(0)} \\
&= \mathrm{Tr} A \ee^{-\tau \mathcal{K}} B \ee^{-(\beta-\tau)\mathcal{K}} \\
&=  \mathrm{Tr} B \ee^{-(\beta-\tau)\mathcal{K}} A \ee^{-\tau \mathcal{K}} \\
&=  \mathrm{Tr} B \ee^{-(\beta-\tau)\mathcal{K}} A \ee^{-\beta \mathcal{K}} \ee^{(\beta-\tau)\mathcal{K}} \\
&= \mathrm{Tr} \ee^{-\beta \mathcal{K}} \ee^{(\beta-\tau)\mathcal{K}} B \ee^{-(\beta-\tau)\mathcal{K}} A \\
&= \ee^{-\beta\Omega} \braket{B(\beta-\tau)A(0)}.
\end{split}
\end{eqnarray*}

In Sec.~\ref{spinwave_correlators}, we obtained the final expression Eq.~\eqref{eq_phi2} of the $\dot{j}\textrm{-}\dot{j}$ correlator. Next, we have to carry out Matsubara frequency summation with bosonic weight $n_B(z)+1$ to get $\mathcal{G}_{\mu\nu} (q,\tau)$, which, via residue theorem, is transformed to a summation of \[- \mathrm{Res}\left[\mathcal{G}_{\mu\nu} (q,z)\right] \mathcal{F}(\tau,z)\] over the six simple poles $\pm z_1,\pm z_2,\pm z_3$ of $\mathcal{G}_{\mu\nu} (q,z)$, wherein \[\mathcal{F}(\tau,z) = \ee^{-z\tau} (n_B(z)+1).\] We can further define $\mathcal{F}_\pm(\tau,z) = \mathcal{F}(\tau,z) \pm \mathcal{F}(\tau,-z)$, whose parity under the substitution $\tau\rightarrow\beta-\tau$ of the imaginary time is $\mp 1$. Careful inspection of $\mathcal{G}_{\mu\nu} (q,z)$ shows that $\Re \mathrm{Res} \mathcal{G}_{\mu\nu} (q,z)|_{z_i(q)} = -\Re \mathrm{Res} \mathcal{G}_{\mu\nu} (q,z)|_{-z_i(q)}$ and $\Im \mathrm{Res} \mathcal{G}_{\mu\nu} (q,z)|_{z_i(q)} = \Im \mathrm{Res} \mathcal{G}_{\mu\nu} (q,z)|_{-z_i(q)}$ for $i=1,2,3$, whereupon $\Re \mathcal{D}(q,\tau) = \sum_{i=1}^3 {\Re \mathrm{Res} \mathcal{D}}(q,z)|_{z_i} \mathcal{F}_-(\tau,z_i)$ for any $\mathcal{D}_{b_\alpha b_\beta},\mathcal{D}_{VV}$ and $\Im \mathcal{D}(q,\tau) = \sum_{i=1}^3 {\Im \mathrm{Res} \mathcal{D}}(q,z)|_{z_i} \mathcal{F}_+(\tau,z_i)$ for any $\mathcal{D}_{b_\alpha V},\mathcal{D}_{V b_\alpha}$ follow.
These properties, together with the symmetry Eq.~\eqref{eq_DeSymmetry} of electron Green's function $\mathcal{D}_{\textrm{e}}$ in Sec.~\ref{electron_correlators} and the fact that summations on $\vec{k},\vec{k}+\vec{q}$ are on the same footing, reassures us of the symmetry Eq.~\eqref{eq_phiSymmetry} $\phi_{\alpha\alpha}(\tau) = \phi_{\alpha\alpha}(\beta-\tau)$.

\section{derivation of the $\dot{j}\textrm{-}\dot{j}$ correlator}\label{App:jdot}
We absorb the gauge charge $q_\mathrm{e}$ into $\vec{a}$ in Hamiltonian Eq.~\eqref{eq_Heff} and define a gauge covariant velocity operator $\vec{v} \equiv v_i \hat{i} = \frac{1}{m} (\vec{p} - \vec{a})$ together with its variant $\cev{v} \equiv \bar{v}_i\hat{i}$, which differs only in that it acts to the left side. For simplicity, we omit hats on operators henceforth except otherwise stated. Needless to make any gauge choice, by deriving the continuity equation from the time-dependent Schr\"odinger equation for Hamiltonian Eq.~\eqref{eq_Heff}, we can get the gauge covariant current density
\begin{eqnarray*}
\begin{split}
\vec{j} &= \frac{1}{2m} \left( \psi^* \vec{p} \psi - \psi \vec{p} \psi^* \right) - \frac{1}{m}\vec{a}\psi^*\psi 
= \frac{1}{2} \left( \psi^* \vec{v} \psi + \psi \vec{v}^* \psi^* \right) \\
&= \frac{1}{2} \psi^* \left(  \vec{v} + \cev{v}^* \right) \psi 
= \mathrm{Re} \left( \psi^* \vec{v} \psi \right).
\end{split}
\end{eqnarray*}
Straightforward calculation gives
\begin{eqnarray*}\label{eq_v1}
\begin{split}
\red{\psi^*} [v_\alpha,v_\beta] \red{\psi} 
&= \red{\psi^*} [\bar{v}_\alpha^*,v_\beta] \red{\psi} \red{= \psi^* \bar{v}_\alpha^* v_\beta \psi - \psi^* v_\beta \bar{v}_\alpha^* \psi \\
&= \frac{1}{m^2} \psi^* \left[ 
(\bar{p}_\alpha^* p_\beta - \bar{p}_\alpha^* a_\beta - a_\alpha p_\beta + a_\alpha a_\beta )\right. \\
&\left.
-( p_\beta \bar{p}_\alpha^* - a_\beta \bar{p}_\alpha^* - p_\beta a_\alpha + a_\alpha a_\beta ) 
\vphantom{p_\beta \bar{p}_\alpha^*}
\right] \psi \\
&= \frac{1}{m^2} \psi^* \left( [a_\beta, \bar{p}_\alpha^*] + [p_\beta, a_\alpha ]\right) \psi \\
&}= \frac{1}{m^2} \red{\psi^*} \left[ -\left( p_\alpha a_\beta \right) + \left( p_\beta a_\alpha \right) \right] \red{\psi}.
\end{split}
\end{eqnarray*}
 Then we have
\begin{eqnarray*}
\begin{split}
[v_\alpha,v^2] &= [\bar{v}_\alpha^*,v^2] 
\red{= [\bar{v}_\alpha^*,v_{\mu}v^{\mu}] 
= v_{\mu} [\bar{v}_\alpha^*,v^{\mu}] + [\bar{v}_\alpha^*,v_{\mu}] v^{\mu} \\
&= \frac{1}{m^2} \left\{ v_{\mu} , -\left( p_\alpha a_\beta \right) + \left( p_\beta a_\alpha \right) \right\} \\
&= \frac{1}{m^3} \left( \left( -\left( p_\mu p_\alpha a^\mu \right) + \left( p_\mu p^\mu a_\alpha \right) \right) )\right. \\
&\left.
+ 2\left( -\left( p_\alpha a^\mu \right)p_\mu  + \left( p^\mu a_\alpha \right)p_\mu \right) 
\vphantom{p^\mu a_\alpha}
\right) \\
&}= \frac{-\ii \red{\hbar} q_\mathrm{e}}{m^3}  \epsilon_{\alpha\beta\gamma} \left( -p^\beta b^\gamma + 2 b^\beta p^\gamma \right).
\end{split}
\end{eqnarray*}
Similarly, we have
\begin{eqnarray*}\label{eq_v3}
\begin{split}
\red{\psi^*} [v_\alpha,\partial^\beta n_i] \red{\psi} &= \red{\psi^*} [\bar{v}_\alpha^*,\partial^\beta n_i] \red{\psi} = \frac{1}{m} \red{\psi^*} (p_\alpha\partial^\beta n_i) \red{\psi},
\end{split}
\end{eqnarray*}
and
\begin{eqnarray*}
\begin{split}
[v_\alpha,V] &= [\bar{v}_\alpha^*,V] = \frac{\red{\hbar^2}1}{8m}[v_\alpha,\sum_i{(\nabla n_i)^2}] \red{\\
&= \frac{\red{\hbar^2}1}{8m}\sum_i{(\partial_\beta n_i[v_\alpha,\partial^\beta n_i] + [v_\alpha,\partial_\beta n_i]\partial^\beta n_i)}\\
&}= \frac{1}{m} (p_\alpha V).
\end{split}
\end{eqnarray*}
Therefore, we can obtain the following concise expression 
\begin{eqnarray}\label{eq_jH}
\begin{split}
[\vec{j},\mathcal{H}_{\textrm{eff}}] &= [\frac{1}{2}(\vec{v}+\cev{v}^*),\frac{1}{2}m\vec{v}^2+V] \\
&= \frac{m}{4}\left( [\vec{v},v^2] + [\cev{v}^*,v^2] \right) + \frac{1}{m} (\vec{p}V)\\
&= \frac{\ii\red{\hbar} q_\mathrm{e}}{2m} \left( \vec{v}\times \vec{b} - \vec{b}\times\vec{v} \right) + \frac{1}{m} (\vec{p}V)\\
&= \frac{\ii\red{\hbar} q_\mathrm{e}}{2m^2} \left( (\vec{p}\times \vec{b}) - 2\vec{b}\times\vec{p} + 2\vec{b}\times\vec{a} \right) + \frac{1}{m} (\vec{p}V)
\end{split}
\end{eqnarray}
Here we neglect the $\vec{b}\times\vec{a}$ term because we only concern about first order effect due to $\vec{b}$ or $\vec{a}$. Consequently, only gauge invariant quantities are present. 

Now let's calculate the $\dot{j}\textrm{-}\dot{j}$ correlator Eq.~\eqref{eq_jdotjdot}.
Plugging Eq.~\eqref{eq_jH} into Eq.~\eqref{eq_jdotjdot}, we get
\begin{widetext}
\begin{eqnarray}\label{eq_phi0}
\begin{split}
\phi_{\alpha\mu}(\tau) &= \left( \frac{\ii\red{\hbar} q_\mathrm{e}}{2m^2} \right)^2 (-1) \braket{\mathrm{T}_\tau [(\vec{p}\times \vec{b}) - 2\vec{b}\times\vec{p}]_\alpha(\tau)[(\vec{p}\times \vec{b}) - 2\vec{b}\times\vec{p}]_\mu(0)} \\
& + \frac{\ii\red{\hbar} q_\mathrm{e}}{2m^2} (-1) (  \braket{\mathrm{T}_\tau [(\vec{p}\times \vec{b}) - 2\vec{b}\times\vec{p}]_\alpha(\tau)\frac{1}{m} (\vec{p}V)_\mu(0)} + \braket{\mathrm{T}_\tau \frac{1}{m} (\vec{p}V)_\alpha(\tau)[(\vec{p}\times \vec{b}) - 2\vec{b}\times\vec{p}]_\mu(0)} ) \\
& + \frac{1}{2m^2} (-1) \braket{\mathrm{T}_\tau (\vec{p}V)_\alpha(\tau) (\vec{p}V)_\mu(0)}
\end{split}
\end{eqnarray}
\end{widetext}

Now we promote all above to field operator representation by replacing wavefunction $\psi^{(*)}(\vec{r})$ by $\psi^{(\dag)}(\vec{r}) = \int{ \dd\vec{k} \psi_{\vec{k}}^{(*)}(\vec{r}) c_{\vec{k}}^{(\dag)} }$ wherein $\psi_{\vec{k}}^{(*)}(\vec{r})$ is the eigenfunction of momentum $\vec{k}$ and $c_{\vec{k}}^{(\dag)}$ is the corresponding electron annihilation (creation) operator. Note that here we neglect the spin degree of freedom of electrons since it is already incorporated via the construction of the EEMF model \eqref{eq_Heff}. Then, using partial integration, for instance, we have
\begin{equation*}
\begin{split}
&\int{\dd \vec{r} \psi^\dag(\vec{r}) \partial_\alpha V \psi(\vec{r}) } = \int{ \dd \vec{k}\dd\vec{q} \;\ii q_\alpha V(\vec{q},\tau) c_{\vec{k}+\vec{q}}^\dag(\tau) c_{\vec{k}}(\tau) },\\
&\int{\dd \vec{r} \psi^\dag(\vec{r}) \partial_\beta b_\gamma \psi(\vec{r}) } = \int{\dd \vec{k}\dd\vec{q} \; \ii q_\beta b_\gamma(\vec{q},\tau) c_{\vec{k}+\vec{q}}^\dag(\tau) c_{\vec{k}}(\tau)},\\
&\int{\dd \vec{r} \psi^\dag(\vec{r}) b_\beta p_\gamma(t) \psi(\vec{r}) } = \int{\dd \vec{k}\dd\vec{q} \; \red{\hbar} k_\gamma b_\beta(\vec{q},\tau) c_{\vec{k}+\vec{q}}^\dag(\tau) c_{\vec{k}}(\tau)}.
\end{split}
\end{equation*}
We then calculate one correlation function as an example of various terms appearing in Eq.~\eqref{eq_phi0}.
\begin{eqnarray}
\begin{split}
&-\braket{\mathrm{T}_\tau (\vec{p}\times \vec{b})_\alpha(\tau) (\vec{p}V)_\mu(0)}\\
& = -(-\ii\red{\hbar})^2 \braket{ \varepsilon_{\alpha\beta\gamma} \partial_\beta b_\gamma(\tau) \partial_\mu V(0)} \\
& = \red{(-\ii\red{\hbar})^2} \varepsilon_{\alpha\beta\gamma} \sum_{\vec{k}_1\vec{q}_1\vec{k}_2\vec{q}_2}{ \ii q_{1\beta} \ii q_{2\mu} \braket{ b_\gamma(\vec{q}_1,\tau) D_1(\tau)  V(\vec{q}_2,0) D_2(0)  } } \\
& = \red{(-\ii\red{\hbar})^2} \varepsilon_{\alpha\beta\gamma} \sum_{\vec{k}_1\vec{q}_1\vec{k}_2\vec{q}_2}{ \ii q_{1\beta} \ii q_{2\mu} \braket{ b_\gamma(\vec{q}_1,\tau)  V(\vec{q}_2,0)} \braket{ D_1(\tau)  D_2(0)  } } \\
& = \red{(-\ii\red{\hbar})^2} \varepsilon_{\alpha\beta\gamma} \sum_{\vec{k}\vec{q}}{  q_{\beta} q_{\mu} \mathcal{D}_{b_\gamma V}(\vec{q},\tau) \mathcal{D}_{\textrm{e}}(\vec{k},\vec{q},\tau)   },
\end{split}
\end{eqnarray}
wherein we define two Matsubara Green's functions, \[\mathcal{D}_{b_\gamma V}(\vec{q},\tau) = -\braket{\mathrm{T}_\tau b_\gamma(\vec{q},\tau)  V(-\vec{q},0)}\] for the fluctuations of EEMF $b_\gamma$ and potential $V$ and \[\mathcal{D}_{\textrm{e}}(\vec{k},\vec{q},\tau) = -\braket{ \mathrm{T}_\tau D_1(\tau)  D_2(0)  }\] for electrons with $D_1(\tau) = c_{\vec{k}_1+\vec{q}_1}^\dag(\tau) c_{\vec{k}_1}(\tau) , D_2(0) = c_{\vec{k}_2+\vec{q}_2}^\dag(0) c_{\vec{k}_2}(0) $. Four bosonic operators commute with each other in the second equality. The aforesaid non-interacting approximation herein justifies the decoupling from the second to third equality. Fourth equality follows from momentum conservation, i.e., the electron correlator yields $\vec{q}_1=-\vec{q}_2,\vec{k}_2=\vec{k}_1+\vec{q}_1$ (see Sec.~\ref{electron_correlators}). And similarly, we also define $\mathcal{D}_{b_\alpha b_\beta},\mathcal{D}_{V b_\alpha},\mathcal{D}_{VV}$. Thus, Eq.~\eqref{eq_phi0} can be expressed as
\begin{widetext}
\begin{eqnarray}\label{eq_phi1_append}
\begin{split}
\\
\\
\\
\phi_{\alpha\mu}(\tau) &= \frac{1}{m^2} (-1) \braket{\mathrm{T}_\tau (\vec{p}V)_\alpha(\tau) (\vec{p}V)_\mu(0)} \\
&+ \left( \frac{\ii\red{\hbar} q_\mathrm{e}}{2m^2} \right)^2 (-1) \braket{\mathrm{T}_\tau [(\vec{p}\times \vec{b}) - 2\vec{b}\times\vec{p}]_\alpha(\tau)[(\vec{p}\times \vec{b}) - 2\vec{b}\times\vec{p}]_\mu(0)} \\
& + \frac{\ii\red{\hbar} q_\mathrm{e}}{2m^2} (-1) (  \braket{\mathrm{T}_\tau [(\vec{p}\times \vec{b}) - 2\vec{b}\times\vec{p}]_\alpha(\tau)\frac{1}{m} (\vec{p}V)_\mu(0)} + \braket{\mathrm{T}_\tau \frac{1}{m} (\vec{p}V)_\alpha(\tau)[(\vec{p}\times \vec{b}) - 2\vec{b}\times\vec{p}]_\mu(0)} ) \\
& = \sum_{\vec{k}\vec{q}}  \mathcal{D}_{\textrm{e}}(\vec{k},\vec{q},\tau) \times 
\left\{ 
\frac{1\red{\hbar^2}}{m^2} q_\alpha q_\mu  \mathcal{D}_{VV}(\vec{q},\tau) \right. \\
&\left. + \left( \frac{\red{\hbar^2} q_\mathrm{e}}{2m^2} \right)^2  \varepsilon^{\alpha\beta\gamma}\varepsilon^{\mu\nu\sigma} \left[ -q_\beta q_\nu \mathcal{D}_{b_\gamma b_\sigma}(\vec{q},\tau) + 4k_\gamma(k+q)_\sigma \mathcal{D}_{b_\beta b_\nu}(\vec{q},\tau) 
-2q_\beta (k+q)_\sigma \mathcal{D}_{b_\gamma b_\nu}(\vec{q},\tau) + 2q_\nu k_\gamma \mathcal{D}_{b_\beta b_\sigma}(\vec{q},\tau) \right] \right. \\
&\left. + \frac{\ii\red{\hbar^3} q_\mathrm{e}}{2m^2} \left[ \varepsilon^{\alpha\beta\gamma} \left( q_\beta q_\mu \mathcal{D}_{b_\gamma V}(\vec{q},\tau) - 2k_\gamma q_\mu \mathcal{D}_{b_\beta V}(\vec{q},\tau) \right) + \varepsilon^{\mu\nu\sigma} \left( q_\alpha q_\nu \mathcal{D}_{Vb_\sigma}(\vec{q},\tau) + 2q_\alpha (k+q)_\sigma  \mathcal{D}_{Vb_\nu}(\vec{q},\tau) \right) \right]  
\vphantom{\frac{\red{\hbar^2}q}{m^2}}
\right\} \\
& = \sum_{\vec{k}\vec{q}}  \mathcal{D}_{\textrm{e}}(\vec{k},\vec{q},\tau) \times 
\left\{ 
\frac{\red{\hbar^2}1}{m^2} q_\alpha q_\mu  \mathcal{D}_{VV}(\vec{q},\tau) \right. \\
&\left. - \left( \frac{\red{\hbar^2} q_\mathrm{e}}{2m^2} \right)^2  \varepsilon^{\alpha\beta\gamma}\varepsilon^{\mu\nu\sigma} (2k+q)_\beta (2k+q)_\sigma \mathcal{D}_{b_\gamma b_\nu}(\vec{q},\tau)
 + \frac{\ii\red{\hbar^3} q_\mathrm{e}}{2m^2}  \varepsilon^{\alpha\beta\gamma} q_\alpha (2k+q)_\beta \left[ \mathcal{D}_{b_\gamma V}(\vec{q},\tau) - \mathcal{D}_{V b_\gamma}(\vec{q},\tau) \right]  
\vphantom{\frac{\hbar^2}{m^2}}
\right\},
\end{split}
\end{eqnarray}
\end{widetext}
wherein the last equality, i.e., Eq.~\eqref{eq_phi1} in the main text, follows from some algebraic manipulations when $\alpha=\mu$.

\label{Bibliography}
\bibliography{reference.bib}  

\begin{thebibliography}{71}%
\makeatletter
\providecommand \@ifxundefined [1]{%
 \@ifx{#1\undefined}
}%
\providecommand \@ifnum [1]{%
 \ifnum #1\expandafter \@firstoftwo
 \else \expandafter \@secondoftwo
 \fi
}%
\providecommand \@ifx [1]{%
 \ifx #1\expandafter \@firstoftwo
 \else \expandafter \@secondoftwo
 \fi
}%
\providecommand \natexlab [1]{#1}%
\providecommand \enquote  [1]{``#1''}%
\providecommand \bibnamefont  [1]{#1}%
\providecommand \bibfnamefont [1]{#1}%
\providecommand \citenamefont [1]{#1}%
\providecommand \href@noop [0]{\@secondoftwo}%
\providecommand \href [0]{\begingroup \@sanitize@url \@href}%
\providecommand \@href[1]{\@@startlink{#1}\@@href}%
\providecommand \@@href[1]{\endgroup#1\@@endlink}%
\providecommand \@sanitize@url [0]{\catcode `\\12\catcode `\$12\catcode
  `\&12\catcode `\#12\catcode `\^12\catcode `\_12\catcode `\%12\relax}%
\providecommand \@@startlink[1]{}%
\providecommand \@@endlink[0]{}%
\providecommand \url  [0]{\begingroup\@sanitize@url \@url }%
\providecommand \@url [1]{\endgroup\@href {#1}{\urlprefix }}%
\providecommand \urlprefix  [0]{URL }%
\providecommand \Eprint [0]{\href }%
\providecommand \doibase [0]{http://dx.doi.org/}%
\providecommand \selectlanguage [0]{\@gobble}%
\providecommand \bibinfo  [0]{\@secondoftwo}%
\providecommand \bibfield  [0]{\@secondoftwo}%
\providecommand \translation [1]{[#1]}%
\providecommand \BibitemOpen [0]{}%
\providecommand \bibitemStop [0]{}%
\providecommand \bibitemNoStop [0]{.\EOS\space}%
\providecommand \EOS [0]{\spacefactor3000\relax}%
\providecommand \BibitemShut  [1]{\csname bibitem#1\endcsname}%
\let\auto@bib@innerbib\@empty
\bibitem [{\citenamefont {Rajaraman}(1987)}]{Rajaraman}%
  \BibitemOpen
  \bibfield  {author} {\bibinfo {author} {\bibfnamefont {R.}~\bibnamefont
  {Rajaraman}},\ }\href {http://amazon.com/o/ASIN/0444870474/} {\emph {\bibinfo
  {title} {Solitons and Instantons, Volume 15: An Introduction to Solitons and
  Instantons in Quantum Field Theory (North-Holland Personal Library)}}},\
  \bibinfo {edition} {1st}\ ed.\ (\bibinfo  {publisher} {North Holland},\
  \bibinfo {year} {1987})\BibitemShut {NoStop}%
\bibitem [{\citenamefont {Ng}(2009)}]{Ng}%
  \BibitemOpen
  \bibfield  {author} {\bibinfo {author} {\bibfnamefont {T.-K.}\ \bibnamefont
  {Ng}},\ }\href {http://amazon.com/o/ASIN/352740726X/} {\emph {\bibinfo
  {title} {Introduction to Classical and Quantum Field Theory}}},\ \bibinfo
  {edition} {1st}\ ed.\ (\bibinfo  {publisher} {Wiley-VCH},\ \bibinfo {year}
  {2009})\BibitemShut {NoStop}%
\bibitem [{\citenamefont {Skyrme}(1962)}]{Skyrme}%
  \BibitemOpen
  \bibfield  {author} {\bibinfo {author} {\bibfnamefont {T.}~\bibnamefont
  {Skyrme}},\ }\href {\doibase 10.1016/0029-5582(62)90775-7} {\bibfield
  {journal} {\bibinfo  {journal} {Nuclear Physics}\ }\textbf {\bibinfo {volume}
  {31}},\ \bibinfo {pages} {556 } (\bibinfo {year} {1962})}\BibitemShut
  {NoStop}%
\bibitem [{\citenamefont {Wright}\ and\ \citenamefont {Mermin}(1989)}]{LQ}%
  \BibitemOpen
  \bibfield  {author} {\bibinfo {author} {\bibfnamefont {D.~C.}\ \bibnamefont
  {Wright}}\ and\ \bibinfo {author} {\bibfnamefont {N.~D.}\ \bibnamefont
  {Mermin}},\ }\href {\doibase 10.1103/RevModPhys.61.385} {\bibfield  {journal}
  {\bibinfo  {journal} {Rev. Mod. Phys.}\ }\textbf {\bibinfo {volume} {61}},\
  \bibinfo {pages} {385} (\bibinfo {year} {1989})}\BibitemShut {NoStop}%
\bibitem [{\citenamefont {Ho}(1998)}]{BEC1}%
  \BibitemOpen
  \bibfield  {author} {\bibinfo {author} {\bibfnamefont {T.-L.}\ \bibnamefont
  {Ho}},\ }\href {\doibase 10.1103/PhysRevLett.81.742} {\bibfield  {journal}
  {\bibinfo  {journal} {Phys. Rev. Lett.}\ }\textbf {\bibinfo {volume} {81}},\
  \bibinfo {pages} {742} (\bibinfo {year} {1998})}\BibitemShut {NoStop}%
\bibitem [{\citenamefont {Ohmi}\ and\ \citenamefont {Machida}(1998)}]{BEC2}%
  \BibitemOpen
  \bibfield  {author} {\bibinfo {author} {\bibfnamefont {T.}~\bibnamefont
  {Ohmi}}\ and\ \bibinfo {author} {\bibfnamefont {K.}~\bibnamefont {Machida}},\
  }\href {\doibase 10.1143/JPSJ.67.1822} {\bibfield  {journal} {\bibinfo
  {journal} {Journal of the Physical Society of Japan}\ }\textbf {\bibinfo
  {volume} {67}},\ \bibinfo {pages} {1822} (\bibinfo {year}
  {1998})}\BibitemShut {NoStop}%
\bibitem [{\citenamefont {Sondhi}\ \emph {et~al.}(1993)\citenamefont {Sondhi},
  \citenamefont {Karlhede}, \citenamefont {Kivelson},\ and\ \citenamefont
  {Rezayi}}]{QHE}%
  \BibitemOpen
  \bibfield  {author} {\bibinfo {author} {\bibfnamefont {S.~L.}\ \bibnamefont
  {Sondhi}}, \bibinfo {author} {\bibfnamefont {A.}~\bibnamefont {Karlhede}},
  \bibinfo {author} {\bibfnamefont {S.~A.}\ \bibnamefont {Kivelson}}, \ and\
  \bibinfo {author} {\bibfnamefont {E.~H.}\ \bibnamefont {Rezayi}},\ }\href
  {\doibase 10.1103/PhysRevB.47.16419} {\bibfield  {journal} {\bibinfo
  {journal} {Phys. Rev. B}\ }\textbf {\bibinfo {volume} {47}},\ \bibinfo
  {pages} {16419} (\bibinfo {year} {1993})}\BibitemShut {NoStop}%
\bibitem [{\citenamefont {Bogdanov}\ and\ \citenamefont
  {Yablonskii}(1989)}]{Skprediction1}%
  \BibitemOpen
  \bibfield  {author} {\bibinfo {author} {\bibfnamefont {A.~N.}\ \bibnamefont
  {Bogdanov}}\ and\ \bibinfo {author} {\bibfnamefont {D.~A.}\ \bibnamefont
  {Yablonskii}},\ }\href {http://jetp.ac.ru/cgi-bin/e/index/e/68/1/p101?a=list}
  {\bibfield  {journal} {\bibinfo  {journal} {JETP}\ }\textbf {\bibinfo
  {volume} {95}},\ \bibinfo {pages} {182} (\bibinfo {year} {1989})}\BibitemShut
  {NoStop}%
\bibitem [{\citenamefont {R\"{o}{\ss}ler}\ \emph {et~al.}(2006)\citenamefont
  {R\"{o}{\ss}ler}, \citenamefont {Bogdanov},\ and\ \citenamefont
  {Pfleiderer}}]{Skprediction2}%
  \BibitemOpen
  \bibfield  {author} {\bibinfo {author} {\bibfnamefont {U.~K.}\ \bibnamefont
  {R\"{o}{\ss}ler}}, \bibinfo {author} {\bibfnamefont {A.~N.}\ \bibnamefont
  {Bogdanov}}, \ and\ \bibinfo {author} {\bibfnamefont {C.}~\bibnamefont
  {Pfleiderer}},\ }\href {\doibase 10.1038/nature05056} {\bibfield  {journal}
  {\bibinfo  {journal} {Nature}\ }\textbf {\bibinfo {volume} {442}},\ \bibinfo
  {pages} {797} (\bibinfo {year} {2006})}\BibitemShut {NoStop}%
\bibitem [{\citenamefont {M\"uhlbauer}\ \emph {et~al.}(2009)\citenamefont
  {M\"uhlbauer}, \citenamefont {Binz}, \citenamefont {Jonietz}, \citenamefont
  {Pfleiderer}, \citenamefont {Rosch}, \citenamefont {Neubauer}, \citenamefont
  {Georgii},\ and\ \citenamefont {B\"oni}}]{MnSi1}%
  \BibitemOpen
  \bibfield  {author} {\bibinfo {author} {\bibfnamefont {S.}~\bibnamefont
  {M\"uhlbauer}}, \bibinfo {author} {\bibfnamefont {B.}~\bibnamefont {Binz}},
  \bibinfo {author} {\bibfnamefont {F.}~\bibnamefont {Jonietz}}, \bibinfo
  {author} {\bibfnamefont {C.}~\bibnamefont {Pfleiderer}}, \bibinfo {author}
  {\bibfnamefont {A.}~\bibnamefont {Rosch}}, \bibinfo {author} {\bibfnamefont
  {A.}~\bibnamefont {Neubauer}}, \bibinfo {author} {\bibfnamefont
  {R.}~\bibnamefont {Georgii}}, \ and\ \bibinfo {author} {\bibfnamefont
  {P.}~\bibnamefont {B\"oni}},\ }\href {\doibase 10.1126/science.1166767}
  {\bibfield  {journal} {\bibinfo  {journal} {Science}\ }\textbf {\bibinfo
  {volume} {323}},\ \bibinfo {pages} {915} (\bibinfo {year}
  {2009})}\BibitemShut {NoStop}%
\bibitem [{\citenamefont {Tonomura}\ \emph {et~al.}(2012)\citenamefont
  {Tonomura}, \citenamefont {Yu}, \citenamefont {Yanagisawa}, \citenamefont
  {Matsuda}, \citenamefont {Onose}, \citenamefont {Kanazawa}, \citenamefont
  {Park},\ and\ \citenamefont {Tokura}}]{MnSi2}%
  \BibitemOpen
  \bibfield  {author} {\bibinfo {author} {\bibfnamefont {A.}~\bibnamefont
  {Tonomura}}, \bibinfo {author} {\bibfnamefont {X.}~\bibnamefont {Yu}},
  \bibinfo {author} {\bibfnamefont {K.}~\bibnamefont {Yanagisawa}}, \bibinfo
  {author} {\bibfnamefont {T.}~\bibnamefont {Matsuda}}, \bibinfo {author}
  {\bibfnamefont {Y.}~\bibnamefont {Onose}}, \bibinfo {author} {\bibfnamefont
  {N.}~\bibnamefont {Kanazawa}}, \bibinfo {author} {\bibfnamefont {H.~S.}\
  \bibnamefont {Park}}, \ and\ \bibinfo {author} {\bibfnamefont
  {Y.}~\bibnamefont {Tokura}},\ }\href {\doibase 10.1021/nl300073m} {\bibfield
  {journal} {\bibinfo  {journal} {Nano Letters}\ }\textbf {\bibinfo {volume}
  {12}},\ \bibinfo {pages} {1673} (\bibinfo {year} {2012})},\ \bibinfo {note}
  {pMID: 22360155}\BibitemShut {NoStop}%
\bibitem [{\citenamefont {Yu}\ \emph {et~al.}(2010{\natexlab{a}})\citenamefont
  {Yu}, \citenamefont {Kanazawa}, \citenamefont {Onose}, \citenamefont
  {Kimoto}, \citenamefont {Zhang}, \citenamefont {Ishiwata}, \citenamefont
  {Matsui},\ and\ \citenamefont {Tokura}}]{FeGe}%
  \BibitemOpen
  \bibfield  {author} {\bibinfo {author} {\bibfnamefont {X.~Z.}\ \bibnamefont
  {Yu}}, \bibinfo {author} {\bibfnamefont {N.}~\bibnamefont {Kanazawa}},
  \bibinfo {author} {\bibfnamefont {Y.}~\bibnamefont {Onose}}, \bibinfo
  {author} {\bibfnamefont {K.}~\bibnamefont {Kimoto}}, \bibinfo {author}
  {\bibfnamefont {W.~Z.}\ \bibnamefont {Zhang}}, \bibinfo {author}
  {\bibfnamefont {S.}~\bibnamefont {Ishiwata}}, \bibinfo {author}
  {\bibfnamefont {Y.}~\bibnamefont {Matsui}}, \ and\ \bibinfo {author}
  {\bibfnamefont {Y.}~\bibnamefont {Tokura}},\ }\href {\doibase
  10.1038/nmat2916} {\bibfield  {journal} {\bibinfo  {journal} {Nature
  Materials}\ }\textbf {\bibinfo {volume} {10}},\ \bibinfo {pages} {106}
  (\bibinfo {year} {2010}{\natexlab{a}})}\BibitemShut {NoStop}%
\bibitem [{\citenamefont {Yu}\ \emph {et~al.}(2010{\natexlab{b}})\citenamefont
  {Yu}, \citenamefont {Onose}, \citenamefont {Kanazawa}, \citenamefont {Park},
  \citenamefont {Han}, \citenamefont {Matsui}, \citenamefont {Nagaosa},\ and\
  \citenamefont {Tokura}}]{FeCoSi1}%
  \BibitemOpen
  \bibfield  {author} {\bibinfo {author} {\bibfnamefont {X.~Z.}\ \bibnamefont
  {Yu}}, \bibinfo {author} {\bibfnamefont {Y.}~\bibnamefont {Onose}}, \bibinfo
  {author} {\bibfnamefont {N.}~\bibnamefont {Kanazawa}}, \bibinfo {author}
  {\bibfnamefont {J.~H.}\ \bibnamefont {Park}}, \bibinfo {author}
  {\bibfnamefont {J.~H.}\ \bibnamefont {Han}}, \bibinfo {author} {\bibfnamefont
  {Y.}~\bibnamefont {Matsui}}, \bibinfo {author} {\bibfnamefont
  {N.}~\bibnamefont {Nagaosa}}, \ and\ \bibinfo {author} {\bibfnamefont
  {Y.}~\bibnamefont {Tokura}},\ }\href {\doibase 10.1038/nature09124}
  {\bibfield  {journal} {\bibinfo  {journal} {Nature}\ }\textbf {\bibinfo
  {volume} {465}},\ \bibinfo {pages} {901} (\bibinfo {year}
  {2010}{\natexlab{b}})}\BibitemShut {NoStop}%
\bibitem [{\citenamefont {M\"unzer}\ \emph {et~al.}(2010)\citenamefont
  {M\"unzer}, \citenamefont {Neubauer}, \citenamefont {Adams}, \citenamefont
  {M\"uhlbauer}, \citenamefont {Franz}, \citenamefont {Jonietz}, \citenamefont
  {Georgii}, \citenamefont {B\"oni}, \citenamefont {Pedersen}, \citenamefont
  {Schmidt}, \citenamefont {Rosch},\ and\ \citenamefont
  {Pfleiderer}}]{FeCoSi2}%
  \BibitemOpen
  \bibfield  {author} {\bibinfo {author} {\bibfnamefont {W.}~\bibnamefont
  {M\"unzer}}, \bibinfo {author} {\bibfnamefont {A.}~\bibnamefont {Neubauer}},
  \bibinfo {author} {\bibfnamefont {T.}~\bibnamefont {Adams}}, \bibinfo
  {author} {\bibfnamefont {S.}~\bibnamefont {M\"uhlbauer}}, \bibinfo {author}
  {\bibfnamefont {C.}~\bibnamefont {Franz}}, \bibinfo {author} {\bibfnamefont
  {F.}~\bibnamefont {Jonietz}}, \bibinfo {author} {\bibfnamefont
  {R.}~\bibnamefont {Georgii}}, \bibinfo {author} {\bibfnamefont
  {P.}~\bibnamefont {B\"oni}}, \bibinfo {author} {\bibfnamefont
  {B.}~\bibnamefont {Pedersen}}, \bibinfo {author} {\bibfnamefont
  {M.}~\bibnamefont {Schmidt}}, \bibinfo {author} {\bibfnamefont
  {A.}~\bibnamefont {Rosch}}, \ and\ \bibinfo {author} {\bibfnamefont
  {C.}~\bibnamefont {Pfleiderer}},\ }\href {\doibase
  10.1103/PhysRevB.81.041203} {\bibfield  {journal} {\bibinfo  {journal} {Phys.
  Rev. B}\ }\textbf {\bibinfo {volume} {81}},\ \bibinfo {pages} {041203}
  (\bibinfo {year} {2010})}\BibitemShut {NoStop}%
\bibitem [{\citenamefont {Seki}\ \emph {et~al.}(2012)\citenamefont {Seki},
  \citenamefont {Yu}, \citenamefont {Ishiwata},\ and\ \citenamefont
  {Tokura}}]{CuOSeO}%
  \BibitemOpen
  \bibfield  {author} {\bibinfo {author} {\bibfnamefont {S.}~\bibnamefont
  {Seki}}, \bibinfo {author} {\bibfnamefont {X.~Z.}\ \bibnamefont {Yu}},
  \bibinfo {author} {\bibfnamefont {S.}~\bibnamefont {Ishiwata}}, \ and\
  \bibinfo {author} {\bibfnamefont {Y.}~\bibnamefont {Tokura}},\ }\href
  {\doibase 10.1126/science.1214143} {\bibfield  {journal} {\bibinfo  {journal}
  {Science}\ }\textbf {\bibinfo {volume} {336}},\ \bibinfo {pages} {198}
  (\bibinfo {year} {2012})}\BibitemShut {NoStop}%
\bibitem [{\citenamefont {Heinze}\ \emph {et~al.}(2011)\citenamefont {Heinze},
  \citenamefont {von Bergmann}, \citenamefont {Menzel}, \citenamefont {Brede},
  \citenamefont {Kubetzka}, \citenamefont {Wiesendanger}, \citenamefont
  {Bihlmayer},\ and\ \citenamefont {Bl{\"u}gel}}]{Fe_film}%
  \BibitemOpen
  \bibfield  {author} {\bibinfo {author} {\bibfnamefont {S.}~\bibnamefont
  {Heinze}}, \bibinfo {author} {\bibfnamefont {K.}~\bibnamefont {von
  Bergmann}}, \bibinfo {author} {\bibfnamefont {M.}~\bibnamefont {Menzel}},
  \bibinfo {author} {\bibfnamefont {J.}~\bibnamefont {Brede}}, \bibinfo
  {author} {\bibfnamefont {A.}~\bibnamefont {Kubetzka}}, \bibinfo {author}
  {\bibfnamefont {R.}~\bibnamefont {Wiesendanger}}, \bibinfo {author}
  {\bibfnamefont {G.}~\bibnamefont {Bihlmayer}}, \ and\ \bibinfo {author}
  {\bibfnamefont {S.}~\bibnamefont {Bl{\"u}gel}},\ }\href@noop {} {\bibfield
  {journal} {\bibinfo  {journal} {Nature Physics}\ }\textbf {\bibinfo {volume}
  {7}},\ \bibinfo {pages} {713} (\bibinfo {year} {2011})}\BibitemShut {NoStop}%
\bibitem [{\citenamefont {Lee}\ \emph {et~al.}(2009)\citenamefont {Lee},
  \citenamefont {Kang}, \citenamefont {Onose}, \citenamefont {Tokura},\ and\
  \citenamefont {Ong}}]{THE1}%
  \BibitemOpen
  \bibfield  {author} {\bibinfo {author} {\bibfnamefont {M.}~\bibnamefont
  {Lee}}, \bibinfo {author} {\bibfnamefont {W.}~\bibnamefont {Kang}}, \bibinfo
  {author} {\bibfnamefont {Y.}~\bibnamefont {Onose}}, \bibinfo {author}
  {\bibfnamefont {Y.}~\bibnamefont {Tokura}}, \ and\ \bibinfo {author}
  {\bibfnamefont {N.~P.}\ \bibnamefont {Ong}},\ }\href {\doibase
  10.1103/PhysRevLett.102.186601} {\bibfield  {journal} {\bibinfo  {journal}
  {Phys. Rev. Lett.}\ }\textbf {\bibinfo {volume} {102}},\ \bibinfo {pages}
  {186601} (\bibinfo {year} {2009})}\BibitemShut {NoStop}%
\bibitem [{\citenamefont {Neubauer}\ \emph {et~al.}(2009)\citenamefont
  {Neubauer}, \citenamefont {Pfleiderer}, \citenamefont {Binz}, \citenamefont
  {Rosch}, \citenamefont {Ritz}, \citenamefont {Niklowitz},\ and\ \citenamefont
  {B\"oni}}]{THE2}%
  \BibitemOpen
  \bibfield  {author} {\bibinfo {author} {\bibfnamefont {A.}~\bibnamefont
  {Neubauer}}, \bibinfo {author} {\bibfnamefont {C.}~\bibnamefont
  {Pfleiderer}}, \bibinfo {author} {\bibfnamefont {B.}~\bibnamefont {Binz}},
  \bibinfo {author} {\bibfnamefont {A.}~\bibnamefont {Rosch}}, \bibinfo
  {author} {\bibfnamefont {R.}~\bibnamefont {Ritz}}, \bibinfo {author}
  {\bibfnamefont {P.~G.}\ \bibnamefont {Niklowitz}}, \ and\ \bibinfo {author}
  {\bibfnamefont {P.}~\bibnamefont {B\"oni}},\ }\href {\doibase
  10.1103/PhysRevLett.102.186602} {\bibfield  {journal} {\bibinfo  {journal}
  {Phys. Rev. Lett.}\ }\textbf {\bibinfo {volume} {102}},\ \bibinfo {pages}
  {186602} (\bibinfo {year} {2009})}\BibitemShut {NoStop}%
\bibitem [{\citenamefont {Zang}\ \emph {et~al.}(2011)\citenamefont {Zang},
  \citenamefont {Mostovoy}, \citenamefont {Han},\ and\ \citenamefont
  {Nagaosa}}]{Zang}%
  \BibitemOpen
  \bibfield  {author} {\bibinfo {author} {\bibfnamefont {J.}~\bibnamefont
  {Zang}}, \bibinfo {author} {\bibfnamefont {M.}~\bibnamefont {Mostovoy}},
  \bibinfo {author} {\bibfnamefont {J.~H.}\ \bibnamefont {Han}}, \ and\
  \bibinfo {author} {\bibfnamefont {N.}~\bibnamefont {Nagaosa}},\ }\href
  {\doibase 10.1103/PhysRevLett.107.136804} {\bibfield  {journal} {\bibinfo
  {journal} {Phys. Rev. Lett.}\ }\textbf {\bibinfo {volume} {107}},\ \bibinfo
  {pages} {136804} (\bibinfo {year} {2011})}\BibitemShut {NoStop}%
\bibitem [{\citenamefont {Everschor}\ \emph {et~al.}(2012)\citenamefont
  {Everschor}, \citenamefont {Garst}, \citenamefont {Binz}, \citenamefont
  {Jonietz}, \citenamefont {M\"uhlbauer}, \citenamefont {Pfleiderer},\ and\
  \citenamefont {Rosch}}]{SkHall}%
  \BibitemOpen
  \bibfield  {author} {\bibinfo {author} {\bibfnamefont {K.}~\bibnamefont
  {Everschor}}, \bibinfo {author} {\bibfnamefont {M.}~\bibnamefont {Garst}},
  \bibinfo {author} {\bibfnamefont {B.}~\bibnamefont {Binz}}, \bibinfo {author}
  {\bibfnamefont {F.}~\bibnamefont {Jonietz}}, \bibinfo {author} {\bibfnamefont
  {S.}~\bibnamefont {M\"uhlbauer}}, \bibinfo {author} {\bibfnamefont
  {C.}~\bibnamefont {Pfleiderer}}, \ and\ \bibinfo {author} {\bibfnamefont
  {A.}~\bibnamefont {Rosch}},\ }\href {\doibase 10.1103/PhysRevB.86.054432}
  {\bibfield  {journal} {\bibinfo  {journal} {Phys. Rev. B}\ }\textbf {\bibinfo
  {volume} {86}},\ \bibinfo {pages} {054432} (\bibinfo {year}
  {2012})}\BibitemShut {NoStop}%
\bibitem [{\citenamefont {Ritz}\ \emph {et~al.}(2013)\citenamefont {Ritz},
  \citenamefont {Halder}, \citenamefont {Wagner}, \citenamefont {Franz},
  \citenamefont {Bauer},\ and\ \citenamefont {Pfleiderer}}]{NFL}%
  \BibitemOpen
  \bibfield  {author} {\bibinfo {author} {\bibfnamefont {R.}~\bibnamefont
  {Ritz}}, \bibinfo {author} {\bibfnamefont {M.}~\bibnamefont {Halder}},
  \bibinfo {author} {\bibfnamefont {M.}~\bibnamefont {Wagner}}, \bibinfo
  {author} {\bibfnamefont {C.}~\bibnamefont {Franz}}, \bibinfo {author}
  {\bibfnamefont {A.}~\bibnamefont {Bauer}}, \ and\ \bibinfo {author}
  {\bibfnamefont {C.}~\bibnamefont {Pfleiderer}},\ }\href {\doibase
  10.1038/nature12023} {\bibfield  {journal} {\bibinfo  {journal} {Nature}\
  }\textbf {\bibinfo {volume} {497}},\ \bibinfo {pages} {231} (\bibinfo {year}
  {2013})}\BibitemShut {NoStop}%
\bibitem [{\citenamefont {Jonietz}\ \emph {et~al.}(2010)\citenamefont
  {Jonietz}, \citenamefont {M\"uhlbauer}, \citenamefont {Pfleiderer},
  \citenamefont {Neubauer}, \citenamefont {M\"unzer}, \citenamefont {Bauer},
  \citenamefont {Adams}, \citenamefont {Georgii}, \citenamefont {B\"oni},
  \citenamefont {Duine}, \citenamefont {Everschor}, \citenamefont {Garst},\
  and\ \citenamefont {Rosch}}]{ultralow1}%
  \BibitemOpen
  \bibfield  {author} {\bibinfo {author} {\bibfnamefont {F.}~\bibnamefont
  {Jonietz}}, \bibinfo {author} {\bibfnamefont {S.}~\bibnamefont
  {M\"uhlbauer}}, \bibinfo {author} {\bibfnamefont {C.}~\bibnamefont
  {Pfleiderer}}, \bibinfo {author} {\bibfnamefont {A.}~\bibnamefont
  {Neubauer}}, \bibinfo {author} {\bibfnamefont {W.}~\bibnamefont {M\"unzer}},
  \bibinfo {author} {\bibfnamefont {A.}~\bibnamefont {Bauer}}, \bibinfo
  {author} {\bibfnamefont {T.}~\bibnamefont {Adams}}, \bibinfo {author}
  {\bibfnamefont {R.}~\bibnamefont {Georgii}}, \bibinfo {author} {\bibfnamefont
  {P.}~\bibnamefont {B\"oni}}, \bibinfo {author} {\bibfnamefont {R.~A.}\
  \bibnamefont {Duine}}, \bibinfo {author} {\bibfnamefont {K.}~\bibnamefont
  {Everschor}}, \bibinfo {author} {\bibfnamefont {M.}~\bibnamefont {Garst}}, \
  and\ \bibinfo {author} {\bibfnamefont {A.}~\bibnamefont {Rosch}},\ }\href
  {\doibase 10.1126/science.1195709} {\bibfield  {journal} {\bibinfo  {journal}
  {Science}\ }\textbf {\bibinfo {volume} {330}},\ \bibinfo {pages} {1648}
  (\bibinfo {year} {2010})}\BibitemShut {NoStop}%
\bibitem [{\citenamefont {Yu}\ \emph {et~al.}(2012)\citenamefont {Yu},
  \citenamefont {Kanazawa}, \citenamefont {Zhang}, \citenamefont {Nagai},
  \citenamefont {Hara}, \citenamefont {Kimoto}, \citenamefont {Matsui},
  \citenamefont {Onose},\ and\ \citenamefont {Tokura}}]{ultralow2}%
  \BibitemOpen
  \bibfield  {author} {\bibinfo {author} {\bibfnamefont {X.}~\bibnamefont
  {Yu}}, \bibinfo {author} {\bibfnamefont {N.}~\bibnamefont {Kanazawa}},
  \bibinfo {author} {\bibfnamefont {W.}~\bibnamefont {Zhang}}, \bibinfo
  {author} {\bibfnamefont {T.}~\bibnamefont {Nagai}}, \bibinfo {author}
  {\bibfnamefont {T.}~\bibnamefont {Hara}}, \bibinfo {author} {\bibfnamefont
  {K.}~\bibnamefont {Kimoto}}, \bibinfo {author} {\bibfnamefont
  {Y.}~\bibnamefont {Matsui}}, \bibinfo {author} {\bibfnamefont
  {Y.}~\bibnamefont {Onose}}, \ and\ \bibinfo {author} {\bibfnamefont
  {Y.}~\bibnamefont {Tokura}},\ }\href {\doibase 10.1038/ncomms1990} {\bibfield
   {journal} {\bibinfo  {journal} {Nature Communications}\ }\textbf {\bibinfo
  {volume} {3}},\ \bibinfo {pages} {988} (\bibinfo {year} {2012})}\BibitemShut
  {NoStop}%
\bibitem [{\citenamefont {Hamamoto}\ \emph {et~al.}(2015)\citenamefont
  {Hamamoto}, \citenamefont {Ezawa},\ and\ \citenamefont {Nagaosa}}]{Hamamoto}%
  \BibitemOpen
  \bibfield  {author} {\bibinfo {author} {\bibfnamefont {K.}~\bibnamefont
  {Hamamoto}}, \bibinfo {author} {\bibfnamefont {M.}~\bibnamefont {Ezawa}}, \
  and\ \bibinfo {author} {\bibfnamefont {N.}~\bibnamefont {Nagaosa}},\ }\href
  {\doibase 10.1103/PhysRevB.92.115417} {\bibfield  {journal} {\bibinfo
  {journal} {Phys. Rev. B}\ }\textbf {\bibinfo {volume} {92}},\ \bibinfo
  {pages} {115417} (\bibinfo {year} {2015})}\BibitemShut {NoStop}%
\bibitem [{\citenamefont {Romming}\ \emph {et~al.}(2013)\citenamefont
  {Romming}, \citenamefont {Hanneken}, \citenamefont {Menzel}, \citenamefont
  {Bickel}, \citenamefont {Wolter}, \citenamefont {von Bergmann}, \citenamefont
  {Kubetzka},\ and\ \citenamefont {Wiesendanger}}]{memofunc0}%
  \BibitemOpen
  \bibfield  {author} {\bibinfo {author} {\bibfnamefont {N.}~\bibnamefont
  {Romming}}, \bibinfo {author} {\bibfnamefont {C.}~\bibnamefont {Hanneken}},
  \bibinfo {author} {\bibfnamefont {M.}~\bibnamefont {Menzel}}, \bibinfo
  {author} {\bibfnamefont {J.~E.}\ \bibnamefont {Bickel}}, \bibinfo {author}
  {\bibfnamefont {B.}~\bibnamefont {Wolter}}, \bibinfo {author} {\bibfnamefont
  {K.}~\bibnamefont {von Bergmann}}, \bibinfo {author} {\bibfnamefont
  {A.}~\bibnamefont {Kubetzka}}, \ and\ \bibinfo {author} {\bibfnamefont
  {R.}~\bibnamefont {Wiesendanger}},\ }\href {\doibase 10.1126/science.1240573}
  {\bibfield  {journal} {\bibinfo  {journal} {Science}\ }\textbf {\bibinfo
  {volume} {341}},\ \bibinfo {pages} {636} (\bibinfo {year}
  {2013})}\BibitemShut {NoStop}%
\bibitem [{\citenamefont {Koshibae}\ \emph {et~al.}(2015)\citenamefont
  {Koshibae}, \citenamefont {Kaneko}, \citenamefont {Iwasaki}, \citenamefont
  {Kawasaki}, \citenamefont {Tokura},\ and\ \citenamefont
  {Nagaosa}}]{memofunc1}%
  \BibitemOpen
  \bibfield  {author} {\bibinfo {author} {\bibfnamefont {W.}~\bibnamefont
  {Koshibae}}, \bibinfo {author} {\bibfnamefont {Y.}~\bibnamefont {Kaneko}},
  \bibinfo {author} {\bibfnamefont {J.}~\bibnamefont {Iwasaki}}, \bibinfo
  {author} {\bibfnamefont {M.}~\bibnamefont {Kawasaki}}, \bibinfo {author}
  {\bibfnamefont {Y.}~\bibnamefont {Tokura}}, \ and\ \bibinfo {author}
  {\bibfnamefont {N.}~\bibnamefont {Nagaosa}},\ }\href
  {http://stacks.iop.org/1347-4065/54/i=5/a=053001} {\bibfield  {journal}
  {\bibinfo  {journal} {Japanese Journal of Applied Physics}\ }\textbf
  {\bibinfo {volume} {54}},\ \bibinfo {pages} {053001} (\bibinfo {year}
  {2015})}\BibitemShut {NoStop}%
\bibitem [{\citenamefont {Tomasello}\ \emph {et~al.}(2014)\citenamefont
  {Tomasello}, \citenamefont {Martinez}, \citenamefont {Zivieri}, \citenamefont
  {Torres}, \citenamefont {Carpentieri},\ and\ \citenamefont
  {Finocchio}}]{memofunc2}%
  \BibitemOpen
  \bibfield  {author} {\bibinfo {author} {\bibfnamefont {R.}~\bibnamefont
  {Tomasello}}, \bibinfo {author} {\bibfnamefont {E.}~\bibnamefont {Martinez}},
  \bibinfo {author} {\bibfnamefont {R.}~\bibnamefont {Zivieri}}, \bibinfo
  {author} {\bibfnamefont {L.}~\bibnamefont {Torres}}, \bibinfo {author}
  {\bibfnamefont {M.}~\bibnamefont {Carpentieri}}, \ and\ \bibinfo {author}
  {\bibfnamefont {G.}~\bibnamefont {Finocchio}},\ }\href {\doibase
  10.1038/srep06784} {\bibfield  {journal} {\bibinfo  {journal} {Sci. Rep.}\
  }\textbf {\bibinfo {volume} {4}},\ \bibinfo {pages} {6784} (\bibinfo {year}
  {2014})}\BibitemShut {NoStop}%
\bibitem [{\citenamefont {Zhang}\ \emph {et~al.}(2015)\citenamefont {Zhang},
  \citenamefont {Ezawa},\ and\ \citenamefont {Zhou}}]{memofunc3}%
  \BibitemOpen
  \bibfield  {author} {\bibinfo {author} {\bibfnamefont {X.}~\bibnamefont
  {Zhang}}, \bibinfo {author} {\bibfnamefont {M.}~\bibnamefont {Ezawa}}, \ and\
  \bibinfo {author} {\bibfnamefont {Y.}~\bibnamefont {Zhou}},\ }\href {\doibase
  10.1038/srep09400} {\bibfield  {journal} {\bibinfo  {journal} {Sci. Rep.}\
  }\textbf {\bibinfo {volume} {5}},\ \bibinfo {pages} {9400} (\bibinfo {year}
  {2015})}\BibitemShut {NoStop}%
\bibitem [{\citenamefont {Koshibae}\ and\ \citenamefont
  {Nagaosa}(2014)}]{heating}%
  \BibitemOpen
  \bibfield  {author} {\bibinfo {author} {\bibfnamefont {W.}~\bibnamefont
  {Koshibae}}\ and\ \bibinfo {author} {\bibfnamefont {N.}~\bibnamefont
  {Nagaosa}},\ }\href {\doibase 10.1038/ncomms6148} {\bibfield  {journal}
  {\bibinfo  {journal} {Nature Communications}\ }\textbf {\bibinfo {volume}
  {5}},\ \bibinfo {pages} {5148} (\bibinfo {year} {2014})}\BibitemShut
  {NoStop}%
\bibitem [{\citenamefont {Iwasaki}\ \emph {et~al.}(2013)\citenamefont
  {Iwasaki}, \citenamefont {Mochizuki},\ and\ \citenamefont
  {Nagaosa}}]{Iwasaki2}%
  \BibitemOpen
  \bibfield  {author} {\bibinfo {author} {\bibfnamefont {J.}~\bibnamefont
  {Iwasaki}}, \bibinfo {author} {\bibfnamefont {M.}~\bibnamefont {Mochizuki}},
  \ and\ \bibinfo {author} {\bibfnamefont {N.}~\bibnamefont {Nagaosa}},\ }\href
  {\doibase 10.1038/nnano.2013.176} {\bibfield  {journal} {\bibinfo  {journal}
  {Nature Nanotechnology}\ }\textbf {\bibinfo {volume} {8}},\ \bibinfo {pages}
  {742} (\bibinfo {year} {2013})}\BibitemShut {NoStop}%
\bibitem [{\citenamefont {Jiang}\ \emph {et~al.}(2015)\citenamefont {Jiang},
  \citenamefont {Upadhyaya}, \citenamefont {Zhang}, \citenamefont {Yu},
  \citenamefont {Jungfleisch}, \citenamefont {Fradin}, \citenamefont {Pearson},
  \citenamefont {Tserkovnyak}, \citenamefont {Wang}, \citenamefont {Heinonen},
  \citenamefont {te~Velthuis},\ and\ \citenamefont {Hoffmann}}]{creation0}%
  \BibitemOpen
  \bibfield  {author} {\bibinfo {author} {\bibfnamefont {W.}~\bibnamefont
  {Jiang}}, \bibinfo {author} {\bibfnamefont {P.}~\bibnamefont {Upadhyaya}},
  \bibinfo {author} {\bibfnamefont {W.}~\bibnamefont {Zhang}}, \bibinfo
  {author} {\bibfnamefont {G.}~\bibnamefont {Yu}}, \bibinfo {author}
  {\bibfnamefont {M.~B.}\ \bibnamefont {Jungfleisch}}, \bibinfo {author}
  {\bibfnamefont {F.~Y.}\ \bibnamefont {Fradin}}, \bibinfo {author}
  {\bibfnamefont {J.~E.}\ \bibnamefont {Pearson}}, \bibinfo {author}
  {\bibfnamefont {Y.}~\bibnamefont {Tserkovnyak}}, \bibinfo {author}
  {\bibfnamefont {K.~L.}\ \bibnamefont {Wang}}, \bibinfo {author}
  {\bibfnamefont {O.}~\bibnamefont {Heinonen}}, \bibinfo {author}
  {\bibfnamefont {S.~G.~E.}\ \bibnamefont {te~Velthuis}}, \ and\ \bibinfo
  {author} {\bibfnamefont {A.}~\bibnamefont {Hoffmann}},\ }\href {\doibase
  10.1126/science.aaa1442} {\bibfield  {journal} {\bibinfo  {journal}
  {Science}\ }\textbf {\bibinfo {volume} {349}},\ \bibinfo {pages} {283}
  (\bibinfo {year} {2015})}\BibitemShut {NoStop}%
\bibitem [{\citenamefont {Milde}\ \emph {et~al.}(2013)\citenamefont {Milde},
  \citenamefont {K\"ohler}, \citenamefont {Seidel}, \citenamefont {Eng},
  \citenamefont {Bauer}, \citenamefont {Chacon}, \citenamefont {Kindervater},
  \citenamefont {M\"uhlbauer}, \citenamefont {Pfleiderer}, \citenamefont
  {Buhrandt}, \citenamefont {Sch\"utte},\ and\ \citenamefont
  {Rosch}}]{SkMerge}%
  \BibitemOpen
  \bibfield  {author} {\bibinfo {author} {\bibfnamefont {P.}~\bibnamefont
  {Milde}}, \bibinfo {author} {\bibfnamefont {D.}~\bibnamefont {K\"ohler}},
  \bibinfo {author} {\bibfnamefont {J.}~\bibnamefont {Seidel}}, \bibinfo
  {author} {\bibfnamefont {L.~M.}\ \bibnamefont {Eng}}, \bibinfo {author}
  {\bibfnamefont {A.}~\bibnamefont {Bauer}}, \bibinfo {author} {\bibfnamefont
  {A.}~\bibnamefont {Chacon}}, \bibinfo {author} {\bibfnamefont
  {J.}~\bibnamefont {Kindervater}}, \bibinfo {author} {\bibfnamefont
  {S.}~\bibnamefont {M\"uhlbauer}}, \bibinfo {author} {\bibfnamefont
  {C.}~\bibnamefont {Pfleiderer}}, \bibinfo {author} {\bibfnamefont
  {S.}~\bibnamefont {Buhrandt}}, \bibinfo {author} {\bibfnamefont
  {C.}~\bibnamefont {Sch\"utte}}, \ and\ \bibinfo {author} {\bibfnamefont
  {A.}~\bibnamefont {Rosch}},\ }\href {\doibase 10.1126/science.1234657}
  {\bibfield  {journal} {\bibinfo  {journal} {Science}\ }\textbf {\bibinfo
  {volume} {340}},\ \bibinfo {pages} {1076} (\bibinfo {year}
  {2013})}\BibitemShut {NoStop}%
\bibitem [{\citenamefont {Nagaosa}\ and\ \citenamefont {Tokura}(2012)}]{EEMF0}%
  \BibitemOpen
  \bibfield  {author} {\bibinfo {author} {\bibfnamefont {N.}~\bibnamefont
  {Nagaosa}}\ and\ \bibinfo {author} {\bibfnamefont {Y.}~\bibnamefont
  {Tokura}},\ }\href {\doibase 10.1088/0031-8949/2012/t146/014020} {\bibfield
  {journal} {\bibinfo  {journal} {Physica Scripta}\ }\textbf {\bibinfo {volume}
  {2012}},\ \bibinfo {pages} {014020} (\bibinfo {year} {2012})}\BibitemShut
  {NoStop}%
\bibitem [{\citenamefont {Schulz}\ \emph {et~al.}(2012)\citenamefont {Schulz},
  \citenamefont {Ritz}, \citenamefont {Bauer}, \citenamefont {Halder},
  \citenamefont {Wagner}, \citenamefont {Franz}, \citenamefont {Pfleiderer},
  \citenamefont {Everschor}, \citenamefont {Garst},\ and\ \citenamefont
  {Rosch}}]{EEMF1}%
  \BibitemOpen
  \bibfield  {author} {\bibinfo {author} {\bibfnamefont {T.}~\bibnamefont
  {Schulz}}, \bibinfo {author} {\bibfnamefont {R.}~\bibnamefont {Ritz}},
  \bibinfo {author} {\bibfnamefont {A.}~\bibnamefont {Bauer}}, \bibinfo
  {author} {\bibfnamefont {M.}~\bibnamefont {Halder}}, \bibinfo {author}
  {\bibfnamefont {M.}~\bibnamefont {Wagner}}, \bibinfo {author} {\bibfnamefont
  {C.}~\bibnamefont {Franz}}, \bibinfo {author} {\bibfnamefont
  {C.}~\bibnamefont {Pfleiderer}}, \bibinfo {author} {\bibfnamefont
  {K.}~\bibnamefont {Everschor}}, \bibinfo {author} {\bibfnamefont
  {M.}~\bibnamefont {Garst}}, \ and\ \bibinfo {author} {\bibfnamefont
  {A.}~\bibnamefont {Rosch}},\ }\href {\doibase 10.1038/nphys2231} {\bibfield
  {journal} {\bibinfo  {journal} {Nature Physics}\ }\textbf {\bibinfo {volume}
  {8}},\ \bibinfo {pages} {301} (\bibinfo {year} {2012})}\BibitemShut {NoStop}%
\bibitem [{\citenamefont {Nagaosa}\ \emph {et~al.}(2012)\citenamefont
  {Nagaosa}, \citenamefont {Yu},\ and\ \citenamefont {Tokura}}]{EEMF2}%
  \BibitemOpen
  \bibfield  {author} {\bibinfo {author} {\bibfnamefont {N.}~\bibnamefont
  {Nagaosa}}, \bibinfo {author} {\bibfnamefont {X.~Z.}\ \bibnamefont {Yu}}, \
  and\ \bibinfo {author} {\bibfnamefont {Y.}~\bibnamefont {Tokura}},\ }\href
  {\doibase 10.1098/rsta.2011.0405} {\bibfield  {journal} {\bibinfo  {journal}
  {Philosophical Transactions of the Royal Society of London A: Mathematical,
  Physical and Engineering Sciences}\ }\textbf {\bibinfo {volume} {370}},\
  \bibinfo {pages} {5806} (\bibinfo {year} {2012})}\BibitemShut {NoStop}%
\bibitem [{\citenamefont {{Cao}}\ and\ \citenamefont
  {{Jiang}}(2016)}]{monopolegrowth}%
  \BibitemOpen
  \bibfield  {author} {\bibinfo {author} {\bibfnamefont {J.}~\bibnamefont
  {{Cao}}}\ and\ \bibinfo {author} {\bibfnamefont {Y.}~\bibnamefont
  {{Jiang}}},\ }\href {https://arxiv.org/abs/1607.07782} {\bibfield  {journal}
  {\bibinfo  {journal} {ArXiv e-prints}\ } (\bibinfo {year} {2016})},\ \Eprint
  {http://arxiv.org/abs/1607.07782} {arXiv:1607.07782 [cond-mat.mes-hall]}
  \BibitemShut {NoStop}%
\bibitem [{\citenamefont {Takashima}\ and\ \citenamefont
  {Fujimoto}(2014)}]{monopole3}%
  \BibitemOpen
  \bibfield  {author} {\bibinfo {author} {\bibfnamefont {R.}~\bibnamefont
  {Takashima}}\ and\ \bibinfo {author} {\bibfnamefont {S.}~\bibnamefont
  {Fujimoto}},\ }\href {\doibase 10.7566/JPSJ.83.054717} {\bibfield  {journal}
  {\bibinfo  {journal} {Journal of the Physical Society of Japan}\ }\textbf
  {\bibinfo {volume} {83}},\ \bibinfo {pages} {054717} (\bibinfo {year}
  {2014})}\BibitemShut {NoStop}%
\bibitem [{\citenamefont {Sch\"utte}\ and\ \citenamefont
  {Rosch}(2014)}]{monopole1}%
  \BibitemOpen
  \bibfield  {author} {\bibinfo {author} {\bibfnamefont {C.}~\bibnamefont
  {Sch\"utte}}\ and\ \bibinfo {author} {\bibfnamefont {A.}~\bibnamefont
  {Rosch}},\ }\href {\doibase 10.1103/PhysRevB.90.174432} {\bibfield  {journal}
  {\bibinfo  {journal} {Phys. Rev. B}\ }\textbf {\bibinfo {volume} {90}},\
  \bibinfo {pages} {174432} (\bibinfo {year} {2014})}\BibitemShut {NoStop}%
\bibitem [{\citenamefont {Lin}\ and\ \citenamefont {Saxena}(2016)}]{monopole2}%
  \BibitemOpen
  \bibfield  {author} {\bibinfo {author} {\bibfnamefont {S.-Z.}\ \bibnamefont
  {Lin}}\ and\ \bibinfo {author} {\bibfnamefont {A.}~\bibnamefont {Saxena}},\
  }\href {\doibase 10.1103/PhysRevB.93.060401} {\bibfield  {journal} {\bibinfo
  {journal} {Phys. Rev. B}\ }\textbf {\bibinfo {volume} {93}},\ \bibinfo
  {pages} {060401} (\bibinfo {year} {2016})}\BibitemShut {NoStop}%
\bibitem [{\citenamefont {Watanabe}\ and\ \citenamefont
  {Vishwanath}(2016)}]{Watanabemonopole}%
  \BibitemOpen
  \bibfield  {author} {\bibinfo {author} {\bibfnamefont {H.}~\bibnamefont
  {Watanabe}}\ and\ \bibinfo {author} {\bibfnamefont {A.}~\bibnamefont
  {Vishwanath}},\ }\href {\doibase 10.7566/JPSJ.85.064707} {\bibfield
  {journal} {\bibinfo  {journal} {Journal of the Physical Society of Japan}\
  }\textbf {\bibinfo {volume} {85}},\ \bibinfo {pages} {064707} (\bibinfo
  {year} {2016})}\BibitemShut {NoStop}%
\bibitem [{\citenamefont {Binz}\ and\ \citenamefont {Vishwanath}(2006)}]{SkX1}%
  \BibitemOpen
  \bibfield  {author} {\bibinfo {author} {\bibfnamefont {B.}~\bibnamefont
  {Binz}}\ and\ \bibinfo {author} {\bibfnamefont {A.}~\bibnamefont
  {Vishwanath}},\ }\href {\doibase 10.1103/PhysRevB.74.214408} {\bibfield
  {journal} {\bibinfo  {journal} {Phys. Rev. B}\ }\textbf {\bibinfo {volume}
  {74}},\ \bibinfo {pages} {214408} (\bibinfo {year} {2006})}\BibitemShut
  {NoStop}%
\bibitem [{\citenamefont {Park}\ and\ \citenamefont {Han}(2011)}]{SkX2}%
  \BibitemOpen
  \bibfield  {author} {\bibinfo {author} {\bibfnamefont {J.-H.}\ \bibnamefont
  {Park}}\ and\ \bibinfo {author} {\bibfnamefont {J.~H.}\ \bibnamefont {Han}},\
  }\href {\doibase 10.1103/PhysRevB.83.184406} {\bibfield  {journal} {\bibinfo
  {journal} {Phys. Rev. B}\ }\textbf {\bibinfo {volume} {83}},\ \bibinfo
  {pages} {184406} (\bibinfo {year} {2011})}\BibitemShut {NoStop}%
\bibitem [{\citenamefont {Kanazawa}\ \emph {et~al.}(2011)\citenamefont
  {Kanazawa}, \citenamefont {Onose}, \citenamefont {Arima}, \citenamefont
  {Okuyama}, \citenamefont {Ohoyama}, \citenamefont {Wakimoto}, \citenamefont
  {Kakurai}, \citenamefont {Ishiwata},\ and\ \citenamefont
  {Tokura}}]{Kanazawa1}%
  \BibitemOpen
  \bibfield  {author} {\bibinfo {author} {\bibfnamefont {N.}~\bibnamefont
  {Kanazawa}}, \bibinfo {author} {\bibfnamefont {Y.}~\bibnamefont {Onose}},
  \bibinfo {author} {\bibfnamefont {T.}~\bibnamefont {Arima}}, \bibinfo
  {author} {\bibfnamefont {D.}~\bibnamefont {Okuyama}}, \bibinfo {author}
  {\bibfnamefont {K.}~\bibnamefont {Ohoyama}}, \bibinfo {author} {\bibfnamefont
  {S.}~\bibnamefont {Wakimoto}}, \bibinfo {author} {\bibfnamefont
  {K.}~\bibnamefont {Kakurai}}, \bibinfo {author} {\bibfnamefont
  {S.}~\bibnamefont {Ishiwata}}, \ and\ \bibinfo {author} {\bibfnamefont
  {Y.}~\bibnamefont {Tokura}},\ }\href {\doibase
  10.1103/PhysRevLett.106.156603} {\bibfield  {journal} {\bibinfo  {journal}
  {Phys. Rev. Lett.}\ }\textbf {\bibinfo {volume} {106}},\ \bibinfo {pages}
  {156603} (\bibinfo {year} {2011})}\BibitemShut {NoStop}%
\bibitem [{\citenamefont {Kanazawa}\ \emph {et~al.}(2012)\citenamefont
  {Kanazawa}, \citenamefont {Kim}, \citenamefont {Inosov}, \citenamefont
  {White}, \citenamefont {Egetenmeyer}, \citenamefont {Gavilano}, \citenamefont
  {Ishiwata}, \citenamefont {Onose}, \citenamefont {Arima}, \citenamefont
  {Keimer},\ and\ \citenamefont {Tokura}}]{Kanazawa2}%
  \BibitemOpen
  \bibfield  {author} {\bibinfo {author} {\bibfnamefont {N.}~\bibnamefont
  {Kanazawa}}, \bibinfo {author} {\bibfnamefont {J.-H.}\ \bibnamefont {Kim}},
  \bibinfo {author} {\bibfnamefont {D.~S.}\ \bibnamefont {Inosov}}, \bibinfo
  {author} {\bibfnamefont {J.~S.}\ \bibnamefont {White}}, \bibinfo {author}
  {\bibfnamefont {N.}~\bibnamefont {Egetenmeyer}}, \bibinfo {author}
  {\bibfnamefont {J.~L.}\ \bibnamefont {Gavilano}}, \bibinfo {author}
  {\bibfnamefont {S.}~\bibnamefont {Ishiwata}}, \bibinfo {author}
  {\bibfnamefont {Y.}~\bibnamefont {Onose}}, \bibinfo {author} {\bibfnamefont
  {T.}~\bibnamefont {Arima}}, \bibinfo {author} {\bibfnamefont
  {B.}~\bibnamefont {Keimer}}, \ and\ \bibinfo {author} {\bibfnamefont
  {Y.}~\bibnamefont {Tokura}},\ }\href {\doibase 10.1103/PhysRevB.86.134425}
  {\bibfield  {journal} {\bibinfo  {journal} {Phys. Rev. B}\ }\textbf {\bibinfo
  {volume} {86}},\ \bibinfo {pages} {134425} (\bibinfo {year}
  {2012})}\BibitemShut {NoStop}%
\bibitem [{\citenamefont {Tanigaki}\ \emph {et~al.}(2015)\citenamefont
  {Tanigaki}, \citenamefont {Shibata}, \citenamefont {Kanazawa}, \citenamefont
  {Yu}, \citenamefont {Onose}, \citenamefont {Park}, \citenamefont {Shindo},\
  and\ \citenamefont {Tokura}}]{Kanazawa3}%
  \BibitemOpen
  \bibfield  {author} {\bibinfo {author} {\bibfnamefont {T.}~\bibnamefont
  {Tanigaki}}, \bibinfo {author} {\bibfnamefont {K.}~\bibnamefont {Shibata}},
  \bibinfo {author} {\bibfnamefont {N.}~\bibnamefont {Kanazawa}}, \bibinfo
  {author} {\bibfnamefont {X.}~\bibnamefont {Yu}}, \bibinfo {author}
  {\bibfnamefont {Y.}~\bibnamefont {Onose}}, \bibinfo {author} {\bibfnamefont
  {H.~S.}\ \bibnamefont {Park}}, \bibinfo {author} {\bibfnamefont
  {D.}~\bibnamefont {Shindo}}, \ and\ \bibinfo {author} {\bibfnamefont
  {Y.}~\bibnamefont {Tokura}},\ }\href {\doibase 10.1021/acs.nanolett.5b02653}
  {\bibfield  {journal} {\bibinfo  {journal} {Nano Letters}\ }\textbf {\bibinfo
  {volume} {15}},\ \bibinfo {pages} {5438} (\bibinfo {year} {2015})},\ \bibinfo
  {note} {pMID: 26237493}\BibitemShut {NoStop}%
\bibitem [{\citenamefont {Kanazawa}\ \emph {et~al.}(2016)\citenamefont
  {Kanazawa}, \citenamefont {Nii}, \citenamefont {Zhang}, \citenamefont
  {Mishchenko}, \citenamefont {Filippis}, \citenamefont {Kagawa}, \citenamefont
  {Iwasa}, \citenamefont {Nagaosa},\ and\ \citenamefont {Tokura}}]{Nii2}%
  \BibitemOpen
  \bibfield  {author} {\bibinfo {author} {\bibfnamefont {N.}~\bibnamefont
  {Kanazawa}}, \bibinfo {author} {\bibfnamefont {Y.}~\bibnamefont {Nii}},
  \bibinfo {author} {\bibfnamefont {X.-X.}\ \bibnamefont {Zhang}}, \bibinfo
  {author} {\bibfnamefont {A.~S.}\ \bibnamefont {Mishchenko}}, \bibinfo
  {author} {\bibfnamefont {G.~D.}\ \bibnamefont {Filippis}}, \bibinfo {author}
  {\bibfnamefont {F.}~\bibnamefont {Kagawa}}, \bibinfo {author} {\bibfnamefont
  {Y.}~\bibnamefont {Iwasa}}, \bibinfo {author} {\bibfnamefont
  {N.}~\bibnamefont {Nagaosa}}, \ and\ \bibinfo {author} {\bibfnamefont
  {Y.}~\bibnamefont {Tokura}},\ }\href {\doibase 10.1038/ncomms11622}
  {\bibfield  {journal} {\bibinfo  {journal} {Nature Communications}\ }\textbf
  {\bibinfo {volume} {7}},\ \bibinfo {pages} {11622} (\bibinfo {year}
  {2016})}\BibitemShut {NoStop}%
\bibitem [{\citenamefont {Nagaosa}\ and\ \citenamefont
  {Tokura}(2013)}]{Review}%
  \BibitemOpen
  \bibfield  {author} {\bibinfo {author} {\bibfnamefont {N.}~\bibnamefont
  {Nagaosa}}\ and\ \bibinfo {author} {\bibfnamefont {Y.}~\bibnamefont
  {Tokura}},\ }\href {\doibase 10.1038/nnano.2013.243} {\bibfield  {journal}
  {\bibinfo  {journal} {Nature Nanotechnology}\ }\textbf {\bibinfo {volume}
  {8}},\ \bibinfo {pages} {899} (\bibinfo {year} {2013})}\BibitemShut {NoStop}%
\bibitem [{\citenamefont {Altland}\ and\ \citenamefont {Simons}(2010)}]{CMFT}%
  \BibitemOpen
  \bibfield  {author} {\bibinfo {author} {\bibfnamefont {A.}~\bibnamefont
  {Altland}}\ and\ \bibinfo {author} {\bibfnamefont {B.~D.}\ \bibnamefont
  {Simons}},\ }\href {http://amazon.com/o/ASIN/0521769752/} {\emph {\bibinfo
  {title} {Condensed Matter Field Theory}}},\ \bibinfo {edition} {2nd}\ ed.\
  (\bibinfo  {publisher} {Cambridge University Press},\ \bibinfo {year}
  {2010})\BibitemShut {NoStop}%
\bibitem [{\citenamefont {Dzyaloshinskii}(1958)}]{Dzyaloshinskii}%
  \BibitemOpen
  \bibfield  {author} {\bibinfo {author} {\bibfnamefont {I.}~\bibnamefont
  {Dzyaloshinskii}},\ }\href {\doibase 10.1016/0022-3697(58)90076-3} {\bibfield
   {journal} {\bibinfo  {journal} {Journal of Physics and Chemistry of Solids}\
  }\textbf {\bibinfo {volume} {4}},\ \bibinfo {pages} {241 } (\bibinfo {year}
  {1958})}\BibitemShut {NoStop}%
\bibitem [{\citenamefont {Moriya}(1960)}]{Moriya}%
  \BibitemOpen
  \bibfield  {author} {\bibinfo {author} {\bibfnamefont {T.}~\bibnamefont
  {Moriya}},\ }\href {\doibase 10.1103/PhysRev.120.91} {\bibfield  {journal}
  {\bibinfo  {journal} {Phys. Rev.}\ }\textbf {\bibinfo {volume} {120}},\
  \bibinfo {pages} {91} (\bibinfo {year} {1960})}\BibitemShut {NoStop}%
\bibitem [{\citenamefont {Fert}\ and\ \citenamefont
  {Levy}(1980)}]{Fert&LevyDMI}%
  \BibitemOpen
  \bibfield  {author} {\bibinfo {author} {\bibfnamefont {A.}~\bibnamefont
  {Fert}}\ and\ \bibinfo {author} {\bibfnamefont {P.~M.}\ \bibnamefont
  {Levy}},\ }\href {\doibase 10.1103/PhysRevLett.44.1538} {\bibfield  {journal}
  {\bibinfo  {journal} {Phys. Rev. Lett.}\ }\textbf {\bibinfo {volume} {44}},\
  \bibinfo {pages} {1538} (\bibinfo {year} {1980})}\BibitemShut {NoStop}%
\bibitem [{\citenamefont {Wu}\ and\ \citenamefont {Yang}(1975)}]{CNYang1}%
  \BibitemOpen
  \bibfield  {author} {\bibinfo {author} {\bibfnamefont {T.~T.}\ \bibnamefont
  {Wu}}\ and\ \bibinfo {author} {\bibfnamefont {C.~N.}\ \bibnamefont {Yang}},\
  }\href {\doibase 10.1103/PhysRevD.12.3845} {\bibfield  {journal} {\bibinfo
  {journal} {Phys. Rev. D}\ }\textbf {\bibinfo {volume} {12}},\ \bibinfo
  {pages} {3845} (\bibinfo {year} {1975})}\BibitemShut {NoStop}%
\bibitem [{\citenamefont {Kim}\ \emph {et~al.}(1994)\citenamefont {Kim},
  \citenamefont {Furusaki}, \citenamefont {Wen},\ and\ \citenamefont
  {Lee}}]{PatrickLee}%
  \BibitemOpen
  \bibfield  {author} {\bibinfo {author} {\bibfnamefont {Y.~B.}\ \bibnamefont
  {Kim}}, \bibinfo {author} {\bibfnamefont {A.}~\bibnamefont {Furusaki}},
  \bibinfo {author} {\bibfnamefont {X.-G.}\ \bibnamefont {Wen}}, \ and\
  \bibinfo {author} {\bibfnamefont {P.~A.}\ \bibnamefont {Lee}},\ }\href
  {\doibase 10.1103/PhysRevB.50.17917} {\bibfield  {journal} {\bibinfo
  {journal} {Phys. Rev. B}\ }\textbf {\bibinfo {volume} {50}},\ \bibinfo
  {pages} {17917} (\bibinfo {year} {1994})}\BibitemShut {NoStop}%
\bibitem [{\citenamefont {G\"otze}\ and\ \citenamefont
  {W\"olfle}(1972)}]{MemoFunc}%
  \BibitemOpen
  \bibfield  {author} {\bibinfo {author} {\bibfnamefont {W.}~\bibnamefont
  {G\"otze}}\ and\ \bibinfo {author} {\bibfnamefont {P.}~\bibnamefont
  {W\"olfle}},\ }\href {\doibase 10.1103/PhysRevB.6.1226} {\bibfield  {journal}
  {\bibinfo  {journal} {Phys. Rev. B}\ }\textbf {\bibinfo {volume} {6}},\
  \bibinfo {pages} {1226} (\bibinfo {year} {1972})}\BibitemShut {NoStop}%
\bibitem [{\citenamefont {Mori}(1965{\natexlab{a}})}]{Mori1}%
  \BibitemOpen
  \bibfield  {author} {\bibinfo {author} {\bibfnamefont {H.}~\bibnamefont
  {Mori}},\ }\href {\doibase 10.1143/PTP.33.423} {\bibfield  {journal}
  {\bibinfo  {journal} {Progress of Theoretical Physics}\ }\textbf {\bibinfo
  {volume} {33}},\ \bibinfo {pages} {423} (\bibinfo {year}
  {1965}{\natexlab{a}})}\BibitemShut {NoStop}%
\bibitem [{\citenamefont {Mori}(1965{\natexlab{b}})}]{Mori2}%
  \BibitemOpen
  \bibfield  {author} {\bibinfo {author} {\bibfnamefont {H.}~\bibnamefont
  {Mori}},\ }\href {\doibase 10.1143/PTP.34.399} {\bibfield  {journal}
  {\bibinfo  {journal} {Progress of Theoretical Physics}\ }\textbf {\bibinfo
  {volume} {34}},\ \bibinfo {pages} {399} (\bibinfo {year}
  {1965}{\natexlab{b}})}\BibitemShut {NoStop}%
\bibitem [{\citenamefont {Mahan}(2000)}]{Mahan}%
  \BibitemOpen
  \bibfield  {author} {\bibinfo {author} {\bibfnamefont {G.~D.}\ \bibnamefont
  {Mahan}},\ }\href {http://amazon.com/o/ASIN/0306463385/} {\emph {\bibinfo
  {title} {Many-Particle Physics (Physics of Solids and Liquids)}}},\ \bibinfo
  {edition} {3rd}\ ed.\ (\bibinfo  {publisher} {Springer},\ \bibinfo {year}
  {2000})\BibitemShut {NoStop}%
\bibitem [{\citenamefont {Jarrell}\ and\ \citenamefont
  {Gubernatis}(1996)}]{BayesianInference}%
  \BibitemOpen
  \bibfield  {author} {\bibinfo {author} {\bibfnamefont {M.}~\bibnamefont
  {Jarrell}}\ and\ \bibinfo {author} {\bibfnamefont {J.}~\bibnamefont
  {Gubernatis}},\ }\href {\doibase
  http://dx.doi.org/10.1016/0370-1573(95)00074-7} {\bibfield  {journal}
  {\bibinfo  {journal} {Physics Reports}\ }\textbf {\bibinfo {volume} {269}},\
  \bibinfo {pages} {133 } (\bibinfo {year} {1996})}\BibitemShut {NoStop}%
\bibitem [{\citenamefont {Mishchenko}\ \emph {et~al.}(2000)\citenamefont
  {Mishchenko}, \citenamefont {Prokof'ev}, \citenamefont {Sakamoto},\ and\
  \citenamefont {Svistunov}}]{Mishchenko1}%
  \BibitemOpen
  \bibfield  {author} {\bibinfo {author} {\bibfnamefont {A.~S.}\ \bibnamefont
  {Mishchenko}}, \bibinfo {author} {\bibfnamefont {N.~V.}\ \bibnamefont
  {Prokof'ev}}, \bibinfo {author} {\bibfnamefont {A.}~\bibnamefont {Sakamoto}},
  \ and\ \bibinfo {author} {\bibfnamefont {B.~V.}\ \bibnamefont {Svistunov}},\
  }\href {\doibase 10.1103/PhysRevB.62.6317} {\bibfield  {journal} {\bibinfo
  {journal} {Phys. Rev. B}\ }\textbf {\bibinfo {volume} {62}},\ \bibinfo
  {pages} {6317} (\bibinfo {year} {2000})}\BibitemShut {NoStop}%
\bibitem [{\citenamefont {Mishchenko}\ \emph {et~al.}(2015)\citenamefont
  {Mishchenko}, \citenamefont {Nagaosa}, \citenamefont {De~Filippis},
  \citenamefont {de~Candia},\ and\ \citenamefont {Cataudella}}]{Mishchenko2}%
  \BibitemOpen
  \bibfield  {author} {\bibinfo {author} {\bibfnamefont {A.~S.}\ \bibnamefont
  {Mishchenko}}, \bibinfo {author} {\bibfnamefont {N.}~\bibnamefont {Nagaosa}},
  \bibinfo {author} {\bibfnamefont {G.}~\bibnamefont {De~Filippis}}, \bibinfo
  {author} {\bibfnamefont {A.}~\bibnamefont {de~Candia}}, \ and\ \bibinfo
  {author} {\bibfnamefont {V.}~\bibnamefont {Cataudella}},\ }\href {\doibase
  10.1103/PhysRevLett.114.146401} {\bibfield  {journal} {\bibinfo  {journal}
  {Phys. Rev. Lett.}\ }\textbf {\bibinfo {volume} {114}},\ \bibinfo {pages}
  {146401} (\bibinfo {year} {2015})}\BibitemShut {NoStop}%
\bibitem [{\citenamefont {Mishchenko}(2012)}]{Mishchenko3}%
  \BibitemOpen
  \bibfield  {author} {\bibinfo {author} {\bibfnamefont {A.~S.}\ \bibnamefont
  {Mishchenko}},\ }in\ \href
  {http://www.cond-mat.de/events/correl12/manuscripts/} {\emph {\bibinfo
  {booktitle} {Correlated Electrons: From Models to Materials}}},\ \bibinfo
  {editor} {edited by\ \bibinfo {editor} {\bibfnamefont {E.}~\bibnamefont
  {Pavarini}}, \bibinfo {editor} {\bibfnamefont {E.}~\bibnamefont {Koch}},
  \bibinfo {editor} {\bibfnamefont {F.}~\bibnamefont {Anders}}, \ and\ \bibinfo
  {editor} {\bibfnamefont {M.}~\bibnamefont {Jarrell}}}\ (\bibinfo  {publisher}
  {Forschungszentrum Julich},\ \bibinfo {address} {Julich},\ \bibinfo {year}
  {2012})\BibitemShut {NoStop}%
\bibitem [{\citenamefont {Prokof'ev}\ and\ \citenamefont
  {Svistunov}(2013)}]{ConsistentConstraint}%
  \BibitemOpen
  \bibfield  {author} {\bibinfo {author} {\bibfnamefont {N.~V.}\ \bibnamefont
  {Prokof'ev}}\ and\ \bibinfo {author} {\bibfnamefont {B.~V.}\ \bibnamefont
  {Svistunov}},\ }\href {\doibase 10.1134/s002136401311009x} {\bibfield
  {journal} {\bibinfo  {journal} {Jetp Lett.}\ }\textbf {\bibinfo {volume}
  {97}},\ \bibinfo {pages} {649} (\bibinfo {year} {2013})}\BibitemShut
  {NoStop}%
\bibitem [{\citenamefont {Nagaosa}(1999)}]{Nagaosa}%
  \BibitemOpen
  \bibfield  {author} {\bibinfo {author} {\bibfnamefont {N.}~\bibnamefont
  {Nagaosa}},\ }\href {http://amazon.com/o/ASIN/3540655379/} {\emph {\bibinfo
  {title} {Quantum Field Theory in Condensed Matter Physics (Theoretical and
  Mathematical Physics)}}},\ \bibinfo {edition} {1999th}\ ed.\ (\bibinfo
  {publisher} {Springer},\ \bibinfo {year} {1999})\BibitemShut {NoStop}%
\bibitem [{\citenamefont {Press}\ \emph {et~al.}(1992)\citenamefont {Press},
  \citenamefont {Flannery}, \citenamefont {Teukolsky},\ and\ \citenamefont
  {Vetterling}}]{NumericalRecipes}%
  \BibitemOpen
  \bibfield  {author} {\bibinfo {author} {\bibfnamefont {W.~H.}\ \bibnamefont
  {Press}}, \bibinfo {author} {\bibfnamefont {B.~P.}\ \bibnamefont {Flannery}},
  \bibinfo {author} {\bibfnamefont {S.~A.}\ \bibnamefont {Teukolsky}}, \ and\
  \bibinfo {author} {\bibfnamefont {W.~T.}\ \bibnamefont {Vetterling}},\ }\href
  {http://amazon.com/o/ASIN/0521431085/} {\emph {\bibinfo {title} {Numerical
  Recipes in C: The Art of Scientific Computing, Second Edition}}},\ \bibinfo
  {edition} {2nd}\ ed.\ (\bibinfo  {publisher} {Cambridge University Press},\
  \bibinfo {year} {1992})\BibitemShut {NoStop}%
\bibitem [{\citenamefont {Gough}(2009)}]{GSL}%
  \BibitemOpen
  \bibinfo {editor} {\bibfnamefont {B.}~\bibnamefont {Gough}},\ ed.,\ \href
  {http://amazon.com/o/ASIN/0954612078/} {\emph {\bibinfo {title} {GNU
  Scientific Library Reference Manual - Third Edition}}},\ \bibinfo {edition}
  {3rd}\ ed.\ (\bibinfo  {publisher} {Network Theory Ltd.},\ \bibinfo {year}
  {2009})\BibitemShut {NoStop}%
\bibitem [{\citenamefont {Genz}\ and\ \citenamefont {Malik}(1980)}]{Genz1}%
  \BibitemOpen
  \bibfield  {author} {\bibinfo {author} {\bibfnamefont {A.}~\bibnamefont
  {Genz}}\ and\ \bibinfo {author} {\bibfnamefont {A.}~\bibnamefont {Malik}},\
  }\href {\doibase http://dx.doi.org/10.1016/0771-050X(80)90039-X} {\bibfield
  {journal} {\bibinfo  {journal} {Journal of Computational and Applied
  Mathematics}\ }\textbf {\bibinfo {volume} {6}},\ \bibinfo {pages} {295 }
  (\bibinfo {year} {1980})}\BibitemShut {NoStop}%
\bibitem [{\citenamefont {Berntsen}\ \emph {et~al.}(1991)\citenamefont
  {Berntsen}, \citenamefont {Espelid},\ and\ \citenamefont {Genz}}]{Genz2}%
  \BibitemOpen
  \bibfield  {author} {\bibinfo {author} {\bibfnamefont {J.}~\bibnamefont
  {Berntsen}}, \bibinfo {author} {\bibfnamefont {T.~O.}\ \bibnamefont
  {Espelid}}, \ and\ \bibinfo {author} {\bibfnamefont {A.}~\bibnamefont
  {Genz}},\ }\href {\doibase 10.1145/210232.210233} {\bibfield  {journal}
  {\bibinfo  {journal} {ACM Trans. Math. Softw.}\ }\textbf {\bibinfo {volume}
  {17}},\ \bibinfo {pages} {437} (\bibinfo {year} {1991})}\BibitemShut
  {NoStop}%
\bibitem [{\citenamefont {Johnson}(2015)}]{Cubature}%
  \BibitemOpen
  \bibfield  {author} {\bibinfo {author} {\bibfnamefont {S.~G.}\ \bibnamefont
  {Johnson}},\ }\href {http://ab-initio.mit.edu/cubature} {\enquote {\bibinfo
  {title} {Cubature package},}\ } (\bibinfo {year} {accessed Apr.,
  2015})\BibitemShut {NoStop}%
\bibitem [{\citenamefont {Watanabe}\ \emph {et~al.}(2014)\citenamefont
  {Watanabe}, \citenamefont {Parameswaran}, \citenamefont {Raghu},\ and\
  \citenamefont {Vishwanath}}]{WatanabeSkX}%
  \BibitemOpen
  \bibfield  {author} {\bibinfo {author} {\bibfnamefont {H.}~\bibnamefont
  {Watanabe}}, \bibinfo {author} {\bibfnamefont {S.~A.}\ \bibnamefont
  {Parameswaran}}, \bibinfo {author} {\bibfnamefont {S.}~\bibnamefont {Raghu}},
  \ and\ \bibinfo {author} {\bibfnamefont {A.}~\bibnamefont {Vishwanath}},\
  }\href {\doibase 10.1103/PhysRevB.90.045145} {\bibfield  {journal} {\bibinfo
  {journal} {Phys. Rev. B}\ }\textbf {\bibinfo {volume} {90}},\ \bibinfo
  {pages} {045145} (\bibinfo {year} {2014})}\BibitemShut {NoStop}%
\bibitem [{\citenamefont {Watanabe}\ and\ \citenamefont
  {Vishwanath}(2014)}]{WatanabeNFL}%
  \BibitemOpen
  \bibfield  {author} {\bibinfo {author} {\bibfnamefont {H.}~\bibnamefont
  {Watanabe}}\ and\ \bibinfo {author} {\bibfnamefont {A.}~\bibnamefont
  {Vishwanath}},\ }\href {\doibase 10.1073/pnas.1415592111} {\bibfield
  {journal} {\bibinfo  {journal} {Proceedings of the National Academy of
  Sciences}\ }\textbf {\bibinfo {volume} {111}},\ \bibinfo {pages} {16314}
  (\bibinfo {year} {2014})}\BibitemShut {NoStop}%
\bibitem [{\citenamefont {Slichter}(1996)}]{Slichter}%
  \BibitemOpen
  \bibfield  {author} {\bibinfo {author} {\bibfnamefont {C.~P.}\ \bibnamefont
  {Slichter}},\ }\href {http://amazon.com/o/ASIN/3540501576/} {\emph {\bibinfo
  {title} {Principles of Magnetic Resonance (Springer Series in Solid-State
  Sciences) (v. 1)}}},\ \bibinfo {edition} {3rd}\ ed.\ (\bibinfo  {publisher}
  {Springer},\ \bibinfo {year} {1996})\BibitemShut {NoStop}%
\end{thebibliography}%

\end{document}